\documentclass[prx,reprint,twocolumn,showpacs,superscriptaddress]{revtex4-1}
\usepackage{amsmath,amssymb,bm,mathrsfs,graphicx, braket,amsthm,bbm}
\usepackage[colorlinks=true,citecolor=blue,linkcolor=blue]{hyperref}
\usepackage[all,cmtip]{xy}
\usepackage[usenames,dvipsnames]{color}			% Only used for author comments
\usepackage{ulem}
\renewcommand{\vec}[1]{{\bf{#1}}}
\newcommand{\vechat}[1]{\hat{\bm{#1}}}

\begin{document}
\title{Chiral Floquet Phases of Many-body Localized Bosons}

\author{Hoi Chun Po}
\affiliation{Department of Physics, University of California, Berkeley, CA 94720, USA}
\affiliation{Department of Physics, Harvard University, Cambridge MA 02138, USA}

\author{Lukasz Fidkowski}
\affiliation{Department of Physics and Astronomy, Stony Brook University, Stony Brook, NY 11794, USA}

\author{Takahiro Morimoto}
\affiliation{Department of Physics, University of California, Berkeley, CA 94720, USA}

\author{Andrew C. Potter}
\affiliation{Department of Physics, University of Texas at Austin, Austin, TX 78712, USA}

\author{Ashvin Vishwanath}
\affiliation{Department of Physics, University of California, Berkeley, CA 94720, USA}
\affiliation{Department of Physics, Harvard University, Cambridge MA 02138, USA}

\begin{abstract} 
We construct and classify chiral topological phases in driven (Floquet) systems of strongly interacting bosons, with finite-dimensional site Hilbert spaces, in two spatial dimensions. The construction proceeds by introducing exactly soluble models with chiral edges, which in the presence of many-body localization (MBL) in the bulk are argued to lead to stable chiral phases. These chiral phases do not require any symmetry, and in fact owe their existence to the absence of energy conservation in driven systems.  Surprisingly, we show that they are classified by a quantized many-body index, which is well defined for any MBL Floquet system.  The value of this index, which is always the logarithm of a positive rational number, can be interpreted as the entropy per Floquet cycle pumped along the edge, formalizing the notion of quantum-information flow.  We explicitly compute this index for specific models, and show that the nontrivial topology leads to edge thermalization, which provides an interesting link between bulk topology and chaos at the edge. We also discuss chiral Floquet phases in interacting fermionic systems and their relation to chiral bosonic phases.
\end{abstract}

\maketitle

\section{Introduction}
\label{sec:Intro}
Topological phases are typically discussed in terms of the ground state properties of gapped  Hamiltonians. One of the earliest examples is the integer quantum Hall (IQH) effect , where a quantized Hall conductance is established on cooling to low temperatures that are well below the gap scale.  The IQH insulator is an example of a short-range entangled (SRE) topological phase, defined as having a unique ground state on closed manifolds. \footnote{
Some authors define SRE phases as systems whose ground state can be deformed to a product state while maintaining a bulk gap.\cite{XieGuLiuWen} The two definitions are inequivalent, with the mismatch in 2d being the chiral phases.
}  
A well known class of SRE phases are the symmetry-protected topological (SPT) phases, which require the presence of a protecting symmetry, examples of which include the electronic topological insulators. In contrast, the IQH insulators exemplify a more basic class of 2d `chiral' SRE phases whose nontrivial nature persists even in the absence of any symmetry.  These nontrivial chiral phases are fully characterized by the existence of chiral edge modes, and can be diagnosed by their quantized thermal Hall conductance,\cite{FisherKane} which is proportional to the chiral central charge `c' of the effective edge conformal field theory. In particular, the chiral central charge is quantized in units of $c_* = 1/2$ ($c_*=8$) for fermions (bosons) with no additional symmetry.\cite{KitaevHoneycomb,LuVishwanath, MulliganNayak}

More recently it has been realized that, outside of ground state physics, topological phases can also appear in the highly excited states of a many-body system.\cite{Bahri,Chandran,Nandkishore}  An essential ingredient in this scenario is many-body localization, which allows for a description of states in terms of conserved local integrals of motion.\cite{VoskAltman,Serbyn,HuseOganesyan}  This prevents thermalization and endows the excited states with properties similar to those of the ground states of gapped systems.\footnote{
Note that the robustness of many-body localization in 2d is still an open conjecture. Regardless of the validity of the conjecture, MBL phases will nonetheless display non-thermal dynamics in a parametrically long time scale, which by itself is of great physical interest.
}
These distinctive properties can be observed in the dynamics of simple initial states, without the need for cooling, so that in addition to their intrinsic interest, many-body localized (MBL) topological phases also possess practical advantages in terms of realizability.  Although many SPT phases have been realized in this MBL context,\cite{Bahri,Chandran,TarantinoFidkowski,PotterAV} the more fundamental 2d chiral topological phases have so far evaded such a realization.  Indeed, it was argued in Refs.~[\onlinecite{KitaevHoneycomb}], [\onlinecite{PotterAV}] and [\onlinecite{LevinThermalHall}] that such chiral phases are not permitted in excited states, because the resulting system would then be unstable to thermalization.  

A key ingredient in the argument of Refs.~[\onlinecite{KitaevHoneycomb}], [\onlinecite{PotterAV}] and [\onlinecite{LevinThermalHall}] is the inability of a system characterized by local integrals of motion to support a quantized thermal Hall conductance.  On the other hand, it has also been understood recently that MBL systems can be stable to the introduction of a time-dependent, periodic Floquet drive,\cite{FloquetMBL2, FloquetMBL1, FloquetMBL3} and this opens up the possibility for the realization of new topological phases in MBL Floquet systems.\cite{Khemani,Keyserlingk, ElseNayak, PotterMorimotoVishwanath,Roy}
Such a periodic drive renders the thermal Hall conductance ambiguous, since the energy is not conserved, and allows for the possibility of realizing chiral MBL phases in Floquet systems.

Here we will argue that chiral phases, which are intrinsically dynamical in nature and do not have static MBL counterparts, can indeed be realized in periodically driven MBL systems. 
This should be differentiated from various Floquet engineering proposals, which aim at effectively realizing equilibrium phases through periodic driving.\cite{Lindner_FTI, Lindner_3DTF}
The key observation we need is that, despite the absence of a chiral central charge or any other conserved quantity, the notion of a chiral edge is still tenable in such systems.  The quantity being transported at the edge is quantum information, which is omnipresent in quantum systems and can be pumped much like conserved charge or energy.\cite{GNVW}  Specifically, we construct a class of exactly solved spin models (termed `bosonic chiral Floquet' models) which, despite being localized in the bulk, possess such chiral edges.

These models are bosonic analogues of the `Anomalous Floquet-Anderson Insulator' (AFAI), introduced in the insightful works of Refs.~[\onlinecite{Demler,RudnerLevin,TitumLindner,Lindner}], where unidirectional edge states appear in conjunction with an Anderson localized bulk.  However, the stability of these free fermion systems in the presence of interactions remains uncertain, since previous works have not identified an intrinsically many-body index that is quantized in these models; in particular, the topology of these models was linked to certain winding numbers, which are explicitly single-particle properties.  By contrast we will characterize our bosonic chiral Floquet models by an intrinsically interacting many-body topological index.
The topological invariant we use was first discussed in Ref.~[\onlinecite{GNVW}] (henceforth referred to as `GNVW'), in which the index was developed to exhaustively classify 1d quantum cellular automata.  In this work, we will bridge the quantum information perspective in GNVW to the physical problems of bosonic Floquet systems in 2d.  Interestingly, the topological invariant is not just an integer, as one might expect from the familiar IQH classification, but rather takes the form $\nu = \log (p/q)$, where $p/q$ is a positive rational number and depends on the dimensions of the local Hilbert spaces.

A feature of the MBL Floquet problems is the notion of a generic edge phase.  In contrast to MBL Floquet SPT phases, where the edges can be localized following  spontaneous symmetry breaking, the absence of symmetry requirements implies that the edge of an MBL chiral Floquet phase can never be localized.  This indicates that bulk topology can enforce chaotic dynamics at the edge, which is supported by our numerical computation on a model of the chiral Floquet edge.

While our discussion will mostly focus on bosonic systems, where rigorous results are available, we will also shed some light on fermionic problems. We present strong evidence that minimal fermionic chiral Floquet phase is topologically equivalent to the bosonic counterpart. This is in stark contrast to the equilibrium quantum Hall phases, where the minimal bosonic integer chiral phase is equivalent to $8$ copies of the minimal fermionic one.

The plan of attack in this paper will be as follows. We will introduce a class of exactly soluble bosonic Floquet models (the `SWAP models'), and construct a topological model of interacting bosons that features the 1d translation operator at the edge (Sec.~\ref{sec:Model}). Leveraging GNVW's results on the classification of 1d locality-preserving unitary operators, we argue that this edge dynamics cannot be realized in any purely 1d system undergoing finite-time Hamiltonian evolution, and hence signifies a topological 2d bulk.  Having analyzed the SWAP models, we will abstract their essential features and show that they serve as a complete set of representatives for 2d bosonic MBL chiral Floquet phases. This is achieved by first developing a sharp notion of bulk-boundary correspondence in MBL Floquet systems (Sec.~\ref{sec:Edge}), including subtleties associated with the robustness of many-body localization in 2d, and then discussing in depth the interpretation, properties, and classification structure of the topological index in GNVW, which will be referred to as the `chiral unitary index' and denoted by $\nu$ (Sec.~\ref{sec:GNVW}).  After addressing the formal aspects of the MBL chiral Floquet phases, we will demonstrate the explicit computability of $\nu$ via a matrix-product representation (Sec.~\ref{sec:Num_MPU}), discuss an experimental proposal for realization using hardcore bosons in a shaken optical lattice (Sec.~\ref{sec:ExpP}), and elaborate on the physical consequences associated with the anomalous chiral edges of these models (Sec.~\ref{sec:PhysCon}). 
We will conclude by making connection between the bosonic and fermionic problems (Sec.~\ref{sec:Fermions}), and then discussing future directions of research motivated by the results in this work (Sec.~\ref{sec:Dis}).

\section{Bosonic Chiral Floquet Models}
\label{sec:Model}
The goal of this section is to construct an infinite set of Floquet models in 2d bosonic spin systems, which we will refer to as `bosonic chiral Floquet' models.  
For concreteness, we will first analyze in detail a model of spin-1/2's in subsection \ref{sec:BaseModel}, in which the bulk and edge degrees of freedom (DOF) are manifestly decoupled, and the edge Floquet operator is simply given by the 1d translation operator (also frequently called a `shift', as in GNVW).
We will then generalize the model to an infinite set of bosonic chiral Floquet models in subsection \ref{sec:GenModel}, which show similar edge dynamics as the spin-1/2 model. 
We will then argue heuristically in \ref{sec:tAnomaly} that some of these models are topologically nontrivial, as their edges, the 1d translation operators, are anomalous.

\subsection{Example of a bosonic chiral Floquet model}
\label{sec:BaseModel}
The model we construct can be viewed as a bosonic version of the fermionic AFAI model presented in Refs.~[\onlinecite{RudnerLevin}] and [\onlinecite{Lindner}], where the fermions are replaced by hardcore bosons, and each fermionic `hop' is replaced by a bosonic SWAP gate. 
In anticipation of the generalizations in subsequent subsections, we we will view the hardcore bosons as spin-1/2's.

More concretely, let $\Lambda$ be a checkerboard lattice with two (square) sub-lattices $A$ and $B$.
We will view $\Lambda$ as rotated $45$ degrees from the horizontal, and choose primitive lattice vectors $\vechat x$, $\vechat y$ oriented in the $x$ and $y$ directions respectively.  We can then view $\Lambda$ as a crystal lattice with a two-site unit cell, with basis vectors $\vec 0$ for $A$ and $(\vechat x + \vechat y)/2$ for $B$.  Each site will host a spin-$1/2$.  For concreteness, let us take a large rectangular system with size $N_x \times N_y$, so that the total number of sites is $|\Lambda| \equiv 2 N_x N_y$.

\begin{figure}[h]
\begin{center}
{\includegraphics[width=0.45 \textwidth]{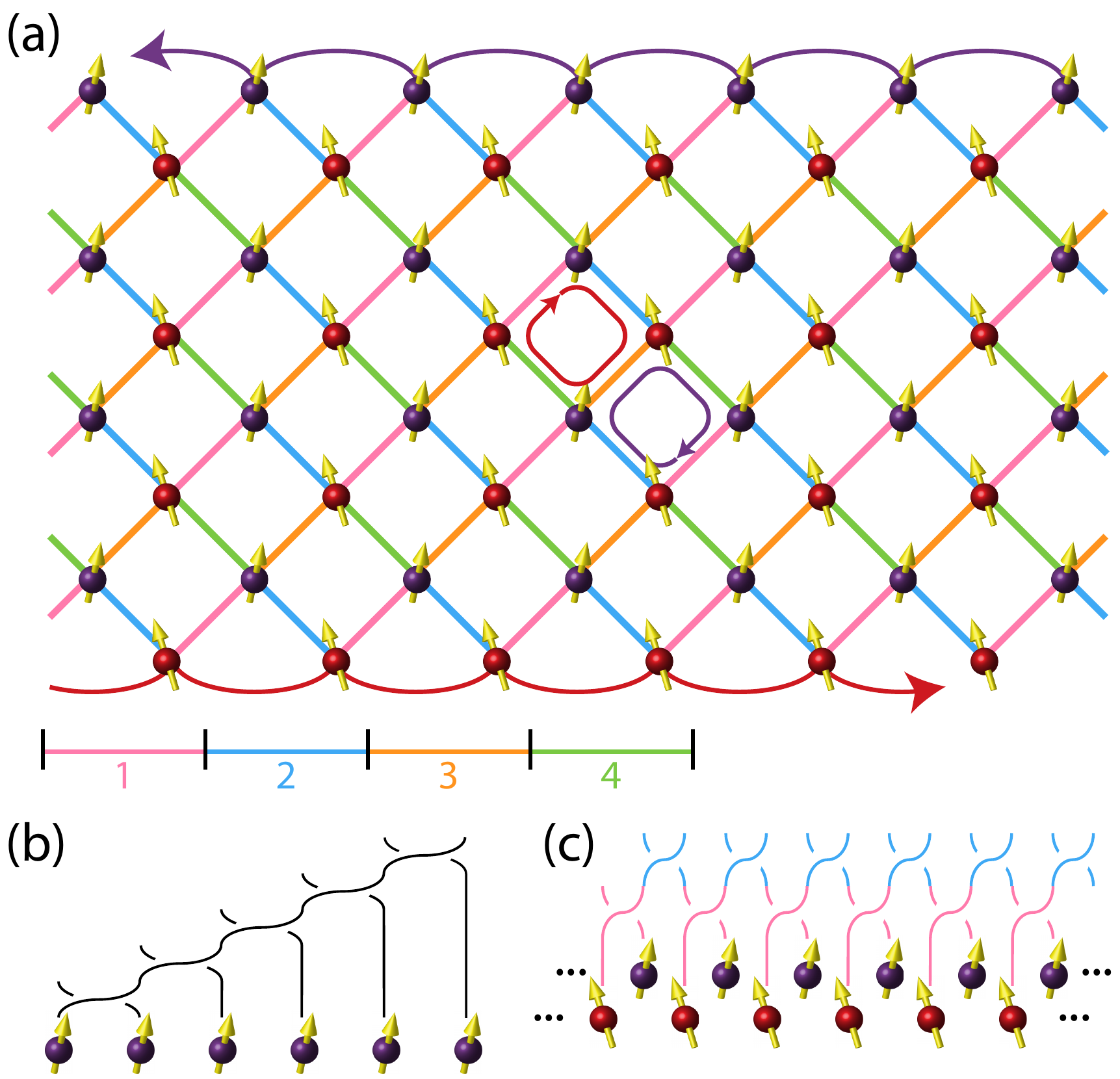}} 
\caption{{\bf Various SWAP circuits considered.} (a) A Floquet phase with counter-propagating translation operators localized at the two edges can be built by applying four layers of SWAP gates, as indicated by $1$ to $4$.
(b) The translation operator for the six-site example has depth five. Generally it will have an infinite circuit depth in the thermodynamic limit. (c) A pair of counter-propagating translation operators, however, can  be realized with depth two.
}
\label{fig:SWAP}
\end{center}

\end{figure}

Our Hamiltonian will be a piecewise constant function of time, with four distinct time steps, each of duration $T/4$.  The four corresponding Hamiltonians will be denoted $\hat H_{(s)}$, $s=1,\ldots, 4$, so that $\hat H(t) = \hat H_{(s)}$ for $(s-1)T/4 \leq t \leq sT/4$.  For each $s$, we define
\begin{align}\label{eq:Hs_Spin}
\hat H&_{(s)} = \frac{  \pi}{ T} \sum_{\vec r }\left(
\hat{\vec{S}}_{\vec r_{\rm B}+\vec b_s}\cdot \hat{\vec{S}}_{\vec r_{\rm A}}
 - \hat 1 \right),
\end{align}
where  $\hat{\vec{S}}_{\vec r_{\rm A}} = \{ \hat X_{\vec x}, \hat Y_{\vec x},\hat Z_{\vec x}\}$ denote the Pauli operators at site $\vec x\in \Lambda$, $\vec r_{\rm A}$ and $\vec r_{\rm B}$ respectively denote the coordinates of the A and B sites in the unit cell $\vec r$, and 
\begin{equation}\begin{split}\label{eq:}
\vec b_{1} = \vec 0; ~~\vec b_{2} = - \vechat x; ~~~ \vec b_{3} = -\vechat  x - \vechat y;~~~ \vec b_{4} = -\vechat  y.
\end{split}\end{equation}
By construction, all terms in $\hat H_{(s)}$ commute and so the time-evolution operator $\exp(-i \hat H_{(s)} T/4)$ for a single time step can be readily computed:
\begin{equation}\begin{split}\label{eq:SWAPdef}
\hat U_{(s)}\equiv&
\exp\left (-i \hat H_{(s)} T/4 \right) = \prod_{\vec r} \hat \chi_{\vec r_{\rm B}+\vec b_s, \vec r_{\rm A}};\\
\hat \chi_{\vec x, \vec y}=&
\frac{1}{2} \left( \hat 1+\hat{\vec{S}}_{\vec x}\cdot \hat{\vec{S}}_{\vec y}\right),
\end{split}\end{equation}
where $\hat \chi_{\vec x,\vec y}$ is simply the `SWAP' gate between the spin-1/2 DOF at sites $\vec x$ and $\vec y$. 
Hence, the total Floquet operator can be viewed as a quantum circuit built entirely of SWAP gates, and is merely a permutation of the lattice sites. 
We denote it by $\hat U_{\rm F} \equiv \hat P_{\rm F} \equiv \hat P_{(4)}\ldots \hat P_{(1)} $ to emphasize that each step is a site permutation. 
Due to its permutation nature, a SWAP circuit is exactly soluble, as we discuss in Appendix \ref{sec:SWAP}.
In particular we will see below that with periodic boundary conditions (PBC), $\hat U_{\rm F}$ is just the identity.

To compute the Floquet operator, which is a site permutation, it is enough to work in the `single-spin-flip sector', defined as the subspace spanned by the orthonormal set of single-spin-flip states $\{ | \vec r_\mu \rangle \equiv \hat S^{+}_{\vec r_\mu}| 0\rangle\}$, where $|0  \rangle \equiv |\downarrow \downarrow \dots \downarrow\rangle$, and $S^{+}_{\vec r_\mu} = (X_{\vec r_\mu} + i Y_{\vec r_\mu})$ is the spin-raising operator at site $\vec r_\mu$.  Restricted to this subspace, $P_{\rm F} \equiv \langle \vec r_\mu|\hat P_{\rm F} |\vec r'_{\mu'}\rangle$  is a $|\Lambda|$-dimensional matrix, and can be efficiently analyzed.
Intuitively, one can compute $P_{\rm F}$ by simply `hopping' the spin-flip following the SWAP gates in the Floquet cycle.
With PBC, one sees that any spin-flip circles an adjacent plaquette and returns to its starting position after the fourth step (Fig.~\ref{fig:SWAP}a).  This implies $P_{\rm F} = 1_{|\Lambda|}$, and hence $\hat U_{\rm F}=\hat P_{\rm F}=\hat 1$.

Although the Floquet operator is apparently trivial with PBC, it could still be topologically nontrivial and display protected, anomalous edge dynamics when the system is under open boundary conditions (OBC).
Explicitly, we consider a cylindrical geometry periodic in $x$ but open in $y$. 
This opens up two circular edges respectively consisting of the A sites at $y = 1$ and B sites at $y = N_y$.
The computation of the Floquet operator with OBC, $\hat U_{\rm F}' \equiv \hat P_{\rm F}'$, proceeds as before.
In the bulk, i.e.~for $y\neq 1$ and $N_y$, one simply finds $\hat P'_{\rm F} | _{\text{Bulk}} = \hat 1$, as none of the SWAP gates acting on the bulk sites have been affected.  On the other hand, for a spin flip starting at, say, $(x,y=1)_A$, the SWAP gates for the third and fourth time step have been `deleted', so in the single-spin-flip sector
\begin{equation}\begin{split}\label{eq:Bose_Edge}
P'_{\rm F} | (x, 1)_{\rm A}\rangle = P'_{(2)}  P'_{(1)} | (x, 1)_{\rm A}\rangle = | (x+1, 1)_{\rm A}\rangle,
\end{split}\end{equation}
and hence $\hat P'_{\rm F}$ permutes the sites $(x,1)_{A}$ in the same way as the unit right-translation operator along the boundary, $\hat t_{y=1;A}$. Similarly, for a spin flip starting at $(x, N_y)_B$, we have
\begin{equation}\begin{split}\label{eq:}
P'_{\rm F} | (x,  N_y)_{\rm B}\rangle = | (x-1,  N_y)_{\rm B}\rangle,
\end{split}\end{equation}
which corresponds to the action of the unit left-translation operator $(\hat t_{y= N_y;B})^{-1}$.
Altogether, one finds
\begin{equation}\begin{split}\label{eq:Pprime_F}
\hat U'_{\rm F} =\hat P'_{\rm F} = \hat t_{y=1;A} \otimes (\hat t_{y= N_y;B})^{-1},
\end{split}\end{equation}
where the bulk and edge DOF are explicitly decoupled. Hence, despite the apparently trivial bulk Floquet operator, the dynamics of the system is nontrivial at the edge, and is governed by the 1d translation operator.

\subsection{General classes of bosonic chiral Floquet models}
\label{sec:GenModel}
The only place we used the spin-1/2 nature of the model in the previous analysis was Eq.~\eqref{eq:SWAPdef}, which can be easily generalized to the case of spin $(p-1)/2$ sites for any $p\geq1$.
In other words, we can replace the two-dimensional site DOF with $p$-dimensional ones and obtain an infinite class of bosonic chiral Floquet models. At the lower edge ($y=1$), each of these models gives the edge unitary operator $\hat Y = \hat t^{(p)}$, the unit right-translation operator for a chain with $p$ dimensional site Hilbert spaces.
Note, however, that spin-0 DOF ($p=1$) are special, as in that case the full many-body Hilbert space is one-dimensional and hence any model is necessarily trivial.
(For simplicity, we will drop the superscript $(p)$ when the Hilbert space dimension is not emphasized, with $\hat t$ understood to be $\hat t^{(p)}$ for some $p>1$.)

A further generalization is obtained by taking any of the above models and performing the four time steps in reverse.  This results in a mirror image model, with the edge unitary operator $\hat Y$ now given by the reverse translations.  Even more generally, one can consider a `stacked' model with two independent layers: a model with sites of dimension $p$ stacked on top of the mirror image of a model with sites of dimension $q$. This gives $\hat Y = \hat t^{(p)} \otimes \overline{ \hat t^{(q)}}$, where we let $\overline{ \hat t^{(q)}} \equiv (\hat t^{(q)})^{-1}$ for clarity.

After the construction of this infinite set of models, a natural question is to ask if these models are all distinct, i.e.~given positive integers $p,q,p', q'$, is the $(p,q)$ model equivalent to the $(p',q')$ model? 
At this stage, however, it is not even clear what the word `equivalent' means, since the usual notion of bulk-boundary correspondence in equilibrium systems cannot be applied to our Floquet setup. To even pose the question, one must first develop a corresponding notion in Floquet systems.

We will undertake this task in Sec.~\ref{sec:Edge}, where we show that the bulk and boundary dynamics can be systematically decoupled in any MBL Floquet system. Granted such decoupling, one can classify MBL Floquet systems by studying smooth deformations among their boundaries. 
Before proceeding with this analysis, however, it is instructive to first examine the classification in a heuristic fashion, using the fact that the $(p,q)$ models constructed here showcase explicit bulk-edge decoupling.  
Specifically, in Sec.~\ref{sec:GNVW}, we will see that in general the $(p,q)$ model is equivalent to the $(p',q')$ model whenever $p/q=p'/q'$. Before that, we will present below a heuristic argument suggesting that the $(p,1)$ models are nontrivial for $p>1$, while the $(p,p)$ models are always trivial.

\subsection{Anomaly of the translation operator}
\label{sec:tAnomaly}
Our intuition with the bulk-boundary correspondence dictates that we should identify obstructions to `smoothly' deforming the boundary operator $\hat Y$ to the trivial one $\hat Y = \hat 1$, i.e.~obstructions in interpreting $\hat Y$ as the finite-time evolution of a local Hamiltonian defined only near the boundary.
We say $\hat Y$ is `locally genreated' when no obstruction is present, and is `anomalous' whenever such interpretation is impossible.  Such an anomalous boundary signifies, at a heuristic level, the presence of a topologically nontrivial bulk.  As the $(p,1)$ models give $\hat Y = \hat t^{(p)}$, we will argue they are nontrivial by showing $ \hat t^{(p)}$ is anomalous when $p>1$.  The actual proof of the claim is deferred to section \ref{sec:GNVW}, where we define the quantized `chiral unitary index' proposed in GNVW and discuss its implications.

We will look for an obstruction to interpreting $\hat t^{(p)}$ as a locally generated unitary.
For our purpose, the class of `locally generated' unitaries is basically the same as the class of finite-depth quantum circuits of local unitaries (FDLUs),\cite{XieGuLiuWen} so one should look for an obstruction to designing an FDLU that implements $\hat t$ in a 1d system.\footnote{
Note that our usage of the phrase `local unitary' means that the unitary operator has a local finite support. In the language of GNVW, our `FDLU' is a finite depth circuit of `partitioned unitaries'. This is slightly different from the definition in Ref.~[\onlinecite{XieGuLiuWen}], where our FDLU is what they simply called a `local unitary'.
}
Since $\hat t^{(p)}$ is a site permutation operation, it can be built using only local SWAP gates, as we show in Fig.~\ref{fig:SWAP}b. However, in this construction $L-1$ layers of SWAPs are used on a ring of size $L$, and therefore the circuit depth diverges in the thermodynamic limit.
This suggests the impossibility of simulating $\hat t^{(p)}$ using any FDLU, which, when proven, will lead to our claim that the $(p,1)$ models are topologically nontrivial.

For the $(p,p)$ models, however, the edge operator $\hat Y = \hat t^{(p)} \otimes \overline{\hat t^{(p)}}$ can be realized in a purely 1d system using a nearest-neighbor SWAP circuit of depth two (Fig.~\ref{fig:SWAP}c). This construction immediately shows that the $(p,p)$ models are topologically trivial for all values of $p$.

\section{Bulk-boundary correspondence in MBL Floquet systems}
\label{sec:Edge}
We have seen that the bosonic chiral Floquet models, featuring the 1d translation operators at the edge, are suggestively nontrivial.  We will now provide a strong theoretical grounding for this claim. The first step is to establish the notion of bulk-boundary correspondence in Floquet systems.

The bulk-boundary correspondence is frequently applied in the classification of gapped equilibrium phases of matter, where the ability to cleanly distinguish between bulk and boundary DOF in the presence of a gap allows one to classify distinct bulk phases in terms of the anomalous properties of their edges, which are symmetry-protected to be gapless in $d=2$.  In a Floquet system, however, the notion of a ground state, and hence an excitation gap, is meaningless.  Nonetheless, if a Floquet system is MBL, so that one can identify mutually commuting quasi-local DOF,\cite{VoskAltman,Serbyn,HuseOganesyan} a version of the bulk-boundary correspondence becomes possible.  While similar approaches have been adopted for 1d systems in Refs.~[\onlinecite{Khemani, Keyserlingk,ElseNayak,PotterMorimotoVishwanath,Roy}], in the following we generalize the previous discussions and present a self-contained construction applicable to Floquet systems in any spatial dimension $d$, and then specialize to $d=2$ and apply it to the bosonic chiral Floquet models we constructed.

\subsection{MBL Floquet systems in any dimension}
We begin by first clarifying what we mean by `MBL' and `distinct' in our Floquet context.  We will call a Floquet system MBL if its Floquet operator $\hat U_{\rm F}$ can be written in the bulk as a product of mutually commuting quasi-local unitary operators.
Here, we say an operator is `quasi-local' if its nontrivial action is exponentially localized.
(See precise definition of `MBL' in Eq.~\eqref{eq:MBLF} below).
Two MBL Floquet systems are said to be distinct if it is impossible to interpolate between their Hamiltonians $\hat H_0(t)$ and $\hat H_1(t)$ within the class of MBL Floquet systems.  More precisely, we say $\hat H_0(t)$ and $\hat H_1(t)$ are distinct if for any continuous interpolation 
\begin{equation}\begin{split}\label{eq:}
\{ \hat H_s(t)~:~ \hat H_s(t+T) = \hat H_s(t), ~0 \leq s \leq 1\},
\end{split}\end{equation}
the corresponding family of Floquet operators $ \hat U_{\rm F}(s) $ will fail to be MBL for some $s$.

Next we proceed to explain the asserted bulk-boundary correspondence in detail.
For concreteness, we work on a large boundary-less 2d geometry, say a torus.  Write the Hamiltonian as
\begin{align}
\hat H (t)=\sum_r \hat H_r (t),
\end{align}
where each local term $\hat H_r (t) = \hat H_r(t+T)$ is nontrivial only within a finite neighborhood $b(r)$ surrounding lattice site $r$.  
Consider the Floquet operator
\begin{align} 
\hat U_{\rm F} = \mathcal T \exp \left(-i \int_0^T dt \,\hat H(t) \right),
\end{align}
which is uniquely-defined up to an arbitrary choice in the reference time $t=0$. We say $\hat U_{\rm F }$ is MBL when it can be written as a product of mutually commuting quasi-local unitaries $\hat U_r$:
\begin{align} \label{eq:MBLF}
\hat U_F = \prod_r \hat U_r.
\end{align}
Here each $\hat U_r$ is a quasi-local operator approximately localized within a finite neighborhood $b'(r)$ of lattice site $r$, and $[\hat U_r, \hat U_{r'}]=0$.  In the `l-bit' notation commonly used to describe MBL systems, each $\hat U_r$ can be represented as $\hat U_r = \exp (- i F_r (\{\hat \tau_{r'}\}))$, where $\{ \hat \tau_{r'}\}$ is a complete set of Hermitian, mutually-commuting, quasi-local and conserved operators (l-bits), and $F_r$ is a local real function.  Identifying the exact form of the l-bits for a given system is a nontrivial task, but in the following discussion we will not need to use this l-bit representation explicitly.

Before we move on to develop the notion of bulk-boundary correspondence in an MBL system, we pause to discuss the distinction between our use of the phrases `MBL Floquet  \emph{systems}' and `MBL Floquet \emph{phases}'. 
To define a phase, one should demonstrate that the defining properties are robust, i.e.~they are stable under all small physical perturbations.  Assuming many-body localization is robust, one can view MBL systems as the out-of-equilibrium analogues of gapped equilibrium systems, and define a phase as a collection of systems that can be smoothly deformed into each other while maintaining their MBL nature.  Although we implicitly assume such robustness of 2d many-body localization throughout this work when discussing MBL Floquet phases, we expect our results to hold in some long pre-thermal regime independent of the robustness of many-body localization.\cite{AbaninPreThermal,Tomotaka,Abanin_Rigor,pt_FloquetTimeCrystal}

\subsection{Bulk-boundary correspondence}
To discuss bulk-boundary correspondence in an MBL Floquet system, it is useful to first define two length scales that are important in our MBL Floquet problem: the `localization length' $\xi$ and the `Lieb-Robinson length' $\ell_\text{LR}$.
Physically, $\xi$ is the maximum radius of $b'(r)$, the neighborhood of sites on which any $\hat U_r$ can act nontrivially. Equivalently, $\hat U_r$ can be viewed as a local operator acting in $b'(r)$ tensored with the identity on the sites outside $b'(r)$.
Due to the mentioned exponentially decaying tails of the l-bits, $\xi$ so-defined is in principle unbounded.  Instead, we will relax our definition and allow $\hat U_r$ to differ from the identity outside $b'(r)$ by a small error.  This allows for a finite $\xi$, with the corresponding error bounded by $\mathcal O(e^{-r'/\xi})$ for sites at a distance $r'$ away from $b'(r)$.

Another important length scale is $\ell_\text{LR} \equiv T v_{\rm LR}$, where $v_{\rm LR}$ is the maximum Lieb-Robinson velocity\cite{LiebRobinson} of $\hat H(t)$ for $0\leq t \leq T$.  The intuition behind this `Lieb-Robinson length' $\ell_\text{LR}$ is that since the Floquet period $T$ is finite and $\hat H(t)$ is local, there should be a finite causal `light-cone' beyond which no communication, i.e.~information exchange, is possible.  Using the Lieb-Robinson theorem, this intuition is formalized as follows: for any localized operator $\hat O$, $\hat U^{\dagger}_{\rm F} \hat O \hat U_{\rm F} $ can act nontrivially only on sites within a distance $\ell_\text{LR}$ from the support of $\hat O$, again up to exponentially small corrections.
Note that, for an MBL system, $\xi$ and $\ell_\text{LR}$ are not independent: From Eq.~\eqref{eq:MBLF}, one sees that $\ell_\text{LR} \leq 2 \xi$, the maximum possible distance between any two points lying in the same neighborhood $b'(r)$.

\begin{figure}[tb]
\begin{center}
{\includegraphics[width=0.45 \textwidth]{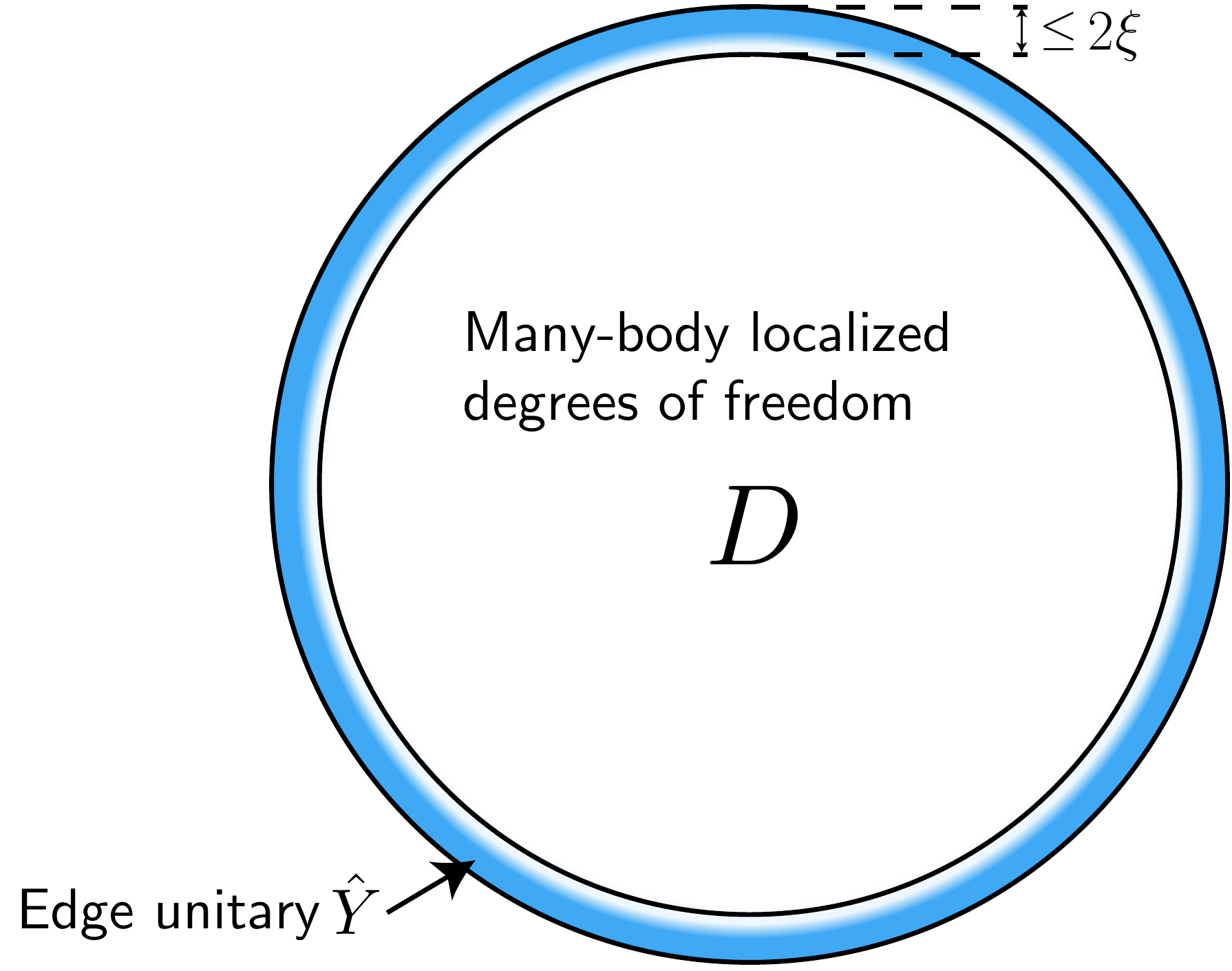}} 
\caption{
\label{fig:DefY}
{\bf Schematic definition of the edge operator $\hat Y$ in a system with an MBL bulk.} Up to an exponentially small error, the edge operator is contained within a distance of $2\xi$ from the edge, where $\xi$ denotes the bulk localization length.
 }
\end{center}
\end{figure}

Now consider a large region $D$ in our lattice (a disk in 2d, or a ball in $d>2$), with diameter much bigger than $2 \xi$.  There are two unitary operators that can be naturally associated to the system on the ball $D$.  One is the Floquet unitary constructed from $\hat H_D$, the Hamiltonian truncated to $D$:
\begin{align}
\hat U'_{\rm F }= \mathcal T \exp \left(-i \int_0^T dt \, \hat H_D(t) \right);~
\hat H_D(t) = \sum_{r: b(r) \subset D} \hat H_r(t).
\end{align}
The other is just the truncation of the original Floquet unitary $\hat U_{\rm F}$ of the full system to the ball $D$:
\begin{align}
\hat U''_{\rm F}=\prod_{r: b'(r) \subset D} \hat U_r .
\end{align}
Note that while $U'_{\rm F}$ can be constructed for any Floquet system, to construct $U''_{\rm F}$ one needs to invoke the MBL assumption. By the Lieb-Robinson bound, these two unitaries $\hat U'_{\rm F}$ and $\hat U''_{\rm F}$ must agree in the interior of $D$, with an error that decreases exponentially for sites away from the boundary.
In other words, their mismatch,
\begin{align} \label{eq:Ydef}
\hat Y = (\hat U''_{\rm F})^{-1} \hat U'_{\rm F} ,
\end{align}
is a unitary operator approximately localized on sites within a distance $2 \xi$ from the boundary of $D$ (Fig.~\ref{fig:DefY}).
By construction, $2 \xi$ is much smaller than the linear size of $D$, and therefore $\hat Y$ is effectively a unitary operator acting on the ($d-1$)-dimensional boundary region of $D$. 
Note that $\hat U'_{\rm F}$ is defined using the time-dependent Hamiltonian $\hat H_{D}(t)$, and hence $\hat Y$ can capture information that is missing in the bulk Floquet Hamiltonian $\hat H_{\rm F}$. In particular, this includes the physical consequences of a possibly nontrivial `micro-motion' within the Floquet period.

The key observation now is that even though $\hat U_{\rm F}''$ is MBL everywhere by construction, $\hat U_{\rm F}'$ may not be MBL at the boundary, where $\hat U_{\rm F}'$  and $\hat U_{\rm F}''$ do not necessarily agree. Hence, $\hat Y$ is not MBL in general.
Rather, we can only argue that $\hat Y$ is \emph{locality-preserving}, in the sense that conjugating by $\hat Y$ takes local operators to nearby local operators (as follows from the Lieb-Robinson bound).
Crucially, there is no guarantee that a locality-preserving unitary can be interpreted as the finite-time evolution operator of a local Hamiltonian.  Specifically, in the case of $\hat Y$, there is no guarantee that $\hat Y$ can be interpreted as a finite-time local Hamiltonian evolution in $d-1$ dimensions.
Whenever there is an obstruction in such reinterpretation, we say $\hat Y$ is anomalous and this signifies the presence of a topologically nontrivial bulk. We emphasize here that the mentioned anomaly makes no reference to symmetries, and is thus a `stronger' anomaly than those associated with the boundary of MBL-SPT Floquet phases recently studied in Refs.~[\onlinecite{Khemani, Keyserlingk, ElseNayak, PotterMorimotoVishwanath, Roy}].

\subsection{Application to the bosonic chiral Floquet models }
\label{sec:dSWAP}
We will now apply the formal construction above to the bosonic chiral Floquet models we described in Sec.~\ref{sec:Model}, corresponding to $d=2$. For simplicity, we will specialize the discussion to the $p=2$ model presented in Sec.~\ref{sec:BaseModel}, although it generalizes immediately to the general classes of $(p,q)$ models discussed in Sec.~\ref{sec:GenModel}.

Recall that, with PBC, the Floquet operator of the model is simply $\hat U_{\rm F} = \hat P_{\rm F} = \hat 1$. Now imagine appending to the original Floquet evolution a fifth time step, corresponding to the time evolution operator $\hat U_{(5)}$. The bulk Floquet operator is then modified into $\hat U_{\rm F} = \hat U_{(5)} \hat P_{\rm F} = \hat U_{(5)}$, and in particular the system is MBL if $\hat U_{(5)}$ is MBL. 
In fact, even without this fifth time step, i.e.~even if we had set $\hat U_{(5)}=\hat 1$, we would have also obtained a localized model, since, as noted above, the time evolution operator for the first four time steps is simply $\hat P_{\rm F}=\hat 1$.  The purpose of the disorder introduced in the fifth time step is to render the MBL property robust against small arbitrary perturbations of the Hamiltonian, and hence allow our system to represent a stable Floquet MBL phase.  The particular form of the disorder is unimportant, as long as the resulting unitary can be deformed into the model we constructed while remaining MBL, i.e.~we stay within the same MBL Floquet phase.  In the following, we will choose a particular form of $\hat U_{(5)}$ for concreteness and convenience.

First we replace time intervals $T/4$ with $T/5$ in the clean system and simultaneously increase the strength of the Hamiltonian such  that the single-step evolution operator, $\hat U_{(s)}$ for $s=1,\dots,4$, is unchanged (Eq. \ref{eq:SWAPdef}), and then consider the disordering Hamiltonian
\begin{equation}\begin{split}\label{eq:H5}
\hat H_{(5)}=\sum_{\vec x \in \Lambda} J_{\vec x} \hat Z_{\vec x},
\end{split}\end{equation}
where the coupling constants $J_{\vec x} \in [0, 5 \pi /T]$ are independent random variables.  Again, $H_{(5)}$ is exactly soluble, with
\begin{align} \label{eq:step_five}
\hat U_{(5)}\equiv
\exp\left(-i \hat H_{(5)} T/5 \right) = \prod_{\vec x \in \Lambda} \exp\left(- i J_{\vec x} \hat Z_{\vec x} \right).
\end{align}
Note that with this choice of $\hat U_{(5)}$, the model also has a $U(1)$ symmetry generated by spin rotation about the $z$-axis. This symmetry is inessential to our construction. Once again, one can consider a more general disordering Hamiltonian breaking this symmetry, and as we will argue in Sec.~\ref{sec:GNVW} the nontrivial nature of the phase survives.

To compute the edge operator $\hat Y$ as defined in Eq.~\eqref{eq:Ydef}, we have to consider two Floquet operators defined on an open geometry: the operator $\hat U_{\rm F}'$ corresponding to the evolution of the truncated Hamiltonian, and $\hat U_{\rm F}''$, the direct truncation of the bulk Floquet operator.
Although we previously chose the edge to be the boundary of a disk $D$, in the present construction it is more convenient to take the cylindrical geometry adopted in Sec.~\ref{sec:Model}, which corresponds to taking OBC in one of the two torus directions.  

To evaluate $\hat U_{\rm F}'$, recall from Eq.~\eqref{eq:Pprime_F} that, with this OBC, the first four time steps in the Floquet cycle give rise to the evolution operator $\hat P'_{\rm F} =  \hat t_{y=1;A} \otimes (\hat t_{y= N_y;B})^{-1}$. Since the disordering Hamiltonian Eq.~\eqref{eq:H5} is on-site, it is unaffected by the change in boundary condition, and therefore we find 
\begin{equation}\begin{split}\label{eq:}
\hat U_{\rm F}' = \hat U_{(5)} \left(  \hat t_{y=1;A} \otimes (\hat t_{y=N_y;B})^{-1}\right).
\end{split}\end{equation}
By the same token, $\hat U_{\rm F}'' = \hat U_{(5)}  \hat P_{\rm F}''$, where $\hat P_{\rm F}''$ is the truncation of the PBC permutation operator $\hat P_{\rm F}$ to the cylinder defined by the OBC, i.e.~we should discard all terms in $\hat P_{\rm F}$ that cross the boundary between $y=1$ and $y= N_y$. However, as $\hat P_{\rm F} = \hat 1$ it is again unaffected by the truncation. This gives $\hat U''_{\rm F} = \hat U_{(5)} $, and therefore
\begin{equation}\begin{split}\label{eq:Yt}
\hat Y = (\hat U''_{\rm F})^{-1} \hat U'_{\rm F}   =  \hat t_{y=1;A} \otimes (\hat t_{y=N_y;B})^{-1},
\end{split}\end{equation}
giving the same edge operator as in the clean model in Sec.~\ref{sec:Model}.

Note that the simplicity of $\hat Y$ hinges on our choice of a simple on-site disordering Hamiltonian $\hat H_{(5)}$. For a more general disordering Hamiltonian, $\hat U_{(5)}'$ and $\hat U_{(5)}''$ will only cancel exactly in the bulk, and modify $\hat Y$ by a unitary operator bi-local on the two circular edges. Nevertheless, the tensor-product factorization in Eq.~\eqref{eq:Yt} will still hold with an accuracy $\mathcal O (e^{-N_y/\xi})$.
We will see in Sec.~\ref{sec:GNVW} that, as long as we stay in the same MBL Floquet phase, such modification does not affect the nontrivial nature of the model, which is captured by the chiral unitary index of a single edge. 

\begin{figure}[h]
\begin{center}
{\includegraphics[width=0.45\textwidth]{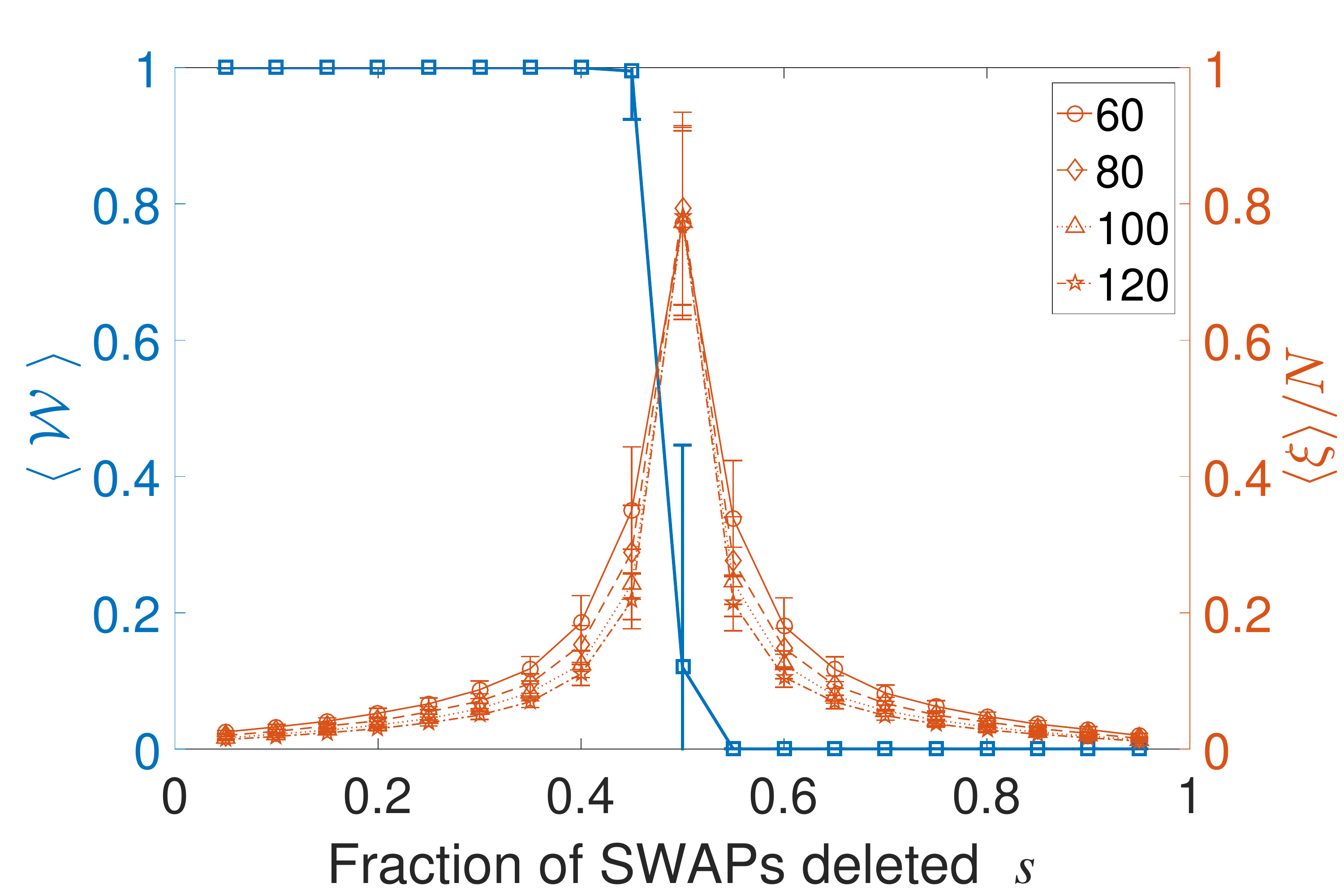}} 
\caption{{\bf Edge and bulk characterization of the `diluted' SWAP model.} Each data point corresponds to the disorder average of $200$ disorder realization, with the standard deviation indicated as error bars. 
Plotted against the right axis is the localization length $\xi$.
$\langle \xi \rangle$ was computed for $N\times N$ systems with PBC, where $N = 60,80,100$ and $120$.
As shown in the plot, the system is localized for small and large values of $s$, but is delocalized around $s = 0.5$. 
Accompanying the delocalization is a change in the edge behavior,  as can be seen in the data on $\mathcal W$, plotted against the left axis.  $\mathcal W$ is defined as follows: 
The edge operator $\hat Y$ for each disorder realization is computed following the discussion near Eq.~\eqref{eq:Yt}, using a $100\times 100$ system in the cylindrical geometry periodic in $x$ but open in $y$. 
Regardless of $s$, $\hat Y$ can be interpreted as an element of the site permutation group, and hence can be factorized into mutually commuting terms, with each term being an oriented `loop' in the lattice (Appendix \ref{sec:SWAP}). 
The edge operator at a single edge is identified by restricting to the `loops' that are strictly contained within the lower half of the system, and $\mathcal W$ is defined as the net winding number of these `loops' about the cylinder axis. $W = 1$ and $0$ respectively indicate a chiral and non-chiral edge operator. Note that $\mathcal W$ is not well-defined near $s = 0.5$, where $\langle \xi\rangle$ and $N$ become comparable, and therefore $\langle \mathcal W \rangle$ is not quantized in that region.
\label{fig:dSWAP}
 }
\end{center}
\end{figure}

To further support the claim that the MBL chiral Floquet model, characterized by the edge operator $\hat Y = \hat t$ at a single edge, is in a nontrivial phase, it is instructive to study its behavior as we tune the system to a manifestly trivial fixed point. 
More explicitly, consider two 2d MBL Floquet systems defined on the same cylinder geometry, described by the Floquet operators $\hat U_{\rm F}'(0)$ and $\hat U_{\rm F}'(1)$ with respectively anomalous and trivial edges, say $\hat Y(0) = \hat t$ and $\hat Y(1) = \hat 1$. 
Imagine a smooth family of Floquet systems $\{\hat U'_{\rm F}(s) ~:~0\leq s \leq 1 \}$ interpolating between the two.
By our claim, $\hat U'_{\rm F}(s)$ will necessarily become delocalized at some $s_c$ if the MBL chiral Floquet model is indeed nontrivial.
To illustrate this, we define a family of models by randomly deleting a fraction $s$ of the SWAP gates applied in the Floquet period. 
While $\hat U_{\rm F}'(s=0)$ is simply the original chiral Floquet model, very few sites are permuted as $s\rightarrow 1$, and hence $\hat U_{\rm}'(s=1) = \hat U_{(5)}'$ and $\hat Y(1) = \hat 1$. As shown in Fig.~\ref{fig:dSWAP}, the system indeed fails to be MBL around $s = 0.5$, consistent with the argument above.

\section{The chiral unitary index}
\label{sec:GNVW}
In the previous sections we reduced the problem of classifying 2d bosonic MBL Floquet systems to that of classifying 1d locality-preserving unitary operators $\hat Y$, and we constructed an infinite set of models, labeled by two positive integers $(p,q)$, which realizes  $\hat Y = \hat t^{(p)}\otimes \overline{\hat t^{(q)}}$.
Having provided circumstantial evidence that some of these models, e.g. the $(p,1)$ models with $p>1$, are nontrivial, we now argue that the quantized index defined in GNVW classifies bosonic MBL chiral Floquet phases in 2d via bulk-edge correspondence.  We will call this index the `\emph{chiral unitary index}' and denote it by $\nu$.   Roughly speaking, $\nu(\hat Y)$ is a measure of the imbalance in quantum information pumping at the edge under the evolution governed by the locality-preserving unitary operator $\hat Y$.  As we will see below, $\nu$ takes the form $\log (p/q)$, where $p,q$ are relatively prime positive integers. 

In particular, in this section we will justify the use of the adjective `chiral', show that $\nu$ is insensitive to the shape of the edge, and argue that it is in fact robust to any smooth deformation preserving the MBL nature of the bulk.

\subsection{One-dimensional quantum information transport}
The goal of this section is to develop a classification of 1d locality-preserving unitary operators, which we denote by $\hat Y$.  Recall that by `locality-preserving', we mean that conjugation by $\hat Y$ takes local operators to nearby local operators.  Above we have argued heuristically that certain locality-preserving unitaries, like $\hat t$, are anomalous, i.e.~cannot arise as the finite-time evolution of a 1d local Hamiltonian.  Precisely such a class of operators was studied in GNVW, which showed that the anomaly associated with $\hat Y$ is exhaustively captured by a quantized index, with the translation operators playing the role of generators for nontrivial indices.

To get some intuition on what makes a locality-preserving operator $\hat Y$ anomalous, it is useful to first recall the equilibrium classification of 2d gapped quantum phases.  In the equilibrium setting, the effective low energy dynamics of the 1d edge is typically described by a conformal field theory, which is partially characterized by its `chiral central charge'.  The chiral central charge describes the chiral flow of energy along the edge at finite temperature -- roughly, it counts the number of chiral edge modes.  When non-zero, it signals an anomaly, i.e.~an obstruction to realizing the corresponding field theory as the low energy theory of a truly 1d microscopic system.

In the present Floquet setting, instead of a 1d effective low energy field theory, we are given a locality-preserving unitary $\hat Y$ governing the edge dynamics.  The analogue of the anomaly in this setting, namely the inability to realize $\hat Y$ as the Floquet operator of a truly 1d system, turns out to also have an interpretation in terms of chiral transport.  While the appearance of chirality may have been expected given the vital role played by the translation operator in the previous sections, the transport interpretation is now more subtle, as there is no conserved charge or even energy. 

To understand what is being transported, consider again the unit translation operator $\hat t$ on a 1d ring of length $L \gg 1$, and let $\hat \rho$ be the density operator of any quantum state defined on the ring.  For any local measurement $\hat O_x$, the transformed state $\hat \rho' \equiv \hat t \hat \rho \hat t^\dagger$ gives results obeying $\text{Tr}( \hat \rho' \hat O_{x+1} ) = \text{Tr}(\hat \rho \hat O_{x} )$, where $\text{Tr}$ denotes the trace over a set of orthonormal bases in the physical Hilbert space.  Stated in words, the measurement results of $\hat \rho'$ at $x+1$ are identical to those of $\hat \rho$ at $x$.  Such unidirectional, perfect correlation is independent of the choice and structure of $\hat \rho$, and is indeed a property of the operator $\hat t$ itself.  Correlation is information -- what is pumped in a state-independent manner is therefore, suggestively, \emph{quantum information}, which are present in our Floquet setting even when both charge and energy are not conserved.

To formalize this idea, we now give a precise definition of the chiral unitary index of $\nu(\hat Y)$.

\subsection{Definition of the chiral unitary index $\nu$}

Let us work with a finite, but arbitrarily large, 1d ring whose lattice sites, labeled by $x$, host bosonic `spin' Hilbert spaces $\mathcal H_x$ of dimension $p_x$. For concreteness, pick an orthonormal basis $\{ |i_x\rangle\}$ for each $\mathcal H_x$, and let $\mathcal A_x$ be the set of local operators at a site $x$:
\begin{equation}\begin{split}\label{eq:}
\mathcal A_x = \left \{ \sum_{i_x,j_x=1}^{p_x} a_{i_x j_x} | i_x \rangle \langle j_x| ~:~ a_{i_x j_x}\in \mathbb C \right\}.
\end{split}\end{equation} 
Since the elements of $\mathcal A_x$, being operators, can both be added and multiplied, $\mathcal A_x$ forms an algebra.  An explicit orthonormal basis for $\mathcal A_x$ is $\{ \hat e^x_{ij} \equiv | i_x \rangle \langle j_x | ~:~ i,j=1,\dots,p_x\}$. Additionally, one can consider more general operator algebras $\mathcal A_S$ consisting of operators acting on a collection of sites $S$: $\mathcal A_S = \bigotimes_{x \in S} \mathcal A_x$.

Now let $\hat Y$ be a locality-preserving unitary operator in this 1d system.  As described in the previous section, the degree to which $\hat Y$ preserves locality is quantified by a Lieb-Robinson length $\ell_\text{LR}$.
Recall the defining property of $\ell_\text{LR}$ is that for any local operator $\hat O$, $\hat Y^{\dagger} \hat O \hat Y $ can act nontrivially only on sites at most a distance $\ell_\text{LR}$ from the support of $\hat O$, up to exponentially small corrections.

We now set up a `flow gauge' that measures how the locality-preserving unitary $\hat Y$ pumps quantum information across a spatial cut in the system.  Specifically, consider two contiguous intervals of sites $L$ and $R$, residing immediately to the left and right of the cut respectively.  These intervals host independent operator algebras $\mathcal A_L$ and $\mathcal A_R$, in the sense that $[ \hat a_L , \hat a_R]=0$ for all $\hat a_L \in \mathcal A_L$ and $\hat a_R \in \mathcal A_R$. For brevity, we will say $\mathcal A$ and $\mathcal B$ commute and write $[ \mathcal A, \mathcal B] = 0$ when they are independent in this sense.  Now let $Y(\mathcal A_L) \equiv \{ \hat Y \hat a_L \hat Y^\dagger  ~:~ \hat a_L \in \mathcal A_L\}$ and $Y(\mathcal A_R) \equiv \{ \hat Y \hat a_R \hat Y^\dagger  ~:~ \hat a_R \in \mathcal A_R\}$.  
Note that an operator basis for $\mathcal  A_\alpha$, $\alpha = L,R$, can be 
$\{ \hat Y \hat e^\alpha_{ij} \hat Y^\dagger~:~ i,j=1,\dots,p_\alpha\}$.
Qualitatively, a measure of how $[ Y(\mathcal A_L) , \mathcal A_R] \neq 0$ captures the extent to which $\mathcal A_L$ becomes entangled with $\mathcal A_R$ under the action of $\hat Y$, and therefore reflects the amount of quantum information pumped from left to right across the cut.\footnote{
Note that we adopt the `opposite' convention as compared to GNVW, which makes the direction of information flow more intuitive.}  Similarly, a measure of how $[ Y(\mathcal A_R) , \mathcal A_L] \neq 0$ captures the flow of quantum information from right to left.  To fully characterize the chiral character of $\hat Y$, we should look at the difference between these two measurements.

To ground the discussion quantitatively, we will need to introduce a measure of the extent to which two sets of operators $\mathcal A$ and $\mathcal B$, acting on a common Hilbert space $\mathcal H_{\Lambda}$ with $\dim(\mathcal H_\Lambda) = p_\Lambda$, are independent, i.e.~we seek a measure that is minimized when the two sets of operators commute ($[ \mathcal A , \mathcal B] = 0$), and is maximized when the two sets coincide ($\mathcal A = \mathcal B$). A natural candidate is the overlap of the respective orthonormal operator bases, which we define below. Let $\mathcal A$ and $\mathcal B$ be respectively spanned by the bases $\{ \hat e_{ij}^a ~:~i,j=1,\dots,p_a\}$ and $\{ \hat e_{lm}^b  ~:~l,m=1,\dots,p_b \}$, and define
\begin{equation}\begin{split}\label{eq:eta_def}
\eta\left(\mathcal A, \mathcal B \right) \equiv  \frac{\sqrt{p_a p_b}}{p_{\Lambda} } \sqrt{\sum_{i,j=1}^{p_a} \sum_{l,m=1}^{p_b}\left| \text{Tr}_\Lambda \left( \hat e_{ij}^{a \dagger} \hat e_{lm}^{b} \right) \right|^2 },
\end{split}\end{equation}
where the trace $\text{Tr}_\Lambda$ is restricted to $\mathcal H_\Lambda$.
In Appendix \ref{sec:Algebras}, we show that $\eta$ so-defined enjoys the desired properties: (i) it is independent of the orthonormal basis chosen; (ii) $\eta(\mathcal A,\mathcal B) = 1$ when $[\mathcal A,\mathcal B] =0$; and (iii) $\eta(\mathcal A,\mathcal A) = p_a$.

With all these preparations we can now define the index.  We let $\mathcal A_L$ and $\mathcal A_R$ be as above, but with the extra condition that they have sizes $\geq \ell_\text{LR}$. Let $\mathcal H_{\Lambda}$ be the Hilbert space associated with a very large finite region $\Lambda$ containing both $L$ and $R$ (for example $\Lambda$ could be the whole 1d system, assumed to be a finite ring).  We then define the chiral unitary index $\nu(\hat Y)$ as
\begin{equation}\begin{split}\label{eq:nu_def}
\nu (\hat Y) \equiv \log \text{ind}(\hat Y); ~~~
\text{ind}(\hat Y) \equiv \frac{\eta( Y( \mathcal A_L), \mathcal A_R  )}{\eta( \mathcal A_L , Y (\mathcal A_R) )},
\end{split}\end{equation}
where $\text{ind}(\hat Y)$ is equivalent to the index defined in Eq.~(45) of GNVW, and we will refer to it as the `GNVW index'. Although there is no formal distinction between $\nu$ and $\text{ind}$, we nonetheless introduce $\nu$, i.e.~takes the logarithm of the GNVW index, such that the connection of the index with the more familiar equilibrium results will be more transparent, as we elaborate on below.

\subsection{Index of the translation operator}
To develop some intuition for the chiral unitary index $\nu(\hat Y)$, or equivalently, its exponential, the GNVW index $\text{ind}(\hat Y)$, let us first compute it for two simple cases.  First, if $\hat Y = \hat 1$ is the identity operator, then clearly $\nu(\hat 1)=\log(1)=0$, since $[\mathcal A_L,\mathcal A_R] = 0$.  Second, in the case that the spin Hilbert spaces $\mathcal H_x$ all have the same dimension $p_x=p>1$, we can let $\hat Y = \hat t$ be the unit right-translation operator.
The Lieb-Robinson length of $\hat t$ is just $1$ and therefore intervals $L$ and $R$ of length $1$ -- i.e. single sites -- are already sufficiently large to be used in the index computation in Eq.~\eqref{eq:nu_def}.
Taking $\mathcal A_L = \mathcal A_x$, $\mathcal A_R = \mathcal A_{x+1}$, one finds
\begin{equation}\begin{split}\label{eq:indt}
\nu(\hat t) = \log \frac{\eta( \hat t \mathcal A_x \hat t^\dagger  ,  \mathcal A_{x+1} )}{\eta( \hat A_{x}  , \hat t \mathcal A_{x+1} \hat t^\dagger)}
= \log \frac{\eta(\mathcal A_{x+1}, \mathcal A_{x+1}) }{\eta(\mathcal A_{x }, \mathcal A_{x+2})} 
=  \log  p,
\end{split}\end{equation}
which is non-zero as long as $p>1$, and so as claimed the translation operator has a nontrivial index.  In addition, by a similar computation one sees that $\nu(\hat t^\dagger) = - \log p$, consistent with the intuition that $\hat t$ and $\hat t^\dagger$ pump quantum information in opposite directions.

Pictorially, the GNVW index of the translation operator corresponds to the transport of the entire on-site $p$-dimensional Hilbert space $\mathcal H_{x}$ across the cut. Note also that the computation of $\nu(\hat t)$ above is independent of $x$, the location of the cut. In addition, one can check that expanding the sizes of $L$ and $R$ will leave $\nu(\hat t)$ invariant. Hence, $\nu(\hat t)$ is independent of the arbitrariness in defining $\mathcal A_L $ and $\mathcal A_R$, as one would expect for a well-defined index.

Since $\hat t^2$ brings the $p^2$-dimensional Hilbert space $\mathcal H_{x-1} \otimes \mathcal H_{x}$ across the cut, we should expect $\nu(\hat t^2) = \log(p^2) = 2 \log p$.
In addition, for a system stacked with an identical copy of itself, $\hat t \otimes \hat t$ also brings a $p^2$-dimensional Hilbert space across the cut, and similarly we expect $\nu(\hat t \otimes \hat t)  = 2 \log p$. 
These observations suggest that, as claimed, the chiral unitary index is additive (or equivalently the GNVW index is multiplicative) under both composition (multiplication) and tensor product of the unitaries\cite{GNVW} (Appendix \ref{sec:Algebras}):
\begin{equation}\begin{split}\label{eq:nuAdd}
\nu(\hat Y  \hat Y') =\nu(\hat Y \otimes \hat Y') = \nu(\hat Y) + \nu (\hat Y').
\end{split}\end{equation}

\subsection{Interpretation of the index}
We have motivated the definition of the chiral unitary index, Eq.~\eqref{eq:nu_def}, by quantifying how a locality-preserving unitary operator redistributes quantum information. 
This is exemplified by the index computation for the translation operator, which gives $\nu(\hat t^{(p)}) = \log p$, the maximum von Neumann entropy supported by a $p$-level system. These facts make it natural to identify the chiral unitary index with the flow of information entropy per Floquet cycle along an 1d edge of the 2d system.

Such unidirectional transport of physical quantities is a defining feature of chiral phases of matter. 
For instance, equilibrium chiral phases are accompanied by a quantized nontrivial thermal Hall conductance.
Can we then view the MBL chiral Floquet phases as being in one-to-one correspondence to their equilibrium counterpart, with the only difference being a transport of quantum information in lieu of energy quanta?
We will attack this problem by first posing a seemingly different one:  Conventional wisdom holds that chiral phases with counter-propagating edge modes can `cancel' each other and reproduce a trivial phase.  How does this intuition generalize to the bosonic MBL chiral Floquet phases? 

To answer this, it is instructive to first consider the $(p,p)$ model, which should be trivial by our intuition on `edge cancellation'. To see this, we simply note that it has a trivial index: $\nu \left( \hat t^{(p)} \otimes \overline{ \hat t^{(p)}}\right) = \nu ( \hat t^{(p)} ) + \nu (\overline{ \hat t^{(p)}})  = \log p - \log p = 0$. Physically, this implies the operator $ \hat t^{(p)} \otimes \overline{ \hat t^{(p)}}$ is not anomalous, i.e.~it can be well approximated by an FDLU, as we have shown in Sec.~\ref{sec:tAnomaly}.

Importantly, not all translation operators are equal.  For instance, consider the $(2,3)$ model, which has index $\nu(\hat t^{(2)} \otimes \overline{ \hat t^{(3)}} ) = \log(2)- \log 3 \neq 0$.
Physically, this amounts to the observation that the information capacity of a qutrit ($p=3$) is larger than that of a qubit ($p=2$).  Hence, two chiral unitary operators annihilate each other if and only if they are `equal and opposite'. 
The phenomena of `opposite' operators annihilating each other is familiar from the equilibrium classification of SRE chiral phases, but the presence of logarithms in the current Floquet situation is novel.  

Since the $(p,q)$ model has index $\log (p/q)$, under stacking of these models their chiral unitary indices form $(\log \mathbb Q_+,+)$, the group of log-positive-rational number under addition.\cite{GNVW} This suggests that the classification in our MBL Floquet setting is much richer than that of their equilibrium counterparts. 
Nonetheless, if one is restricted to quantum spin systems with isomorphic $p$-dimensional site Hilbert spaces, the realizable indices are reduced to a subgroup of the general case.
For instance, if $p$ is prime then the classification is reduced to $( \log p^{\mathbb Z},+) \simeq (\mathbb Z,+)$, similar to the equilibrium result - this follows immediately from the alternative definition of the index given in section 7.2 of GNVW.  Slightly more generally, if $p$ has $n_p$ prime factors, then each prime factor generates one `dimension' of a `lattice' and the classification becomes $(\mathbb Z^{n_p},+)$.

\subsection{Robustness of the index}
While we have focused on the $(p,q)$ model and their edges, $\hat Y = \hat t^{(p)} \otimes \overline{ \hat t^{(q)}}$, to develop a physical picture for the chiral unitary index $\nu$, the narration thus far is missing a crucial ingredient: the quantization of $\nu$. 
This quantization is vital for its role in the classification,
since it is needed to ensure stability of the classification to any continuous change of the system, e.g. coming from a change in the precise shape of the 1d boundary chosen in the definition of Sec.~\ref{sec:Edge}, or from generic perturbation to the time-periodic Hamiltonian. 
To see this explicitly, if this quantization was absent, one could only claim that the group $(\log \mathbb Q_+,+)$ classified the special class of $(p,q)$ models, but could not view it as the general classification of 2d bosonic MBL chiral Floquet phases.

This quantization property is nontrivial, since a priori a local dressing of the translation operator could already interfere with the transport of quantum information. 
Specifically, although it is not at all obvious from its definition in Eq.~\eqref{eq:nu_def}, the GNVW index $\text{ind}(\hat Y)$ turns out to be a positive rational number for any locality-preserving $\hat Y$.  In GNVW, a proof of this statement is based on an alternate but equivalent definition of the index as the ratio of the dimensions of two certain finite-dimensional operator algebras, which is manifestly a positive rational number.  Mathematically-minded readers are encouraged to consult section 7 of GNVW for further details of this argument.  In the remainder of this subsection, however, we will instead provide a more physical argument for the quantization.

We first present the main idea behind the derivation. 
Intuitively, small physical deformations can be viewed as a dressing of the original operator by an FDLU, which is built from local unitaries. 
However, a local unitary cannot be chiral: Suppose on the contrary that a local unitary operator $\hat U_{I}$, defined on the finite 1d interval $I$ with boundaries $\partial_L$ and $\partial_R$, is chiral. Then under evolution governed by $\hat U_{I}$, quantum information is gradually depleted from (say) $\partial_L$ and accumulates at $\partial_R$. This is at odds with unitarity, since the dimensions of the local Hilbert space at the two edges are fixed. Hence, we see that $\nu(\hat U_I)= 0$, and from Eq.~\eqref{eq:nuAdd} we conclude that any FDLU has $\nu =0$ and therefore the index is robust. 
Note that the contradiction above is evaded by locality-preserving unitaries defined on a boundary-less geometry, say on a ring or on an infinite line. These are precisely the geometries for which the translation operator $\hat t$ is well-defined, and indeed the geometries that arise as the boundary of a 2d system.

With this picture in mind, we proceed to show that Eq.~\eqref{eq:nu_def} is invariant under smooth physical deformations. 
Specifically, consider two locality preserving unitaries $\hat Y(0)$ and $\hat Y(1)$, together with a smooth interpolation $\hat Y(s)$ between them, parametrized by $s \in [0,1]$.  Assume that the Lieb-Robinson lengths $\ell_\text{LR}(s)$ of $\hat Y(s)$ are all uniformly bounded by some $\ell_\text{LR}$: $\ell_\text{LR}(s) \leq \ell_\text{LR}$.    Then $\hat \gamma (s) \equiv \hat Y(s) \, \hat Y(0)^\dagger$ is a family of unitary operators contractible to the identity.  For small $\delta s$, we can expand 
\begin{align}
\hat \gamma(s+\delta s)=\hat \gamma(s) (1- i \,\delta s \, \hat h(s) + \dots),
\end{align}
where $\hat h(s)$ is a Hermitian operator.  Since $\hat \gamma(s)$ is also locality-preserving, $\hat h(s)$ has to be local.  Therefore $\hat \gamma (1)$ can be viewed as generated by a local Hamiltonian evolution, and is hence well approximated by an FDLU.  As $\hat Y(1) = \hat \gamma (1) \hat Y(0)$ and the chiral unitary index $\nu$ is additive under composition (Eq.~\eqref{eq:nuAdd}), it remains to show that $\nu=0$ for any FDLU.  
Using the composition property once again, one simply needs to show the triviality of a single layer of (necessarily commuting) local unitaries.  In this situation, the only local unitary that can potentially lead to a nontrivial index is the one entangling $\mathcal A_L$ and $\mathcal A_R$, i.e.~the unitary that is sliced by the cut.  
To see that $\nu = 0$ for such a unitary operator, we present below an argument originally given in GNVW.

Let $\hat U_{LR}$ be the local unitary sitting at the cut. We can take $L$ and $R$ as large as we please such that $\hat U_{LR} (\mathcal A_L\otimes \mathcal A_R) \hat U_{LR}^\dagger = \mathcal A_L\otimes \mathcal A_R$. 
Taking $\{ |i_\alpha \rangle ~:~ i_\alpha = 1,\dots,p_\alpha \}$ as a set of orthonormal basis for $\mathcal H_\alpha$, $\alpha = L,R$, we can write
\begin{equation}\begin{split}\label{eq:}
\hat U_{LR} = \sum_{i_\alpha,j_\alpha} \left( U_{LR} \right)^{i_L i_R}_{j_L j_R} | i_L i_R\rangle \langle j_L j_R|,
\end{split}\end{equation}
where $\left( U_{LR} \right)^{i_L i_R}_{j_L j_R}$ is a $p_L p_R$-dimensional unitary matrix. Computing explicitly (assuming index summation convention), one finds
\begin{equation}\begin{split}\label{eq:ULR}
&\eta \left ( U_{LR} \left(  \mathcal A_L \right), \mathcal A_R  \right)\\
=& \frac{\sqrt{p_L p_R}}{p_{L} p_{R}}  \left(
\text{Tr} \left( \hat U_{LR} \hat e_{ij}^{L \dagger} \hat U_{LR}^\dagger \hat e_{lm}^{R}  \right) 
\text{Tr} \left( \hat U_{LR}\hat e_{ij}^{L } \hat U_{LR}^\dagger \hat e_{lm}^{R \dagger}   \right) 
\right)^{1/2} \\
=& \frac{1}{\sqrt{p_L p_R} }
\left( 
 (U_{LR})^{am}_{jb} (U^\dagger_{LR})^{ib}_{al}(U_{LR}^\dagger)^{jd}_{cm} (U_{LR})^{cl}_{id}
\right)^{1/2}.
\end{split}\end{equation}
Now observe that $\eta \left ( \mathcal A_L, U_{LR}  \left( \mathcal A_R \right)\right) = \eta \left ( U_{LR}^\dagger \left( \mathcal A_L \right),  \mathcal A_R  \right) $, and interchanging $U_{LR} \leftrightarrow U_{LR}^\dagger$ in the last line of Eq.~\eqref{eq:ULR} amounts to an index relabeling. As all the indices are summed over, the expression is unchanged and hence $\nu(\hat U_{LR})=0$. 

We have thus shown that the chiral unitary index is invariant under small continuous deformations.  
As a corollary of the above arguments, we also see that the index is independent of the location of the cut.
However, since the composition property, Eq.~\eqref{eq:nuAdd}, is not explicitly derived in this work, our arguments do not constitute as a proof of the index quantization -- for a rigorous proof of this fact, see section 7 of GNVW.
Additionally, GNVW proves that if $ \nu (\hat Y) = 0$, $\hat Y$ must necessarily be an FDLU.  Combined, these statements imply that any locality-preserving $\hat Y$ can be continuously connected to a $(p,q)$-edge for some relatively prime positive integers $p,q$, and so the stacked $(p,q)$ models constructed in the previous section form a complete set of representatives for bosonic MBL chiral Floquet phases in 2d.

\section{Numerical index computation via MPUs}
\label{sec:Num_MPU}
We have argued that the quantized chiral unitary index implies the existence of 2d bosonic MBL chiral Floquet phases, and that the SWAP models we presented serve as prototypes. However, since the class of problems under consideration is strongly interacting, one may expect that a general method to numerically compute the index is unavailable -- in fact even specifying the edge unitary operator $\hat Y$ in a concrete manner is technically challenging. Contrary to this expectation, we will demonstrate below that the index is in fact numerically computable through the use of matrix-product representations.

\begin{figure}[h]
\begin{center}
{\includegraphics[width=0.4 \textwidth]{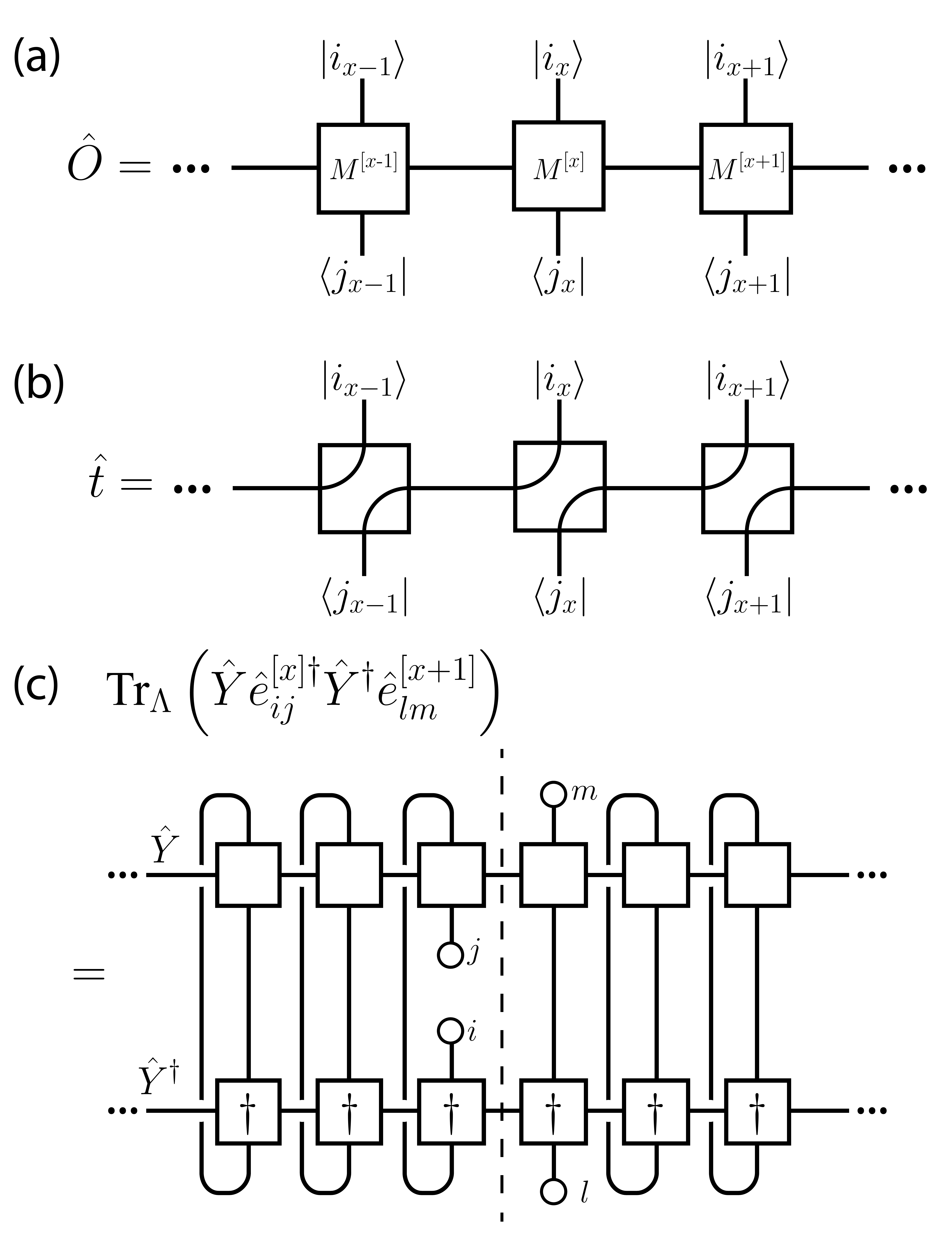}} 
\caption{
\label{fig:MPU_Rep}
{\bf MPU representations.} (a) The matrix-product representation of an operator $\hat O$, Eq.~\ref{eq:MPU_O}, has a simple diagrammatic representation, where each tensor is represented by a box with legs corresponding to the various linear spaces. Connected legs represent contraction.
(b) The tensor $T^{[x]}$ defining the translation operator $\hat t$ can be pictured as a particular connection of the physical and virtual legs, which leads to a `pipeline' shoveling quantum information uniformly to the right.
(c) A similar analogy applies to the diagrammatic representation of $\text{Tr}_\Lambda \left( \hat Y \hat e_{ij}^{[x] \dagger} \hat Y^\dagger \hat e_{lm}^{[x+1]} \right) $, which quantifies the flow of quantum information across the cut (dashed line) under the action of the MPU $\hat Y$.
 }
\end{center}
\end{figure}

One of the most successful frameworks for representing a 1d quantum operator is through the use of a `matrix product' representation,\cite{MPO,DMRG_MPS} in which an operator $\hat O$ is written as
\begin{equation}\begin{split}\label{eq:MPU_O}
\hat O = \sum_{i,j} &\left( \cdots M^{[x]}_{i_x,j_x} M^{[x+1]}_{i_{x+1},j_{x+1} } M^{[x+2]}_{i_{x+2},j_{x+2} }  \cdots \right)\\
&\times | \cdots i_x i_{x+1} i_{x+2} \cdots \rangle \langle  \cdots j_x j_{x+1} j_{x+2} \cdots |,
\end{split}\end{equation}
where for each $x$ and $i_x,j_x = 1,\dots, p_x$, $M^{[x]}_{i_x,j_x}$ is a $\chi_x \times \chi_{x+1}$ matrix, with the maximum value of $\{ \chi_x\}$ known as the bond dimension.
The ellipses may extend to infinity for an infinite chain, terminate at two ends for an open chain, or be subjected to a trace in the bond space for a system with periodic boundary condition. 

In this work we will focus exclusively on matrix-product unitaries (MPUs). In addition, in discussing MPUs it is often convenient to introduce a diagrammatic representation, where each tensor at a site is represented by a box with two `physical' and two `virtual' legs (Fig.~\ref{fig:MPU_Rep}a). The physical legs are attached to `bras' and `kets', corresponding respectively to in-coming and out-going physical states; the virtual legs are contracted with those of the neighboring sites, and provide `room' for quantum information transport, which is necessary when the unitary is not on-site.

FDLUs in 1d are well-described by MPUs with finite bond dimensions, since each local gate in the circuit can be readily represented as an MPU.\cite{MPO,DMRG_MPS} However, to simulate the edge of a 2d MBL Floquet phase, we must relax ourselves from \emph{locally generated} unitaries to \emph{locality-preserving} unitaries. 
The prototypes of locality-preserving, but not locally generated, 1d unitaries are the translation operators. These turn out also to admit a simple MPU representation. 
To see this, assume $\dim(\mathcal H_x) = p$ for all sites $x$, and let the bond dimension be also $p$. Now consider the tensor $ \left( T^{[x]}_{i_x ,j_x }\right)_{\alpha,\beta} = \delta_{i_x, \alpha} \delta_{j_x,\beta}$, which gives an MPU with bond dimension $\chi = p$.
Pictorially, the delta functions act as `connectors' of `quantum information pipes', with the pairing of indices leading to a `pipe' that connects the in-coming states at $x$ to the out-going ones at $x+1$ (Fig.~\ref{fig:MPU_Rep}b). The tensor $T^{[x]}$ therefore defines the unit right-translation operator, as one can explicitly verify.

Having discussed the MPU representations of both the FDLUs and the translation operators, we now make connection with the study of 2d MBL Floquet phases.
Recall that the GNVW classification is exhaustive, i.e.~any such unitary can be written as the product of an FDLU and a (stacked) translation operator with a suitable index. Since both the FDLUs and translation operators admit MPU representations, the GNVW result implies all locality-preserving 1d unitaries can be efficiently simulated by MPUs.
Further using the bulk-boundary correspondence we established in Sec.~\ref{sec:Edge},
one sees that the MPU representation provides a universal framework for describing the edge of any 2d MBL Floquet system. 

In addition, by representing the edge operator $\hat Y$ of a 2d MBL Floquet system as an MPU, one can also compute the chiral unitary index $\nu$ of the system within the MPU formalism. To leverage the power of this framework, we recast the index formula Eq.~\eqref{eq:nu_def} into the MPU language. 
Although a concrete derivation, which we present in Appendix \ref{sec:MPU}, will necessary involve a fair bit of technicality, its diagrammatic representation is quite intuitive, as we discuss below. 

\begin{figure}[h]
\begin{center}
{\includegraphics[width=0.45 \textwidth]{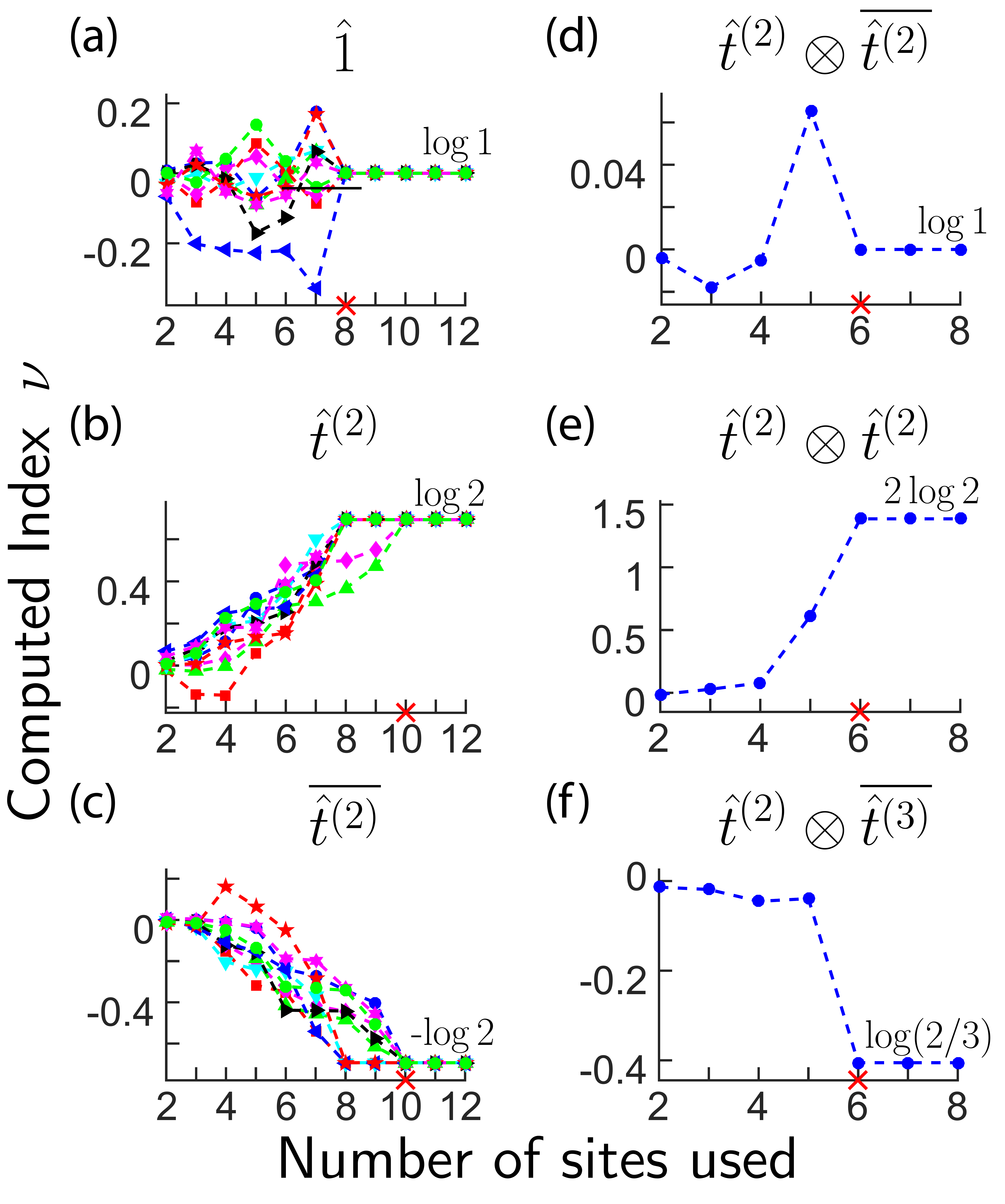}} 
\caption{{\bf Numerical evaluation of the chiral unitary index $\nu$ for different example MPUs.} 
Each panel corresponds to the index calculation of an MPU taking the form $\hat F \hat Y$, where $\hat F$ is a random, disordered 1d FDLU, and $\hat Y$ is an MPU with a known index. The choice of $\hat Y$ is indicated on top of each panel.
The horizontal axes indicate the total number of sites (in both the $L$ and $R$ intervals) used for the computation, and whenever the number is odd, $R$ contains one more site than $L$.
The dashed lines are guides for the eyes, and the red crosses on the horizontal axes indicate where the interval sizes reach the Lieb-Robinson lengths $\ell_\text{LR}$.
For each panel in (a-c), the index of the same disorder realization was computed using 10 randomly chosen cuts, indicated by the different markers. Each panel in (d-f) shows the index computation at a single cut for an MPU.
As shown in all panels, when the size of the intervals reaches $\ell_\text{LR}$, the index computation converges to the expected value.
(Please refer to Appendix \ref{sec:MPU} for details on the MPU construction.)
\label{fig:GNVW}
 }
\end{center}
\end{figure}

As discussed, the in-coming and out-going legs of an MPU can be loosely viewed as the inlets and outlets of a pipeline, with each of the openings labeled by their location on the chain. 
Since $\nu$ is a measure of the net quantum information flow, in the pipeline analogy the index computation amounts to quantifying the rate at which an incompressible fluid in the pipeline flows across a spatial cut.
This rate can be extracted by suitably comparing the inflow and outflow across the cut.
For instance, to measure the left-to-right flow, one can imagine closing all the openings of the pipe, except for the inlet at $x$ and the outlet at $x+1$ . When some test fluid is pumped into the pipe at $x$, it can either get trapped inside the pipe, or emerge out at $x+1$. The portion that flows out from the outlet must have passed through the cut between sites $x$ and $x+1$, and therefore its volume reflects the capacity of the pipe across that cut.
In the MPU language, the analogue of `closing pipe openings' is to contract the free-hanging legs, i.e.~to take traces. This is shown in Fig.~\ref{fig:MPU_Rep}c, where the diagrammatic representation of $\text{Tr}_\Lambda \left( \hat Y \hat e_{ij}^{[x] \dagger} \hat Y^\dagger \hat e_{lm}^{[x+1]} \right) $, which enters the index computation via Eqs.~\eqref{eq:eta_def} and \eqref{eq:nu_def}, takes a form similar to the pipeline picture presented above.

In the analogy above, the inlet at $x-1$ and the outlet at $x+1$ could also be connected, and our measurement was blind to the flow through this connection, which also passes through the specified cut.
Physically, the  range of such `connection' is the range of quantum information redistribution, which is what we defined as the Lieb-Robinson length $\ell_\text{LR}$. To accurately evaluate the flow across the cut, one should therefore increase the number of inlets and outlets open on the two sides until all possible connections have been exposed, i.e.~one needs to choose the intervals $L$ and $R$ in the index computation to be at least as big as the Lieb-Robinson length. This can be observed in Fig.~\ref{fig:GNVW}, where we show the numerical results on the computation of the chiral unitary index using the MPU formula in Appendix ~\ref{sec:MPU}.

\section{Experimental proposal}
\label{sec:ExpP}

Having constructed explicit models and identified a topological index for bosonic MBL chiral Floquet phases in 2d, we now propose an experimental setup that realizes the $\nu = \log 2$ phase using hardcore bosonic ultracold atoms loaded onto a shaken optical lattice.

\begin{figure}[h]
\begin{center}
{\includegraphics[width=0.45 \textwidth]{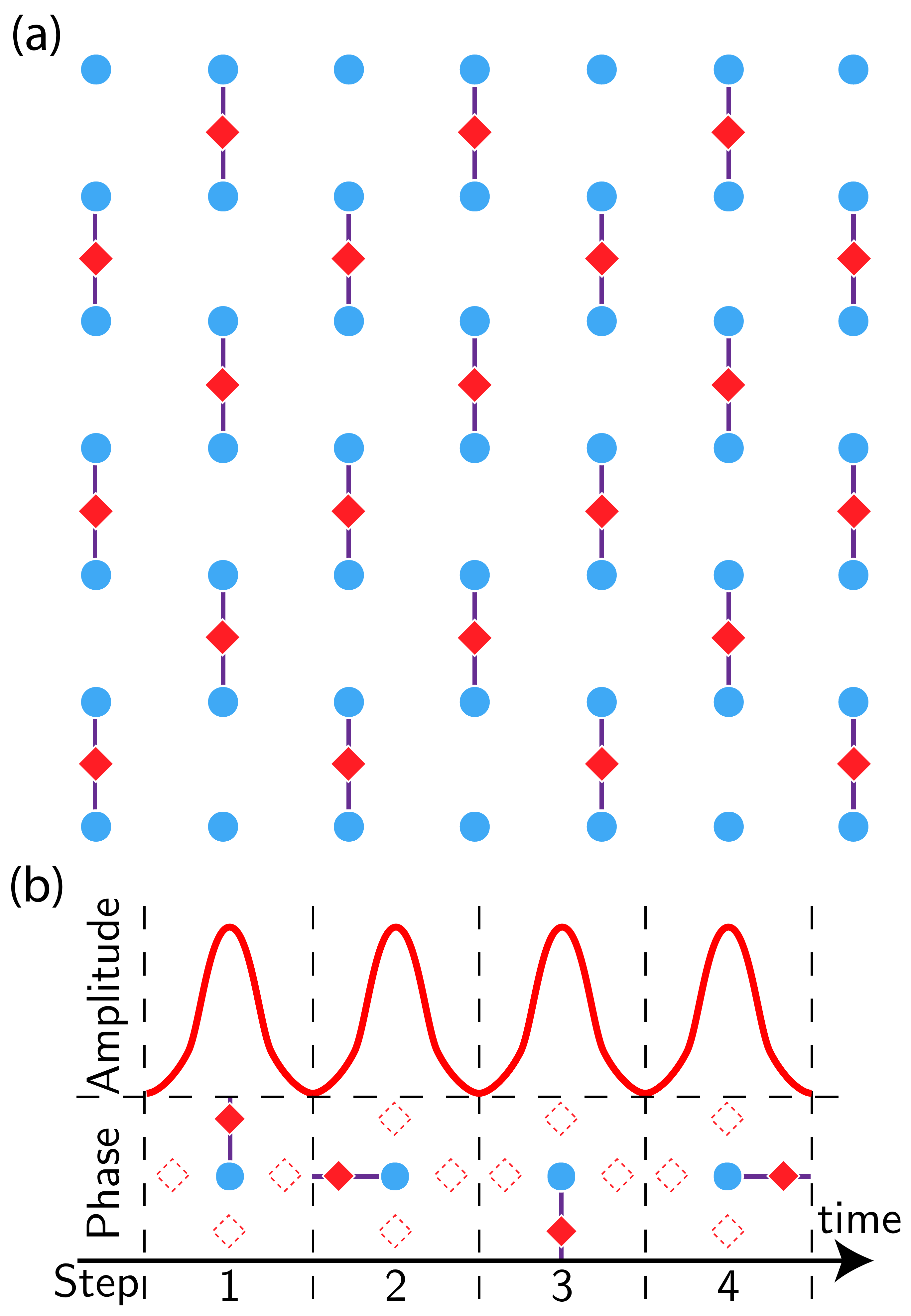}} 
\caption{
\label{fig:OpticalLatt}
{\bf Proposed optical lattice setup for realizing the bosonic MBL chiral Floquet phase using ultracold hardcore bosons.} 
(a) We consider two square lattices, indicated respectively by blue circles and red diamonds, that are rotated by $45$ degrees with respect to each other. We assume the blue lattice is deep and the system is strongly Mott insulating without the red lattice.
The role of the red lattice is to offset the energy barrier between pairs of sites, and thereby turning `on' the bonds between them, indicated by purple lines.
(b) The red lattice is both amplitude- and phase-modulated, and sequentially turns on different sets of bonds in the first four steps of the Floquet cycle.
}
\end{center}
\end{figure}

In the hardcore limit, each site Hilbert space is two-dimensional and can be viewed as an effective spin-1/2.
Inspired by previous experiments in creating more complicated lattice geometries,\cite{KagomeOL,LiebOL} we will consider a pair of short- and long-wavelength optical lattices.
Two square lattices (red and blue), formed by two pairs of retro-reflected laser beams, are used, and a deep vertical lattice is employed to render the system quasi-2d (Fig.~\ref{fig:OpticalLatt}a). 
To create the checkerboard geometry, we take the ratio of the wavelengths to be $\sqrt{2}$, and rotate the two lattices by $45$ degrees. 
We suppose the frequencies of the laser beams are chosen to be respectively blue- and red- detuned from the atomic resonance, and therefore the potential minima of the blue and the maxima of the red are energetically favored.

Suppose the blue lattice is deep such that the system is strongly Mott-insulating when the red lattice is ignored. 
The role of the shaken red lattice is to periodically `turn on' the targeted bonds, which enhances tunneling of particles. 
Intuitively, the tunneling will be  perfect when the shaking profile is appropriately designed, akin to the action of a SWAP gate.

More specifically, we consider a five-step protocol similar to our construction of bosonic chiral Floquet models in Sec.~\ref{sec:BaseModel} and Sec.~\ref{sec:dSWAP}. 
In particular, the last step was again introduced only for incorporating disorder into the system. This can be achieved by turning on a speckle potential, which creates a disordered on-site chemical potential.\cite{Speckle1, Speckle2} 
Similarly, the first four steps will resemble the SWAP models.
In these steps, each of duration $T/5$, the red lattice is ramped on and then off, with the phase of the lasers set to align the potential maxima (energy minima for the atoms) to the locations indicated in Fig.~\ref{fig:OpticalLatt}b. 
The potential maxima of the red lattice mark the `on' bonds, while all the other bonds are considered as `off'.
In a crude approximation, we neglect the tunneling across the `off' bonds, and in each step $s=1,\ldots, 4$ we consider the Bose-Hubbard Hamiltonian $\hat H_s (t) \approx \sum_{\bar{\vec r} \in {\rm RL}_s} \hat H_{\bar{\vec r}} (t)$, where
\begin{equation}\begin{split}\label{eq:}
\hat H_{\bar{\vec r}} (t) =&
J(t) \left( \hat b_{\vec r_{\rm A}}^\dagger \hat b_{\vec r_{\rm B}} + \hat b_{\vec r_{\rm B}}^\dagger \hat b_{\vec r_{\rm A}} \right) + U(t) \sum_{\mu={\rm A,B}} \hat n_{\vec r_\mu} (\hat n_{\vec r_\mu}-1).
\end{split}\end{equation}
${\rm RL}_s$ denotes the set of potential maxima of the red lattice in step $s$, and $\hat b^\dagger_{\vec r_\mu}$ and $\hat n_{\vec r_\mu}$ (for $\mu={\rm A,B}$) are respectively the creation and number operators of the two sites connected by the bond at $\bar{\vec r}$. 
Both $J(t)$ and $U(t)$ are modulated as the lattice is ramped on and off, though that of $U(t)$ is relatively unimportant as we will take the hardcore limit anyway.

By construction, all the terms in $\hat H_{s} (t)$ commute, and we simply solve the dynamics governed by the two-site system $\hat H_{\bar{\vec r}} (t)$.
In the hardcore limit $U(t) \rightarrow \infty$, the effective Hamiltonian in the low-energy subspace, with the particle-number basis $(|00\rangle , |10\rangle , |01\rangle , |11\rangle )$, is a $4\times 4$ matrix
\begin{equation}\begin{split}\label{eq:}
H^{g}_{\bar{\vec r}}(t)
\approx
\left(
\begin{array}{cccc}
0 & 0 & 0  &0\\
0& 0 &  J(t) &0\\
0&J(t) & 0 &0\\
0& 0& 0& 0
\end{array}
\right),
\end{split}\end{equation}
for which the time evolution operator can be easily computed:
\begin{equation}\begin{split}\label{eq:}
U^{g}_{\bar{\vec r}}(t)\approx
\left(
\begin{array}{cccc}
1 & 0 & 0  &0\\
0& \cos\,\phi(t)  & - i \sin\,\phi(t) &0\\
0&- i \sin\,\phi(t) & \cos\,\phi(t) &0\\
0& 0& 0& 1
\end{array}
\right),
\end{split}\end{equation}
where $\phi(t) = \int_0^t \, dt' J(t')$ is the integrated rotation angle in the one-particle subspace. While a more careful estimation will require the computation of $J(t)$ for realistic lattice parameters, it should be possible to design an appropriate amplitude-modulation profile realizing $\phi(T/5)=\pi/2$, which gives a `perfect' tunneling. As such, $U^{g}_{\bar{\vec r}}(T/5)$ is equivalent to the SWAP gate up to the phase rotation $R_{\varphi} \equiv \text{diag}(1,-i,-i,1)$. 

Thanks to the topological nature of the problem, this phase rotation does not affect the realization of the model.
To see this explicitly, note that the phase rotation above can be written as 
\begin{equation}\begin{split}\label{eq:}
\hat R_\varphi = \exp{\left( - i \frac{\pi}{2} \left( \hat n_{\rm A} + \hat n_{\rm B} \right) - i   \pi \, \hat n_{\rm A} \hat n_{\rm B}\right)},
\end{split}\end{equation}
with the number operator $\hat n_\mu = | 1_\mu \rangle \langle 1_\mu |$ in the hardcore limit.
$\hat R_\varphi$ can be interpreted as the evolution of a Hamiltonian diagonal in $\hat n_\mu$. Since all the number operators commute, $\hat U_{\rm F}^{2d}$ remains exactly soluble. In particular, the Floquet operator (together with the fifth, disordering step) takes the MBL form, with $\{ \hat n_{\vec r_\mu} \}$ being the l-bits and the functions $F_j$ only coupling l-bits in the same or adjacent unit cells. 

The argument above can be readily extended to a smooth family of phase rotations connecting $\hat R_{\varphi}$ to the identity, and hence the proposed system is indeed in the same $\nu = \log 2$ MBL chiral Floquet phase as the SWAP model in Sec.~\ref{sec:BaseModel}.
By the same token, restoring the tunneling between the neglected bonds should have a mild effect provided that the system remains MBL, which is anticipated when the blue lattice is sufficiently deep. 

While lattice shaking techniques have already been demonstrated in 2d,\cite{OLShake} simulating MBL models in ultracold atom experiments is still an active research frontier.
Nonetheless, several recent works have already demonstrated their feasibility,\cite{MBL_E2, MBL_E3,MBL_E4} and realization of our proposal could be soon within reach.

\section{Physical consequences}
\label{sec:PhysCon}
In this section, we discuss the physical consequences and experimental signatures associated with the bosonic MBL chiral Floquet phases. We will focus on the thermal behavior of the edge in Sec.~\ref{sec:Chaos}, and the transport of quantum information in Sec.~\ref{sec:AnB}.

\subsection{Topology-enforced edge thermalization}
\label{sec:Chaos}
To this end, it is instructive to first recall the phenomenology associated with SRE topological phases of matter in equilibrium. In 2d, a bulk with SPT order is always paired with an anomalous edge, which cannot be both symmetric and gapped, i.e.~the edge is either gapless, or breaks a symmetry.

In our non-equilibrium setting, the role of an excitation gap is now played by many-body localization, and therefore the edge of a nontrivial 2d MBL Floquet SPT phase cannot be both symmetric and MBL, i.e.~if symmetries are not broken, the edge must be thermal.\footnote{
Note that we are implicitly assuming a non-localized system is generically in a thermal phase. Whether there are any other alternatives is an interesting open question.
}
When the protecting symmetry is broken, however, the boundaries can be localized again.
In particular, some of the proposed phases feature boundaries that become localizable when the discrete time-translation group is reduced to a subgroup, say when sub-harmonic terms are added to the original drive. Interestingly, such symmetry breaking might happen spontaneously and lead to discrete time crystals  \cite{Khemani,Keyserlingk2,FTXstals,Keyserlingk3,Norm_FTXstals} (in Ref.~[\onlinecite{Khemani}] the analogous phase was termed $\pi$ spin glass).

In fact, all the previously-proposed topologically nontrivial MBL Floquet phases are protected by some symmetries, and therefore the boundaries of these systems can \emph{always} be localized by symmetry breaking. The bosonic MBL chiral Floquet phases we discovered, however, correspond to the first established examples that are nontrivial even in the absence of any symmetries, and hence the non-localizability is robust, provided that the system remains Floquet.\footnote{
Note that this is an essential assumption, for if not there may not be a clear way to determine the long-time behavior of the system.}
In particular, the anomalous edge is stable against the introduction of sub-harmonic terms. 
Such robust non-localizability of the chiral edge can be summarized as follows: If $\nu(\hat Y) \neq 0$, then for any 1d local Hamiltonian evolution $\hat F$ (an FDLU), the locality-preserving operator $\hat F \hat Y$ is never localized. This holds even when arbitrarily strong disorder is incorporated into the system via $\hat F$.

\begin{figure}[t]
\begin{center}
{\includegraphics[width=0.48 \textwidth]{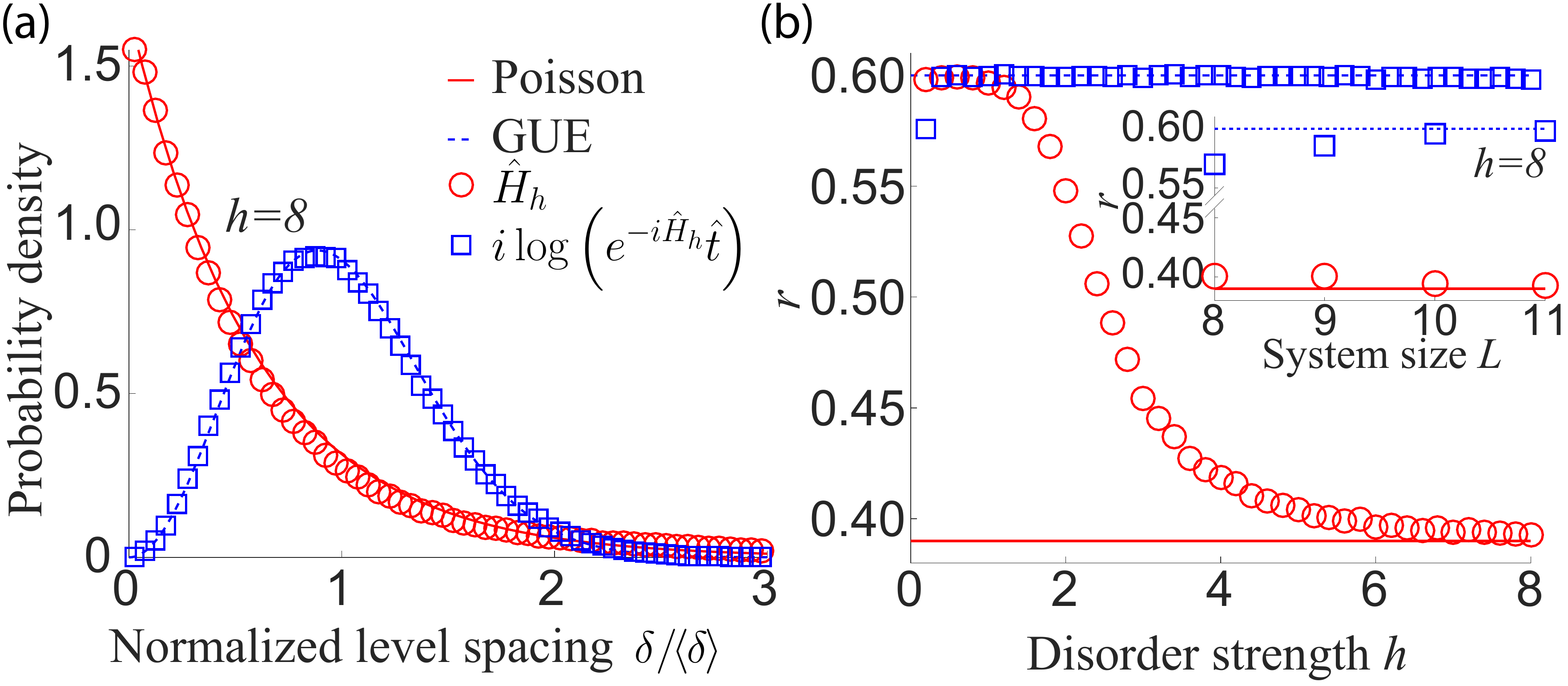}} 
\caption{
\label{fig:loc_therm}
{\bf Chiral edge thermalization.}
The presence or absence of localization is studied using exact diagonalization for a small spin-1/2 chain with size $N\leq11$ sites, and level statistics is extracted using $200$ disorder realizations. Red circles and blue squares respectively indicate level statistics of $\hat H_h$ and $i \log\left( e^{- i \hat H_h} \hat t\right)$.
(a) The level spacing $\delta_n \equiv \epsilon_{n+1}-\epsilon_n$, where $\epsilon_n$ is the $n$-th eigenvalue of the Hamiltonian for one disorder realization, is known to show distinct statistical properties when the system is MBL or thermal. 
The plot shows level statistics  for $h=8$, and the probability density functions of the normalized level spacing $\delta / \langle \delta \rangle$ are well-fitted to the Poisson distribution (red line) and the generalized unitary ensemble (GUE; blue dashed line) respectively for $\hat H_h$ and $i \log\left( e^{- i \hat H_h} \hat t\right)$.
This indicates $\hat H_h$ is localized but $i \log\left( e^{- i \hat H_h} \hat t\right)$ is thermal.
(b) The localization of the system can be further quantified by studying the `$r$ ratio', defined as $ r \equiv \langle \frac{\min (\delta_n, \delta_{n+1} ) }{\max (\delta_n, \delta_{n+1} )} \rangle$.
$r$ is known to converge to $0.39$ (red line) and $0.6$ (blue dashed line) respectively for Poisson and GUE level statistics,\cite{Pal_Huse,DAlessio_Rigol} as observed at large $h$. The inset shows the convergence as a function of system sizes $8\leq L \leq 11$ at a strong disorder of $h=8$.
 }
\end{center}
\end{figure}

To substantiate this claim, we consider a 1d ring of spin-1/2's. Let $\hat F_h \equiv e^{- i \hat H_h}$  be the evolution operator of a disordered 1d Heisenberg Hamiltonian, given by
\begin{align}
\hat H_h \equiv \sum_i \hat {\bm{S}}_i \cdot \hat{\bm{S}}_{i+1} + h\sum_i \bm{w}_i \cdot \hat{\bm{S}}_i,
\end{align}
where $\hat {\bm{S}}_i $  denotes the spin operators at site $i$, $h$ is the strength of on-site random magnetic fields, and $w_{i,\alpha}$ is a random number within the range $[-1,1]$. It is known that $\hat H_h$ is in a thermal phase for $h<2.5$, and an MBL phase for $h>2.5$.\cite{MBL_Ent} 
This is confirmed by looking at the statistics of the level spacing between adjacent energy eigenstates, $\delta_n \equiv \epsilon_{n+1}-\epsilon_n$. As shown in Fig. \ref{fig:loc_therm}a, the level statistics of $\hat H_h$ is Poissonian when $h = 8>2.5$, confirming that it is MBL. 
In addition, for this class of models the `$r$ ratio',  defined by $r=\langle\frac{\min (\delta_n, \delta_{n+1} ) }{\max (\delta_n, \delta_{n+1} )}\rangle$, is known to converge to $0.39$ and $0.6$ when the system is respectively MBL and thermal.\cite{Pal_Huse, DAlessio_Rigol}
This can be seen from the data shown in Fig. 6b, in which the $r$ ratio changes from 0.6 to 0.39 as $h$ is increased, and is consistent with an MBL phase transition at $h=2.5$.\cite{MBL_Ent}

In contrast, when combined with the unit right-translation operator $\hat{t}$, the operator $\hat F_h \hat t$ is never localized regardless of $h$. 
Such distinction between the behavior of $\hat F_h$ and $\hat F_h \hat t$ can be readily seen in the level statistics, as we show in Fig.~\ref{fig:loc_therm}.
In particular, although $\hat F_h$ is MBL at a strong disorder strength of $h = 8 > 2.5$, the level statistics of $\hat F_h \hat t$ agrees with that of a thermal phase (i.e., the $r$ ratio remains at 0.6 in the entire range of $h$).\cite{DAlessio_Rigol}
This suggests that the nontrivial bulk topology of a bosonic MBL chiral Floquet phase can lead to a robust chaotic behavior at the edge.

\subsection{Unidirectional transport and quantum communication}
\label{sec:AnB}
Aside from connection to ergodicity, akin to its equilibrium counterpart a chiral edge operator also gives rise to signature unidirectional transport. 
For instance, in a system with a conserved U(1) charge, one expects an edge operator $\hat Y$, with $\nu(\hat Y) >0$, to generate a right-moving current proportional to the average charge localized at the edge.\cite{Lindner}
However, we stress that the presence of such a conserved charge only bears witness to the chiral nature of the edge, and is not necessary for the existence of a chiral Floquet phase, which was linked to the chiral unitary index $\nu$ defined in the absence of any symmetries. For this reason, a more fundamental characterization of the chiral Floquet phases are in terms of the chiral transport of quantum information, which we investigate below.

\begin{figure}[t]
\begin{center}
{\includegraphics[width=0.45 \textwidth]{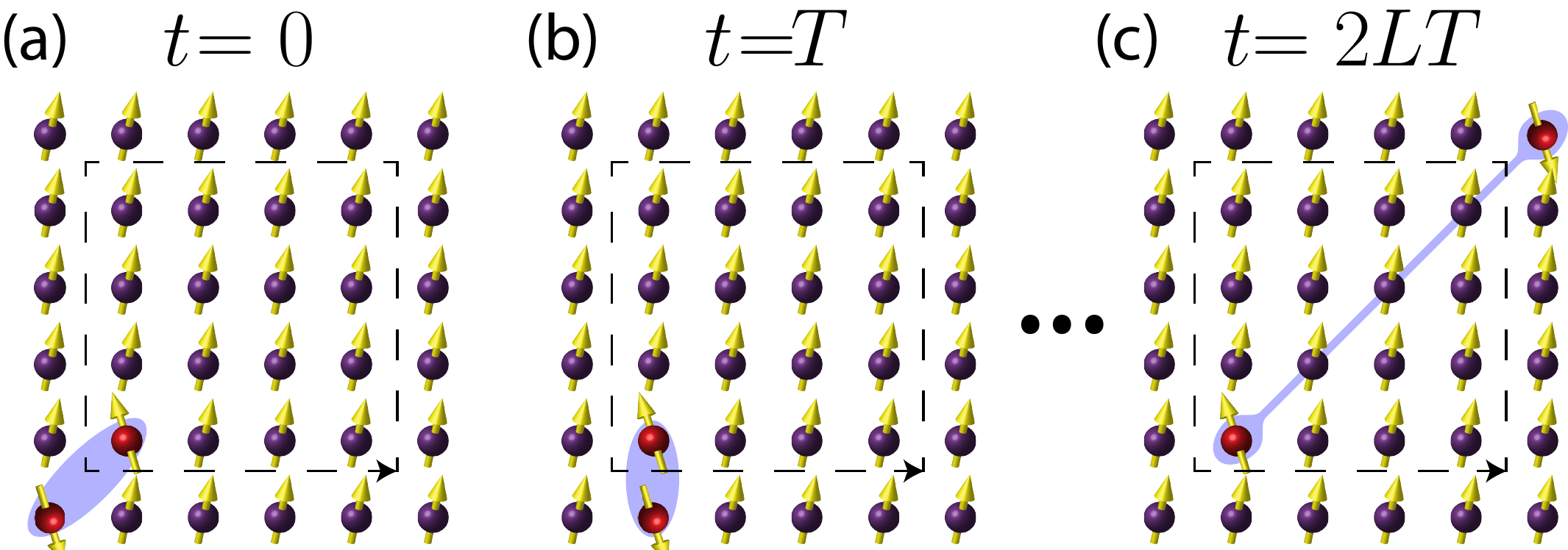}} 
\caption{
\label{fig:AnB}
{\bf Transport of quantum entanglement by the edge of a 2d bosonic MBL chiral Floquet system.} The dashed line indicates schematically the separation between bulk and edge DOF, with the arrow indicating the direction of information transport along the edge. (a) Quantum entanglement (represented by blue shade) is locally created at $t=0$. (b,c) Entanglement is transported through a distance of $\mathcal O(L)$ in time $\mathcal O(LT)$.
 }
\end{center}
\end{figure}

As a thought experiment, imagine Alice and Bob, sitting respectively at $(0,0)$ and  $(L,L)$, are separated by an experimental setup realizing the bosonic MBL chiral Floquet model discussed in Sec.~\ref{sec:BaseModel}. Suppose the system is initialized into a trivial product state, say with all spin-1/2's pointing `up'.
At time $t=0$, Alice can locally create quantum entanglement by entangling a bulk and an edge qubit on her side, respectively at $(1,1)$ and $(0,0)$, and prepare the initial state
\begin{equation}\begin{split}\label{eq:}
|\Psi(0)\rangle =|\Uparrow \rangle
\otimes \frac{1}{\sqrt{2}} \left( |\uparrow \rangle_{(0,0)} |\downarrow \rangle_{(1,1)} - |\downarrow \rangle_{(0,0)} |\uparrow \rangle_{(1,1)}\right),
\end{split}\end{equation}
where $|\Uparrow \rangle$ denotes the `all-up' state for all sites other than those explicitly written out (Fig.~\ref{fig:AnB}a). After one Floquet period ($t=T$), the bulk DOF only pick up an on-site random phase, whereas all the edge DOF are `advanced' by one unit along the edge (Fig.~\ref{fig:AnB}b), and hence the state evolves into
\begin{equation}\begin{split}\label{eq:}
&|\Psi(T)\rangle \\
&\propto  |\Uparrow \rangle
\otimes \frac{1}{\sqrt{2}} \left( |\uparrow \rangle_{(1,0)} |\downarrow \rangle_{(1,1)} - e^{- i \phi }|\downarrow \rangle_{(1,0)} |\uparrow \rangle_{(1,1)}\right),
\end{split}\end{equation}
where $e^{- i \phi}$ is an unimportant phase.
Therefore, after $2L$ Floquet periods (Fig.~\ref{fig:AnB}c), one finds
\begin{equation}\begin{split}\label{eq:}
&|\Psi(2LT)\rangle \\
&\propto  |\Uparrow \rangle
\otimes \frac{1}{\sqrt{2}} \left( |\uparrow \rangle_{(L,L)} |\downarrow \rangle_{(1,1)} - e^{- i \phi' }|\downarrow \rangle_{(L,L)} |\uparrow \rangle_{(1,1)}\right),
\end{split}\end{equation}
i.e.~the `pumping' of quantum information manifests as a transfer of local entanglement across the entire system. Hence, Alice and Bob now share an entangled pair of qubits, transported by the `one-way information highway' at the edge of a 2d bosonic MBL chiral Floquet system, with which they can utilize for quantum communication. 

\section{Fermionic chiral Floquet phases}
\label{sec:Fermions}
We have so far focused on 2d bosonic systems, where we established rigorous results using the GNVW results and an MPU reformulation. However, the basic construction of the SWAP models is very closely related to the noninteracting fermionic AFAI models in Refs.~[\onlinecite{Demler,RudnerLevin,TitumLindner,Lindner}]. In particular, the AFAI model features a chiral edge mode in the clean single-particle Floquet spectrum, and in fact we show in Appendix \ref{sec:AFAI} that, viewed as a many-body operator, the AFAI Floquet evolution acts as the fermion unit translation operator along the edge. As our heuristic arguments apply equally well to fermions, this strongly suggests the AFAI models are also stable to interactions. 

In light of these parallels, it is natural to ask how the bosonic and fermionic chiral Floquet phases are related: Is the fermionic phase equivalent to a bosonic one with a particular chiral topological index $\nu$? In equilibrium, we know that chiral phases are characterized by their chiral central charge $c$, which measures the quantized thermal Hall conductance (aka gravitational anomaly) of the bulk. It is known that without intrinsic bulk topological order, the minimal chiral complex fermion phase is an IQH insulator with $c_{\rm C}=1$, whereas the minimal chiral bosonic phase is the E8 state with $c_{\rm B}=8$.\cite{KitaevHoneycomb} Hence, in equilibrium, one needs to combine $8$ of the minimal complex fermion phases together to equate to a minimal bosonic phase. How is this relation modified in the Floquet setting, where continuous chiral flow of heat is replaced by discrete chiral pumping of quantum information?

Before diving into the details, let us anticipate the results. First, observe that the fermionic chiral Floquet phase (e.g. the AFAI model) pumps a two-state qubit of quantum information ($0$ or $1$ fermion occupation) along the edge. This is the same quantum information capacity as the spin-1/2 bosonic model, which has index $\nu =\log 2$. Hence, we expect the `conversion rate' between the minimal chiral Floquet phases of complex fermion and hardcore bosons to be $1:1$.
We will see that this is indeed the case, which is in stark contrast to the equilibrium result of $c_{\rm B}: c_{\rm C}  = 8:1$.
In the arguments we will assume that, even in the presence of fermions, the chiral Floquet  phases are classified by an index that is additive under stacking.
We will also encounter a Majorana fermion version of the model, which is loosely speaking the square-root of the minimal complex fermion chiral Floquet phase, and hence can be interpreted as having edges that pump a fractional amount of quantum information. 

To establish this, we will first formally decompose complex fermions into pairs of real (Majorana) fermions, and then relate the Majorana chiral edges to complex fermions and subsequently hardcore bosons.
Consider again a checkerboard lattice and a four-step driving protocol similar to the one described in Sec.~\ref{sec:BaseModel}, where instead of hardcore bosons, we let each site $\vec x$ host a complex fermion with creation operator $\hat c_{\vec x}^\dagger$. We then define the Majorana fermions via
\begin{equation}\begin{split}\label{eq:}
\hat \chi_{\vec x} \equiv \hat c_{\vec x} + \hat c_{\vec x}^\dagger ;~~~
\hat {\bar \chi}_{\vec x} \equiv \frac{1}{i} (\hat c_{\vec x} - \hat c_{\vec x}^\dagger ),
\end{split}\end{equation}
and consider a drive that acts nontrivially on $\chi$ but trivially on $\bar \chi$.
The model construction and analysis, detailed in Appendix \ref{sec:Majorana}, is essentially identical to before, with only minor modifications that do not affect the topological nature of the model. At a single edge, the clean model again features $\hat Y = \hat t_{\rm M}$, the unit right-translation operator for the $\chi$-Majorana fermions:
\begin{equation}\begin{split}\label{eq:M_trans}
\hat t_{\rm M} \hat \chi_{x} \hat t_{\rm M}^\dagger = \hat \chi_{x+1},
\end{split}\end{equation}
where we let $x$ indexes the site along the edge, and $\hat t_{\rm M}$ acts trivially on $\bar \chi$.

We can also apply the same driving protocol to the $\bar \chi$ fermions, which is effectively the same as stacking two copies of chiral Floquet models of Majorana fermions.
If $\bar\chi$ are driven in the opposite chirality compared to $\chi$, at the edge we will have a pair of counter-propagating Majorana translation operators, which can arise from a purely 1d local Hamiltonian evolution and the bulk is therefore trivialized (Appendix \ref{sec:Majorana}).
The more interesting case is when $\chi$ and $\bar \chi$ are driven with the same chirality.
This gives the edge operator $\hat Y = \hat t_{\rm M}  \hat t_{\rm \bar M}$, where $\hat t_{\rm \bar M}$ denotes the unit right-translation operator for $\bar \chi$. This gives
\begin{equation}\begin{split}\label{eq:}
(\hat t_{\rm M}  \hat t_{\rm \bar M}) \hat c_{x}^\dagger (\hat t_{\rm M}  \hat t_{\rm \bar M} )^\dagger = \hat c_{x+1}^\dagger,
\end{split}\end{equation}
and hence we identify $\hat t_{\rm M}  \hat t_{\rm \bar M}$ as the translation operator for the complex fermion, i.e.~the conversion ratio between Majorana and complex fermions is $2:1$, as one would expect. 
If we let $\eta_{\rm M}$, $\eta_{\rm C}$ respectively count the copies of unit right-translation operators for Majorana and complex fermions (left-translations are counted as negatives), we can label any edge by $(\eta_{\rm M}, \eta_{\rm C})$, which is additive under stacking. The two-to-one conversion between Majorana and complex fermions described above implies $(2n+m,0) \sim (m,n)$ for any pair of integers $n,m$.

\begin{figure}[b]
\begin{center}
{\includegraphics[width=0.45 \textwidth]{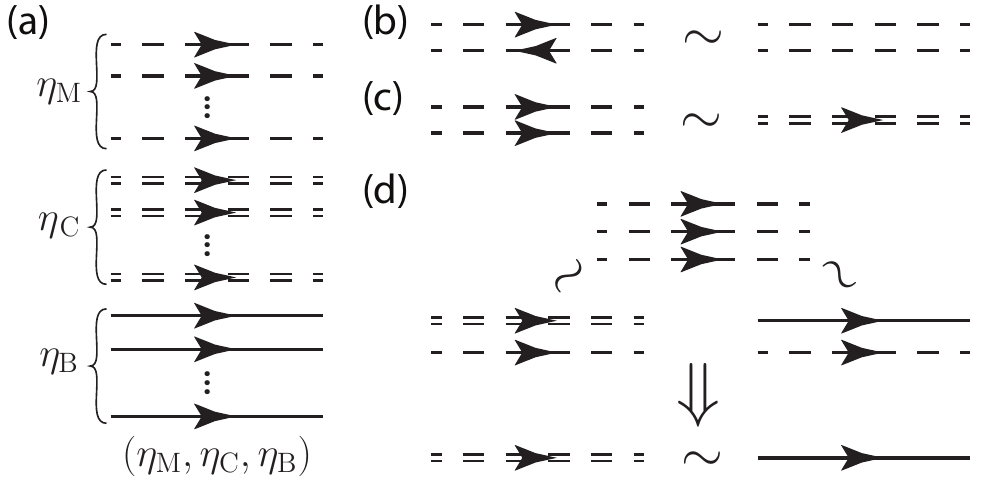}} 
\caption{{\bf Equivalence of fermionic and bosonic chiral edges.}
(a) For a system with Majorana fermions (dashed lines), complex fermions (double dashed lines) and hardcore bosons (solid lines), a chiral edge can be labeled by the tuple of integers $(\eta_{\rm M},\eta_{\rm C},\eta_{\rm B})$, where $|\eta|$ counts the net number of chiral edges for the three particle types, and its sign encodes the chirality. (b) Two counter-propagating Majorana chiral edges will annihilate, corresponding to $(1,0,0) + (-1,0,0)\sim (0,0,0)$. (c) Two co-propagating Majorana edges are equivalent to one complex fermionic edge, implying $(2,0,0)\sim (0,1,0)$. (d) Using a relabeling argument, one sees that $(3,0,0)\sim (1,0,1)$, and therefore $(0,1,0)\sim (0,0,1)$, i.e.~a chiral complex-fermion edge is equivalent to a chiral hardcore-boson edge.
\label{fig:BF_Edge}
 }
\end{center}
\end{figure}

Now we add hardcore bosons, or equivalently spin-1/2's, to the discussion. Their chiral unitary indices are always given by $\nu = \eta_{\rm B} \log 2$, where $\eta_{\rm B} \in \mathbb Z$ can again be viewed as a count of the number of bosonic chiral edge operators. 
The edge characterization is now extended to the tuple of integers $(\eta_{\rm M} , \eta_{\rm C} , \eta_{\rm B} )$. 
To connect fermionic and bosonic models, we will establish $(3,0,0)\sim (1,0,1)$, i.e.~starting with an edge with three copies of chiral Majorana operators, we can convert a pair of them (corresponding to two-dimensional site Hilbert spaces) to a chiral hardcore-boson edge (Fig.~\ref{fig:BF_Edge}). Combined with the conversion between Majorana and complex fermions, we have $(1,1,0) \sim (1,0,1) \Rightarrow (0,1,0) \sim (0,0,1)$.
This implies that, in our MBL Floquet setting, chiral edges of complex fermions and hardcore bosons can be converted to one another in a $1:1$ ratio, which is in stark contrast to the equilibrium result of $8:1$.
Note, however, that one cannot directly convert a fermionic model to a bosonic one, since doing so will violate fermion-parity conservation. As we will see, in canceling $(0,1,0)$ by stacking with $(0,0,-1)$, it is crucial that a trivial pair of counter-propagating Majorana chiral edges is present.
Hence, the statement $(0,1,0) \sim (0,0,1)$ is to be interpreted in the sense of stable equivalence.

To complete this discussion, it remains to show $(3,0,0)\sim (1,0,1)$, which we will demonstrate using a relabeling argument. Let the three co-propagating Majorana translation operators be acting on $\hat \chi_{i}$ for $i=1,2,3$, where the site label is suppressed.
We can formally recast the site Hilbert space as that arising from a hardcore boson, $\hat{\boldsymbol{\tau}}$, tensored with a new Majorana fermion, $\hat{\mu}$, by defining the following operators (on each site):
\begin{equation}\begin{split}\label{eq:}
\hat \mu =& i \hat \chi_1 \hat \chi_2 \hat \chi_3;\\
\hat \tau_1 =& i \hat \chi_1 \hat \chi_2; ~~
\hat \tau_2 = i \hat \chi_2 \hat \chi_3;~~
\hat \tau_3 = i \hat \chi_1 \hat \chi_3,
\end{split}\end{equation}
where the operators $\{ \hat \tau_i\}$ verify the algebra of Pauli matrices and conserve fermion parity. Importantly, $[\hat \mu,\hat \tau_i ] = 0$, and hence $\hat \mu$ and $\hat {\boldsymbol{ \tau}}$ can be regarded as independent degrees of freedom.
Restoring the site indices, the edge evolution simply translates: $\hat{\mu}_x \rightarrow \hat{\mu}_{x+1}$ and $\hat{\boldsymbol{\tau}}_x\rightarrow\hat{\boldsymbol{\tau}}_{x+1}$. We can therefore reinterpret the edge as a chiral Majorana fermion together with a decoupled chiral hardcore boson, which implies $(3,0,0)\sim (1,0,1)$.

Such equivalence implies the fermionic chiral Floquet phases are as stable as their bosonic counterparts. 
While we have not ruled out the logical possibility that the bosonic chiral phases are trivialized in the presence of idle fermions, it is highly unlikely on physical grounds.
Nonetheless, in order to rigorously establish the stability of fermionic chiral phases, one must rule out this possibility by a nontrivial extension of the GNVW classification to fermionic algebras. We will leave this as a challenge for future works.

Finally, we comment on the physical stability of the Majorana phase. While one can construct MBL Majorana models (Appendix \ref{sec:Majorana}), experimentally relevant fermionic systems exhibit charge conservation, and cannot realize the Majorana phase without spontaneous superfluidity or superconductivity. Since such phases feature Goldstone modes (superfluids) or introduce long range interactions (superconductors), they appear to be at odds with many-body localization. Hence, it is more natural to consider the Majorana phases as appearing in fractionalized systems with topological order.\cite{DrewRomain} We leave these generalizations to future works.

\section{Discussion}
\label{sec:Dis}
We showed that the edge dynamics of bosonic 2d systems in an MBL chiral Floquet phase can exhibit a chiral flow of quantum information. These chiral dynamics are anomalous in that they cannot be achieved by any local Hamiltonian evolution in 1d, and are characterized by the quantized chiral unitary index taking the form $\nu = \log (p/q)$, where $p/q \in \mathbb Q_{+}$ is a positive rational number.\cite{GNVW}
These chiral Floquet phases rely on the robustness of MBL phases in 2d, so far an unproven conjecture. However, we reiterate that even in the absence of stable 2d many-body localization, our description will nonetheless provide an accurate universal characterization of parametrically long time pre-thermal dynamics in strongly disordered systems.

Aside from a formal classification via bulk-boundary correspondence, we have also constructed a full set of representative exactly soluble models, developed the concrete numerical framework for index computation, proposed an experimental realization using hardcore ultracold bosons in a shaken optical lattice, elaborated on the physical consequences arising from the nontriviality of the phase, and explored their fermionic counterparts.

Our results suggest that interaction, localization, chirality and periodic driving conspire to realize a novel phase of topological matter -- a new tune that cannot be played if anyone in the quartet is missing. 
As chiral phases are known to be the `root' of many topological phases, the present work stimulates numerous further inquires. We highlight three particular aspects below:

(i) For bosonic systems, we were able to rigorously establish the stability and completeness of the chiral floquet classification via the formal machinery of GNVW. For fermionic systems, we demonstrated the topological equivalence of chiral edges of certain fermionic and bosonic models. However, classification of fermionic (or more general anayonic) Floquet chiral phases, with the same rigor as the purely bosonic case, will require a nontrivial extension of the GNVW results to fermionic or general anyonic operator algebras, which we leave as an important open problem.

(ii) The classification of topological phases is generally richer in the presence of extra symmetry constraints, say global spin rotation symmetries.
Recently, much progress has been made in the classification of Floquet SPT phases of interacting bosons, where once again MBL was invoked as an essential ingredient.\cite{Khemani,Keyserlingk, ElseNayak, PotterMorimotoVishwanath,Roy}
It was proposed that the analogue of cohomology classification in this context of Floquet SPTs in $d$ spatial dimensions can be achieved by identifying the phases with elements of the cohomology group $H^{d+1}(G\times \mathbb Z, U(1))$, where $G$ is a unitary internal symmetry group and $\mathbb Z$ denotes the discrete time translation.\cite{ElseNayak,PotterMorimotoVishwanath}
Although similar cohomology classifications are known to be incapable of capturing the equilibrium chiral phases,\cite{XieGuLiuWen} we point out in Appendix \ref{sec:Coho} an interesting potential link between the two in our Floquet setting. In addition, recall that the edge of an equilibrium SPT phase can be pictured as a pair of counter-propagating chiral edge modes symmetry-protected from cancellation.\cite{XieGuLiuWen} What, if any, are the new phases of matter that arise from a symmetry-enriched version of the present work?

(iii) 
In Sec.~\ref{sec:AnB}, we pointed out how the unidirectional transport of quantum information at the edge of a bosonic MBL chiral Floquet phase can be utilized for entanglement sharing, which is a basic ingredient for any quantum communication protocols.
However, the idealized description there, which involves the exact addressing of the l-bits, cannot be directly applied to a system with generic disorder.
Away from this idealized limit, there will almost certainly be a logarithmically slow spreading of entanglement into the MBL bulk (as for the dephasing in any MBL time evolution), and also to some extent back to the edge. This is in contrast to the ballistic spread in a thermalizing system, and partially quantifies the universal aspect of the `robustness' to imperfect register addressing.
Yet, how the thermal nature of the edge impacts the protocols is less clear.
For instance, in the presence of disorder the chiral information channel is noisy, and would likely decohere the entangled pair. Can this decoherence be efficiently echoed away? Since the entanglement transport is a manifestation of the nontrivial chiral unitary index, can one leverage the quantized nature of the index to design a more robust quantum communication channel?

We leave these questions for future work. 

\emph{Note added --} After posting our manuscript, a related work by Harper and Roy appeared,\cite{HarperRoy} which gives arguments for the stability of the chiral edge of bosonic phases with particle-number conservation, and suggests a similar classification by rational numbers.

\begin{acknowledgements}
The authors would like to acknowledge useful discussions with T.H.~Leung, R.S.K.~Mong, A.~Harrow, M.~Hastings, M. ~Zaletel and Z.~Xiong.  HCP is supported by a Hellman graduate fellowship.  LF is supported by NSF DMR-1519579 and by Sloan FG-2015-65244.  TM and AP are supported by the Gordon and Betty Moore Foundation's EPiQS Initiative through Grant GBMF4307.  AV acknowledges support from a Simons Investigator Award and AFOSR MURI grant FA9550-14-1-0035.
\end{acknowledgements}

\begin{appendix}

\section{Exact solubility of the SWAP model}
\label{sec:SWAP}
In this appendix, we discuss the exact solubility of any quantum circuits composing only of SWAP gates. Intuitively, such solubility can be understood as follows: A SWAP gate amounts to a mere relabeling of the two sites involved, and hence a circuit of SWAP gates can only permute the site labels in the lattice. Such permutation is a classical operation, and so the action of the circuit can be computed efficiently with a cost that is only extensive in the system size, akin to a free fermion problem. 
More specifically, these computations can be done using properties of the symmetric group $\mathcal S_n$ -- the group of permutation of $n$ objects.

\subsection{Correspondence with $\mathcal S_n$}
The main goal of this subsection is to provide a concrete definition of the SWAP gate, and to highlight certain properties that are important to our computation. 
Readers who are already familiar with the SWAP gate can skip this subsection.

To be self-contained, we will first introduce our notations again, and provide a sharper definition of the SWAP gate.
Let $\Lambda$ be a lattice of identical quantum spins, with the number of sites $|\Lambda|$ finite. The full Hilbert space is $\mathcal H_\Lambda = \bigotimes_{i\in\Lambda} \mathcal H_i$, where $p\equiv \dim(\mathcal H_i)$ is the dimension of the site Hilbert spaces. Let $\{ \hat \tau_{ij}~:~i,j\in\Lambda\}$ be the set SWAP gates on the system, which are unitary operators satisfying the following:
\begin{enumerate}
\item $\hat \tau_{ij}^2 = \hat 1$, and hence $\hat \tau_{ij}^\dagger = \hat \tau_{ij}$.
\item For any on-site operator, $\hat \tau_{ij}$ swaps the action of the operator on the sites $i$ and $j$, i.e.~
\begin{equation}\begin{split}\label{eq:}
& \hat \tau_{ij} \left(  \dots \otimes \hat O^i  \otimes\dots \otimes \hat O^j \otimes\dots \right)\hat \tau_{ij} \\
=& \dots \otimes \hat O^j  \otimes\dots \otimes \hat O^i \otimes\dots,
\end{split}\end{equation}
where the tensor-product is written using a fixed ordering of the sites; and
\item $\hat \tau_{ij}$ can be decomposed as 
\begin{equation}\begin{split}\label{eq:SWAPExp}
\hat \tau_{ij} = \sum_{\alpha} \hat M^{\alpha}_{i} \hat M^{\alpha}_{j} = \hat \tau_{ji},
\end{split}\end{equation}
where $\hat M^{\alpha}_{i}$ is nontrivial only at site $i$, i.e.~$ \hat M^{\alpha}_{i} = \dots \hat 1 \otimes \hat 1\otimes \hat O^{\alpha} \otimes \hat 1\otimes \hat 1\dots$, and $\hat O^{\alpha}$ depends only on $\alpha$ but not $i$.
Note that $\hat \tau_{ii}$ is the identity.
\end{enumerate}

While properties [1] and [2] can be viewed as defining properties of the SWAP gate, [3] is a statement on locality and is more subtle. 
Indeed, since fermionic operators are never quite local, the discussion here is restricted to bosonic (spin) systems. 
Besides, a decomposition similar to Eq.~\eqref{eq:SWAPExp} is not unique. Yet, given any such bi-local decomposition, one can rewrite it in the stated symmetric form using properties [1] and [2].

We aim to establish that the group generated by the set of SWAPs, a subgroup of the group of unitary operators on $\mathcal H_\Lambda$, is equivalent to the symmetric group $\mathcal S_{|\Lambda|}$. Loosely, one simply notes that $\mathcal S_{|\Lambda|}$ is the group of site permutations, and $\mathcal S_{|\Lambda|}$ is known to be generated by pair-wise exchanges, which are frequently called `transpositions'. At the `quantum level', these transpositions are given by SWAPs, and hence the claimed equivalence.

More concretely, we introduce the set of objects that are permuted in our context. For each site $i\in \Lambda$, we introduced the local observable algebra, $\mathcal A_{i}$, which is simply the set of local quantum operators, nontrivial only at site $i$, endowed with addition (over $\mathbb C$) and multiplication. Now consider the set $\{ \mathcal A_{i} ~:~ i \in \Lambda\}$, which for $p>1$ is a finite set of $|\Lambda|$ objects. We can therefore define the symmetric group $\mathcal S_{|\Lambda|}$ as the (abstract) group permuting $\mathcal A_i$'s. Now, recall that $\mathcal S_{|\Lambda|}$ is generated by (a subset of) the set of transpositions, and one can check from the listed properties that $\hat \tau_{i,j} \mathcal A_{i} \hat \tau_{i,j} = \mathcal A_j$ and $\hat \tau_{i,j} \mathcal A_{j} \hat \tau_{i,j} = \mathcal A_i$. Hence, we arrive at the stated correspondence.

This correspondence implies that any circuit $\hat {\mathcal C}$ of SWAP gates corresponds to an element $\pi(\hat {\mathcal C}) \in \mathcal S_{|\Lambda|}$, where $\pi$ is the isomorphism sending $\hat \tau_{i,j}$ to the corresponding transposition in (the abstract group) $\mathcal S_{|\Lambda|}$. Two circuits $\hat {\mathcal C}$ and $\hat {\mathcal C}'$ are equivalent iff $\pi(\hat {\mathcal C}) =\pi(\hat {\mathcal C}')$. Hence, to solve $\hat {\mathcal C}$, one simply reduces the corresponding permutation $\pi(\hat {\mathcal C})$ to a simple form, as we will illustrate in the next subsection. 
In addition, note that the only condition we have imposed on the site Hilbert space is $p>1$, and therefore the analysis of a SWAP circuit is largely independent of $p$.

\subsection{Solutions of the SWAP circuits}
As discussed in the main text, the restriction of a SWAP circuit to the single-spin-flip sector is only a $|\Lambda|$-dimensional matrix, and hence can be solved efficiently. The mentioned correspondence between the group of SWAP circuits and the symmetric group implies the full many-body solution can be immediately inferred from this computation: One can show that the restriction to the single-spin-flip sector gives nothing other the `natural permutation representation', which is a faithful representation  of $S_{|\Lambda|}$. Once we have determined $\pi(\hat C)$ using this representation, the entire spectrum and eigenstates of $\hat C$ are determined, as we demonstrate below.

Recall that any element in $S_{|\Lambda|}$ can be written in a unique `cycle decomposition', which follows from the natural permutation action of $S_{|\Lambda|}$ on the set of $|\Lambda|$ objects. For instance, the notation $(a,b,c,d)(e,f)$ refers to a permutation $a\rightarrow b \rightarrow c\rightarrow d \rightarrow a$ and $e \leftrightarrow f$.  In addition, note that the cycles are `disjoint', i.e.~each cycle acts on a different subset of the set of symbols and therefore different disjoint cycles commute.
A cycle involving only two objects, like $(e,f)$, is a `swap' of the two and is called a `transposition'. As discussed, $S_{|\Lambda|}$ is generated by the set of  transpositions, and indeed one can verify that $(a,b,c,d) = (a,b)(b,c)(c,d)$. 

Using the established isomorphism, one can compute $\hat C$ explicitly via the cycle decomposition. For instance, suppose $\pi(\hat C) = (a,b,c,d)(e,f)$, then 
\begin{equation}\begin{split}\label{eq:}
\hat C = \pi^{-1} \circ \pi( \hat C) 
=& \pi^{-1} \left( (a,b)(b,c)(c,d)(e,f) \right)\\
=& \hat \tau_{a,b} \hat \tau_{b,c} \hat \tau_{c,d} \hat \tau_{e,f} .
\end{split}\end{equation}
Hence, the disjoint cycle decomposition of $\pi(\hat C)$ corresponds to a factorization of $\hat C$ into commuting pieces, and any SWAP circuit on $\Lambda$ can be viewed as a single layer of commuting unitary operators (which are not necessarily local). 
In addition, observe that $\pi^{-1} ((a,b,c,d))$ is nothing but the translation operator for the four-site `ring' with sites $a$, $b$, $c$ and $d$. As all the eigenvalues and eigenstates of the translation operator (for any number of sites and $p$) can be readily computed, $\hat C$ can be exactly solved.

\section{The `diluted' SWAP model and mapping to classical percolation}
\label{sec:perco}
In this appendix, we discuss in details the construction and analysis of the `diluted' SWAP model, whose properties are shown in Fig.~\ref{fig:dSWAP} of the main text.

Recall that we define the `diluted' SWAP model as a bond-diluted version of the model introduced in Sec.~\ref{sec:BaseModel}, where any of the SWAP gates are removed at random with probability $s$, and left intact with probability $1-s$. Intuitively, the chiral phase should survive for a very small dilution probability $s\ll 1$, whereas when most bonds are removed, $s\approx 1$, one expects a trivial phase. 

This defines a family of disordered model, labeled by $0 \leq s \leq 1$, that connects the chiral bosonic model ($s=0$) to a trivial model ($s=1$). As detailed in Appendix \ref{sec:SWAP}, a quantum circuit composing only of SWAP gates defined on a lattice $\Lambda$ is exactly soluble, due to its equivalence to a permutation in the symmetric group $\mathcal S_{|\Lambda|}$. One can therefore also analyze the entire family of diluted SWAP models efficiently.

Our main goal is to study the disorder average of $\xi$, the localization length of the Floquet operator (with PBC) defined in Sec.~\ref{sec:Edge} of the main text. As the fifth time step in our model only gives rise to an on-site phase factor, $\xi$ is determined by the permutation factor $\hat P_{\rm F}$ coming from the first four steps of the Floquet cycle. We have established in Appendix \ref{sec:SWAP} that the permutation factor can be factorized into mutually commuting pieces via the cycle decomposition.
Translating into the physical setup, each factor in the decomposition $\hat U_{\rm F} = \prod_{r} \hat U_r$ corresponds to a disjoint cycle, and each `cycle' $r$ is a collection of sites $r = (\vec r_1, \vec r_2,\dots, \vec r_{l_r})$, with the following interpretation: after each Floquet period, a spin flip localized at site $\vec r_{i}$ goes to site $\vec r_{(i + 1 )}$ (mod $l_r$). The radius of the neighborhood $b'(r)$ is then simply $\max(\{ |\vec r_i - \bar {\vec r}| ~:~ i = 1,\dots, l_{r}\})$, where $\bar {\vec r} = \frac{1}{l_r} \sum_{i=1}^{l_r} \vec r_{i} $ is the average location of the sites in the cycle. The localization length $\xi$ is then defined as the maximum of the radii of $b'(r)$ as $r$ runs through all the disjoint cycles. 
As shown in Fig.~\ref{fig:dSWAP}, $\langle \xi\rangle$ diverges around $s=1/2$ and signals the breakdown of many-body localization.

\begin{figure}[tb]
\begin{center}
{\includegraphics[width=0.48 \textwidth]{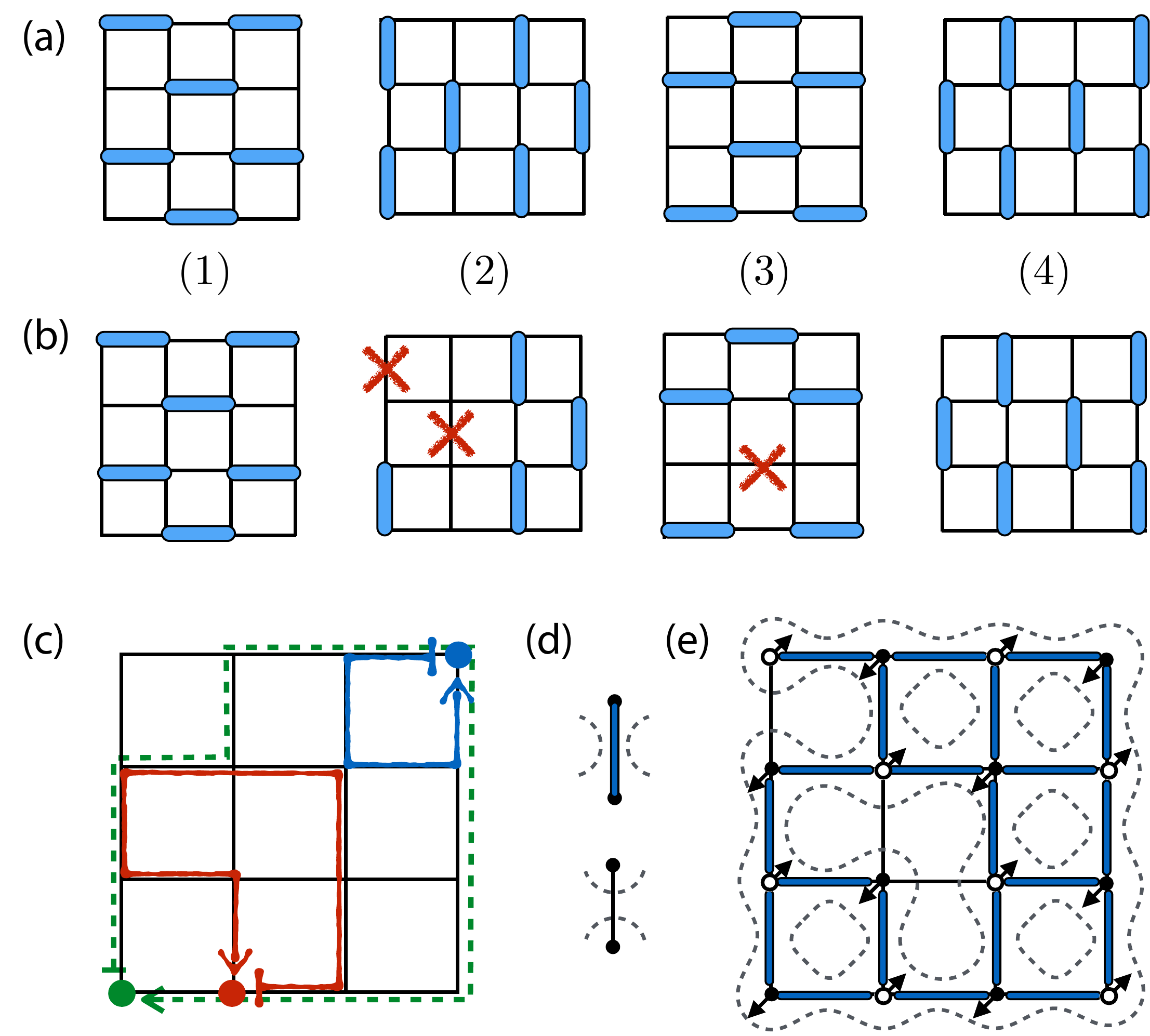}} 
\caption{{\bf Percolation picture of diluted SWAP model} (a) A four-step sequence of SWAP gates (blue dimers) that realizes a chiral Floquet model; (b) A disordered SWAP model obtained by turning off bonds at random (red $\times$'s); (c) Example micro-motion trajectories of three spin flips (red, blue, and green arrows) starting on particular sites (large dots); 
(d) Rules for drawing the dual loops for an occupied (top) and unoccupied (bottom) bonds in the percolation model;
(e) The dual dense-packed loop (dashed lines) representation of a classical percolation cluster, where only the fat blue bonds are occupied.
Spin flips in the corresponding diluted SWAP model follow the loop trajectories indicated by the small arrows.
\label{fig:percolation}
 }
\end{center}
\end{figure}

While $\xi$ is a general diagnostic for many-body localization in any system, in our present construction one can consider another length scale that serves as a proxy of $\xi$. Pictorially, one can imagine tracking the position of a spin flip as it moves under the drive. Following the discussion above, a spin flip must come back to its starting position after $l_r$ Floquet period, where $l_r$ is the length of the cycle the spin flip belongs to. Hence, the trajectories of spin flips in the long-time limit define loops on the lattice, and the length of these loops serves as another length scale differentiating between localized and delocalized behaviors. As we will demonstrate below, utilizing this `loop' picture one can formally map the diluted SWAP model to a classical bond percolation model, where bonds are occupied and unoccupied with probability $s$ or ($1-s$) respectively. The mapping implies the loop lengths will diverge at $s_c= 1/2$, which is consistent with the divergence of $\langle \xi \rangle$.

The mapping proceeds as follows:
For any diluted sequence of SWAP gates (see e.g. Fig.~\ref{fig:percolation}b), one can track the orbits (closed trajectories) of a spin flip initially residing on a particular site (Fig.~\ref{fig:percolation}c). In the chiral phase, edge trajectories circumnavigate the boundary (green dashed line), whereas bulk trajectories form small loops (blue and red lines). In the trivial phase, the chiral edge trajectories are absent, and all bulk loops are again small. To expose the connection to percolation, it is useful to utilize a dual dense-packed loop representations of classical bond percolation clusters. Consider a dense packed loop configuration in which each bond has an accompanying pair of loop segments, that are either oriented parallel to the bond (Fig.~\ref{fig:percolation}d, top panel) if the bond is occupied, or perpendicular to the bond (Fig.~\ref{fig:percolation}d, bottom panel) otherwise. Connecting up the loop segments produces a set of closed loops (Fig.~\ref{fig:percolation}e), which one can easily verify are in one-to-one correspondence with the trajectories of spin flips in the corresponding diluted SWAP model. As indicated by small arrows in (Fig.~\ref{fig:percolation}e), a spin flip initial on a given site executes an orbit corresponding to the loop either down and to the left (A sublattice, filled circles) or up and to the right (B sublattice, open circles).

This mapping immediately shows that the diluted SWAP model has two distinct phases with ($s<1/2$) and without $(s>1/2)$ chiral edges, which survive over a finite range of dilution probability, $s$. The chiral edge is lost sharply at $s_c=1/2$, in which the model is characterized by classical percolation exponents associated with a single diverging length scale $l\sim (s-1/2)^{-\nu}$ with $\nu=4/3$. 

However, we caution that the classical percolation character of the transition is likely special to the perfect SWAP gates used in the model, and that more generic quantum perturbations, such as disordering the local strengths of the Hamiltonian, will likely induce quantum fluctuations that change the nature of the transition. An analogous example is that of the quantum Hall plateau transitions, which can be viewed in a loose sense as percolation of quantum Hall droplets. In that case, tunneling between chiral edge states of the droplets causes the universal scaling properties to differ from those of classical bond percolation.

\section{The GNVW index}
\label{sec:Algebras}
Here we provide an explicit derivation of some of the claims on the properties of the chiral unitary index $\nu$, and discuss why our exposition is non-rigorous. Note that all the claims have been discussed in GNVW, and so are not original. 

We will first show a series of results concerning the `algebra overlap' $\eta$. To this end, recall the definition (with summation convention on repeated indices)
\begin{equation}\begin{split}\label{eq:}
\eta^2\left(\mathcal A, \mathcal B \right) \equiv  \frac{ p_a p_b }{p_{\Lambda}^2 }    \text{Tr}_\Lambda \left( \hat e_{ij}^{a \dagger} \hat e_{lm}^{b} \right) 
\text{Tr}_\Lambda \left( \hat e_{lm}^{b \dagger}  \hat e_{ij}^{a } \right),
\end{split}\end{equation}
where $\Lambda$ denotes a sufficiently large, but finite, set of sites such that any operators $\hat a \in \mathcal A$ and $\hat b \in \mathcal B$ can only act nontrivially within $\Lambda$.\\

$\eta$ has the following properties:\\
\newcounter{PropList}
\stepcounter{PropList}

\noindent (\thePropList) \stepcounter{PropList} $\eta$ is independent of the choice of $\Lambda$.\\

The normalization $p_a p_b / p_{\Lambda}^2$ is precisely chosen for this. Suppose $\Lambda'$ contains one extra site $x$ compared to $\Lambda$, such that $\mathcal H_{\Lambda'} = \mathcal H_{\Lambda} \otimes \mathcal H_{x}$ and $p_{\Lambda'} = p_{\Lambda} p_x$.  Since for any operator $\hat O$ with support $\Lambda$, $\text{Tr}_{\Lambda'}(\hat O|_{\Lambda'}) = \text{Tr}_{\Lambda'}(\hat O|_{\Lambda} \otimes \hat 1_{x})  = p_x\text{Tr}_{\Lambda}(\hat O|_{\Lambda} )  $, the claim follows.\\

\noindent (\thePropList) \stepcounter{PropList} $\eta(\mathcal A, \mathcal B)$ is independent of the arbitrary choice of bases for  $\mathcal A$ and $\mathcal B$.\\

For $\alpha = a,b$, consider a transformation of basis given by the $p_{\alpha}$-dimensional unitary matrix $U^{\alpha}_{ij}$: $\{ \hat e_{ij}^{\alpha} \} \rightarrow \{ U^{\alpha}_{il}\hat e_{lm}^{\alpha}  U^{\alpha *}_{jm} \}$.  Then compute
\begin{equation}\begin{split}\label{eq:}
&   \text{Tr}_\Lambda \left( 
U^{a*}_{iv}\hat e_{vw}^{\alpha \dagger}  U^{a}_{jw}    
U^{b}_{lx} \hat e_{xy}^{b} U^{b*}_{my}
     \right) \times\\
&~~~~~~~~~~~~~~~~~~\text{Tr}_\Lambda \left( 
U^{b*}_{lr}\hat e_{rs}^{b \dagger} U^{b}_{ms}
U^{a}_{it}  \hat e_{tu}^{a } U^{a*}_{ju}
\right)\\
=&  
\left( U^{a*}_{iv}  U^{a}_{it}   \right)
\left(U^{a}_{jw}   U^{a*}_{ju}  \right)
\left( U^{b}_{lx}  U^{b*}_{lr}  \right)
\left( U^{b*}_{my} U^{b}_{ms} \right) \times\\
&~~~~~~~~~~~~~~~~~~
\text{Tr}_\Lambda \left( \hat e_{vw}^{\alpha \dagger}   \hat e_{xy}^{b} \right)  
\text{Tr}_\Lambda \left( 
\hat e_{rs}^{b \dagger} 
\hat e_{tu}^{a } 
\right)\\
=&\text{Tr}_\Lambda \left( \hat e_{vw}^{\alpha \dagger}   \hat e_{xy}^{b} \right)  
\text{Tr}_\Lambda \left( 
\hat e_{xy}^{b \dagger} 
\hat e_{vw}^{a } 
\right),
\end{split}\end{equation}
so indeed $\eta$ is invariant.\\

\noindent (\thePropList) \stepcounter{PropList}  $\eta(\mathcal A, \mathcal B) = 1$ when $[ \mathcal A, \mathcal B]=0$.\\

First note that, by construction, we are interested in $\mathcal A$ and $\mathcal B$ being algebras of local operators, which are finite-dimensional matrix algebras. If $[ \mathcal A, \mathcal B]=0$, one can choose a basis in which $\mathcal A$ is the set of local operators defined on a site $a$, and those in $\mathcal B$ are defined on site $b$. 
By the previous claims, we can then evaluate $\eta$ using $\Lambda = \{ a,b\}$ with $p_{\Lambda} = p_a p_b$. Now $\text{Tr}_{\Lambda}(\hat e_{ij}^{a \dagger} \hat e_{lm}^{b} ) = 
\text{Tr}_{a}(\hat e_{ij}^{a \dagger})\text{Tr}_{b}( \hat e_{lm}^{b} )  =\delta_{ij} \delta_{lm}
$, so
\begin{equation}\begin{split}\label{eq:}
\eta^2(\mathcal A,\mathcal B) = \frac{p_a p_b}{p_a^2 p_b^2} \sum_{i,j=1}^{p_a}\sum_{l,m=1}^{p_b} | \delta_{ij}\delta_{lm}|^2 =1.
\end{split}\end{equation}

\noindent (\thePropList) \stepcounter{PropList} For $\mathcal A = \mathcal B$, $\eta(\mathcal A,\mathcal B)=p_a$.\\

By the arguments in the previous claims, we can choose a basis and take $\Lambda = a$, and see that 
\begin{equation}\begin{split}\label{eq:}
\eta^2(\mathcal A,\mathcal A) = &
\frac{p_a \, p_a}{p_a^2} \sum_{i,j,l,m=1}^{p_a} \left |\text{Tr}_{a}(\hat e^{a \dagger}_{ij} \hat e^a_{lm}) \right|^2\\
=&p_a \sum_{j,m=1}^{p_a} \left |\text{Tr}_{a}(\hat e^{a }_{jm} ) \right|^2 = p_a^2.
\end{split}\end{equation}

Next we proceed to discuss properties of the chiral unitary index $\nu$:\\

\noindent (\thePropList) \stepcounter{PropList} For a `stacked' chain, $\nu(\hat Y_1 \otimes \hat Y_2) = \nu(\hat Y_1) + \nu(\hat Y_2)$.\\

This is an immediate consequence of the following property of the trace: $ \text{Tr} ( a_1 \otimes a_2) =   \text{Tr}_1(a_1) \text{Tr}_{2} (a_2)  $. \\

\noindent (\thePropList) \stepcounter{PropList}  $\nu(\hat Y' \hat Y) = \nu(\hat Y') + \nu(\hat Y)$.\\

The proof of this claim is actually somewhat involved, and we refer the readers to GNVW. In particular, we comment that this property, which is central for the group structure of the classification, is actually quite subtle, and deserves further elaboration.

As we have emphasized throughout, in obtaining a sensible definition of the chiral unitary index $\nu$, it is important that the Lieb-Robinson length $\ell_\text{LR}$ is much smaller than the system size $L$. However, this assumption is unpleasing from a mathematical point of view, since whenever $L$ is finite it is easy to find a finite product of locality-preserving unitary operators, say $\hat t^{L/2}$ with $L$ even, that violates this condition. Therefore, the classification $\nu \in (\log \mathbb Q_+,+)$ could only be sensible if the thermodynamic limit $L \rightarrow \infty$ is taken first. Yet, the many-body Hilbert space is not a mathematically well-defined concept in the thermodynamic limit, and so the formalism presented in the present work is non-rigorous. GNVW overcomes this difficulty by introducing the notion of `quantum cellular automata', which are automorphisms of the C*-closure of the observable algebras on the infinite chain. This can be viewed as a regularization of the problem by focusing exclusively on local operators (observables), as is natural in a physics problems, and studying only the transformations (automorphisms) among local operators. When these transformations are explicitly written out in a finite system, they correspond to conjugation by locality-preserving unitary operators, and hence our emphasis on the Lieb-Robinson length.

\section{Computing the GNVW index of an MPU}
\label{sec:MPU}

\subsection{Index formula in the MPU language}
As can be seen in the definition, our goal will be to compute $\eta(\hat Y(\mathcal A_{ L}) ,\mathcal A_{R})$ and $\eta(\mathcal A_{ L} ,\hat Y( \mathcal A_{ R}) )$  when a locality-preserving $\hat Y$ is given as an MPU.
Similar to discussions of matrix-product states (MPS), it will be most convenient to introduce a graphical representation of the equations, as was done in (Fig.~\ref{fig:MPU_Rep}a).
While we annotated each box by a `$M^{[x]}$' in Fig.~\ref{fig:MPU_Rep}a to emphasize that the tensors are site-dependent, as one would expect in a disordered system, in the following we will drop the annotation for clarity of the diagrams.
We will also represent a basis of the operator algebra at site $x$, $\hat e_{ij}^{[x]} \equiv | i_{x}\rangle \langle j_x |$ for $i_x,j_x = 1,\dots,p_x$, by 
\begin{equation}\begin{split}\label{eq:}
\begin{tabular}{cc}
$\hat e^{[x]}_{ij}$ = & 
\raisebox{-.5\height}{\includegraphics[width=0.02\textwidth]{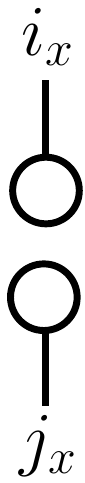}}
\end{tabular}.
\end{split}\end{equation}

Now consider evaluating the index by specifying a cut between sites $x$ and $x+1$, and using $\mathcal A_{ L} = \mathcal A_x$, $\mathcal A_{ R} = \mathcal A_{x+1}$. In evaluating $\eta (\hat Y(\mathcal A_x), \mathcal A_{x+1})$, the nontrivial task is to compute $\text{Tr}_{\Lambda} \left(   \hat Y \hat e_{i_x j_x}^{[x]} \hat Y^\dagger \hat e_{i_{x+1} j_{x+1}}^{[x+1]\dagger }\right)
\text{Tr}_{\Lambda} \left( \hat Y \hat e_{i_x j_x}^{[x] \dagger} \hat Y^\dagger \hat e_{i_{x+1} j_{x+1}}^{[x+1]} \right)
$, where the repeated indices are traced over. We will discuss below how to evaluate this quantity in the MPU language.

To this end, see that (Fig.~\ref{fig:MPU_Rep}c)
\begin{equation}\begin{split}\label{eq:}
&\text{Tr}_{\Lambda} \left( \hat Y \hat e_{i_x j_x}^{[x] \dagger} \hat Y^\dagger \hat e_{i_{x+1} j_{x+1}}^{[x+1]} \right)\\
=& 
\raisebox{-.45\height}{\includegraphics[width=0.4\textwidth]{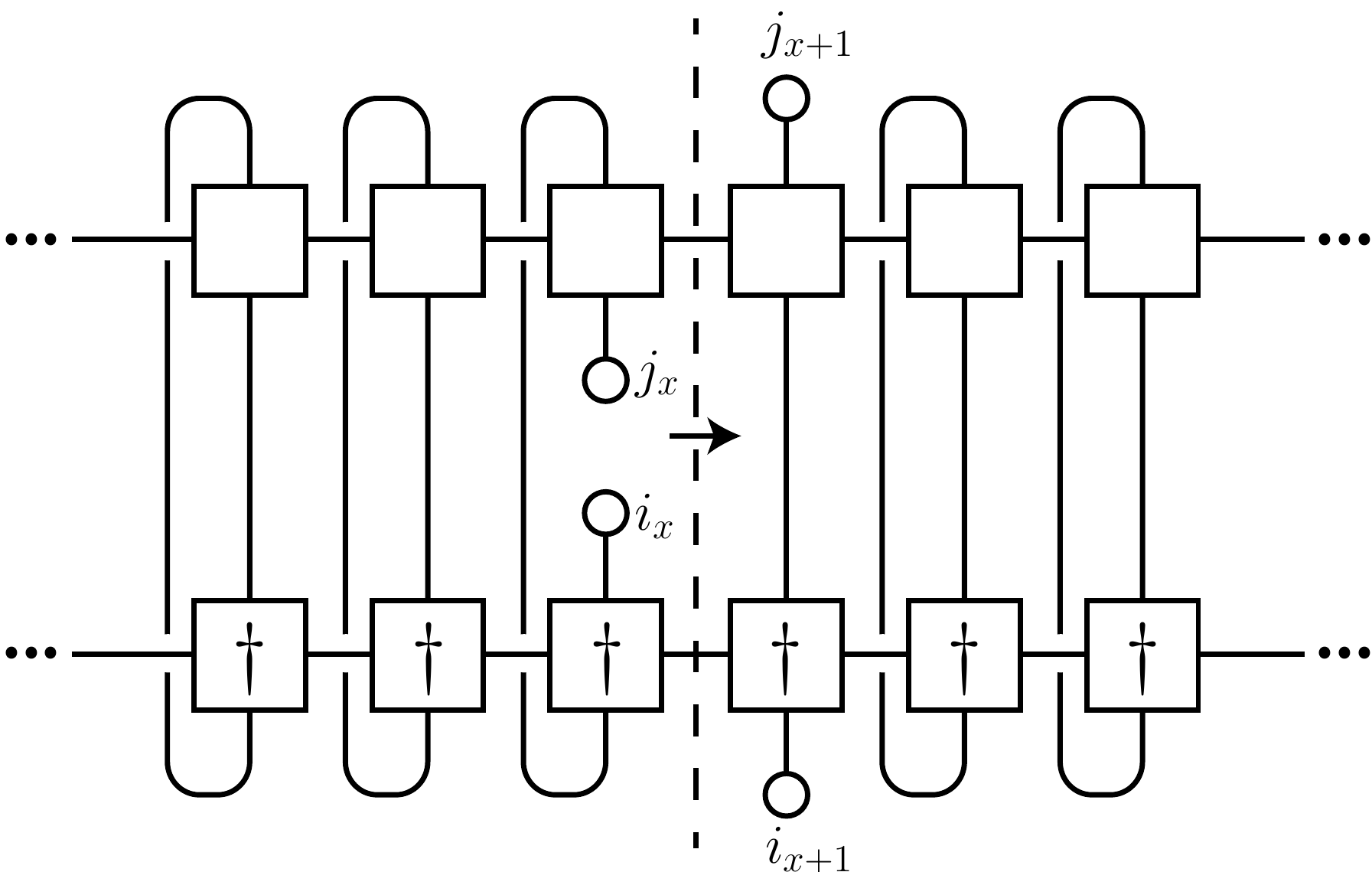}} \,,
\end{split}\end{equation}
where the vertical dashed line with an arrow indicates the location of the cut, and we have chosen $\Lambda$ to be the entire system, assumed to be a finite ring of size $L\gg 1$. Note the $\dagger$ annotation, indicating that the tensors appearing are those corresponding to $\hat Y^\dagger$. Also note that the tensors at different sites are generally distinct, and therefore $\hat Y$ is not translationally invaraint.

One can combine all the tensors for sites other than $x$ and $x+1$ into one single tensor, which only has legs in the `bond space'. Representing this combined tensor as a shaded box, we rewrite
\begin{equation}\begin{split}\label{eq:}
&\text{Tr}_{\Lambda} \left( \hat Y \hat e_{i_x j_x}^{[x] \dagger} \hat Y^\dagger \hat e_{i_{x+1} j_{x+1}}^{[x+1]} \right)\\
=& 
\raisebox{-.45\height}{\includegraphics[width=0.2\textwidth]{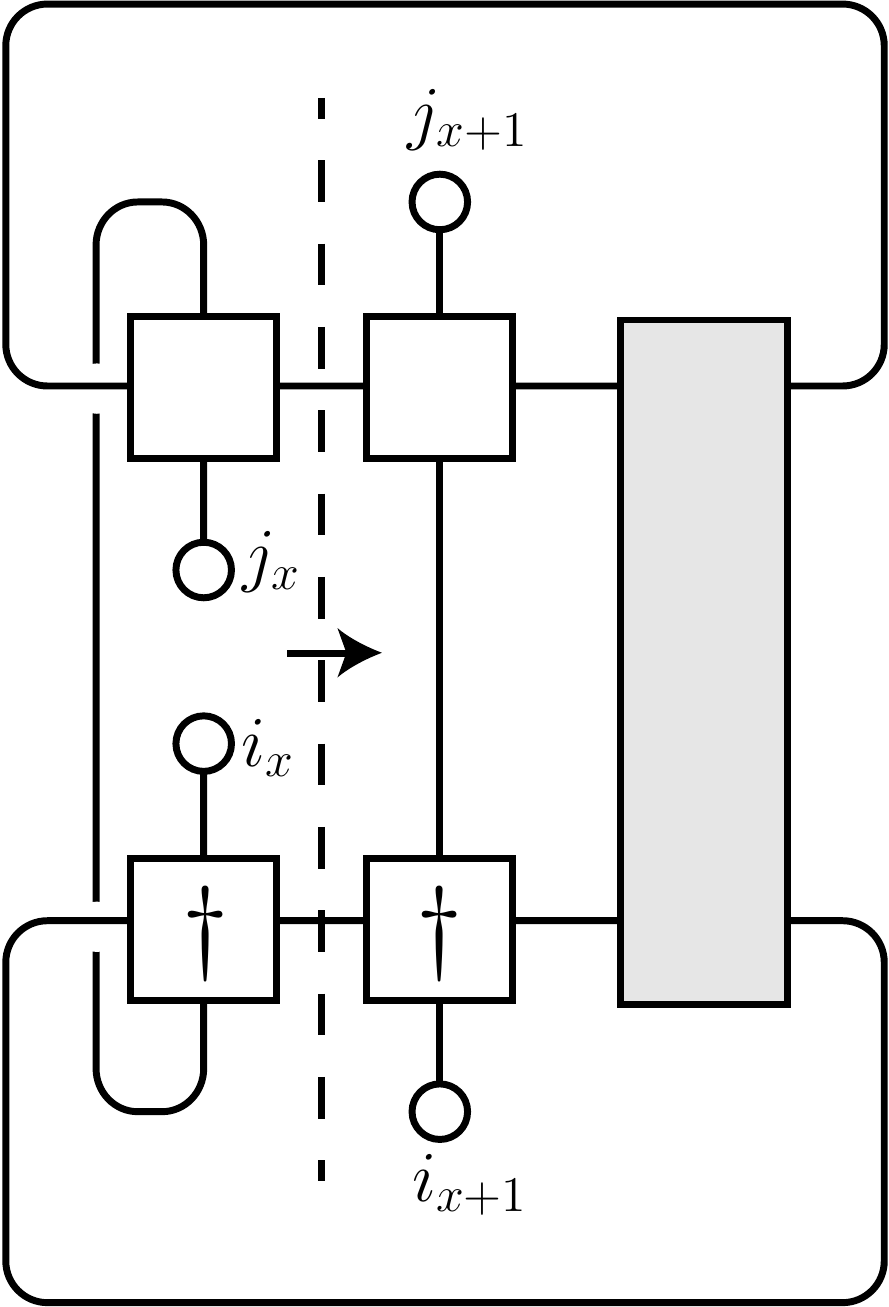}} 
=\sum_{\alpha} \raisebox{-.45\height}{\includegraphics[width=0.2\textwidth]{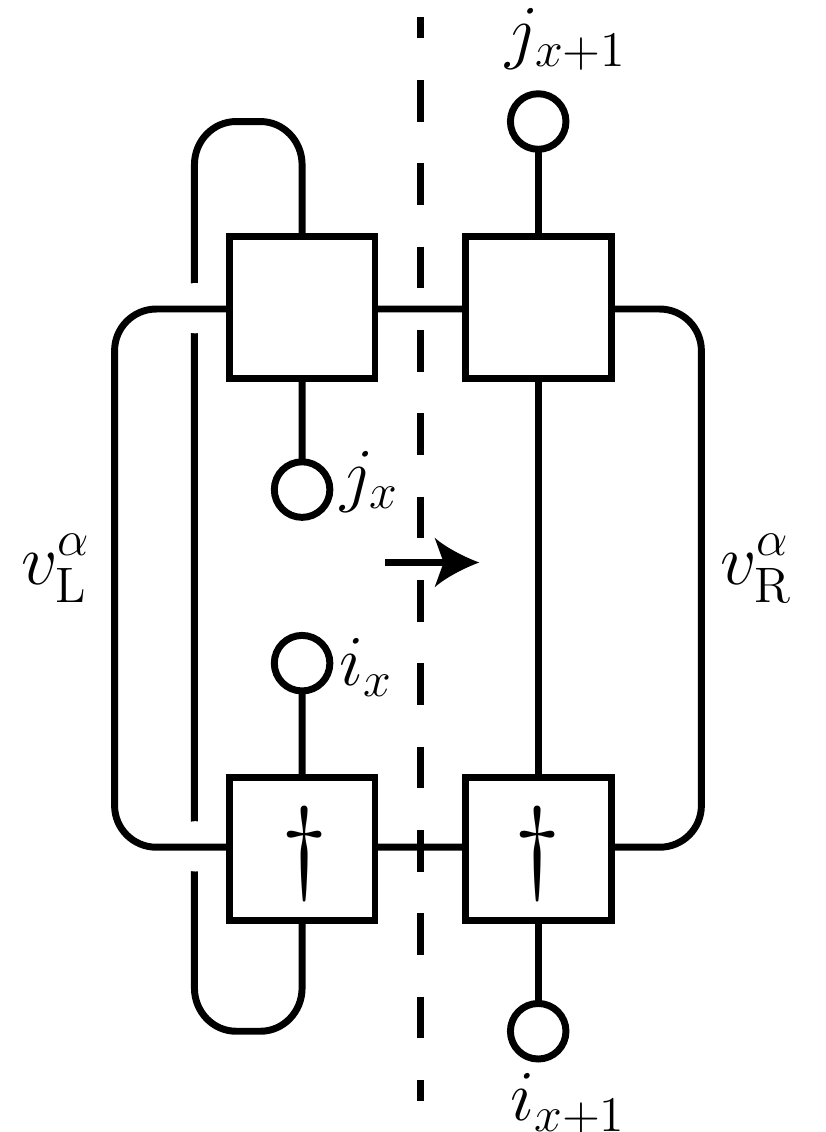}} 
\,,
\end{split}\end{equation}
where in the second equality sign we performed a singular-value decomposition (SVD) on the shaded tensor. In practice, for sufficiently large $L$ there is only one singular value to sum over -- far away from the cut (with distances measured in the Lieb-Robinson length), the operator $\hat Y \hat e_{i_x j_x}^{[x] \dagger} \hat Y^\dagger \hat e_{i_{x+1} j_{x+1}}^{[x+1]} $ is simply the identity (in principle with exponential accuracy), and therefore the evaluation of the trace can be done on a sufficiently large open interval containing the cut. Since such intervals correspond to the OBC, there should only be a single singular value in the SVD of the shaded tensor.

Finally, we evaluate (with summation convention)
\begin{equation}\begin{split}\label{eq:MPU_Contract}
&\text{Tr}_{\Lambda} \left(   \hat Y \hat e_{i_x j_x}^{[x]} \hat Y^\dagger \hat e_{i_{x+1} j_{x+1}}^{[x+1]\dagger }\right)
\text{Tr}_{\Lambda} \left( \hat Y \hat e_{i_x j_x}^{[x] \dagger} \hat Y^\dagger \hat e_{i_{x+1} j_{x+1}}^{[x+1]} \right) \\
=&\,
\raisebox{-.5\height}{\includegraphics[width=0.32\textwidth]{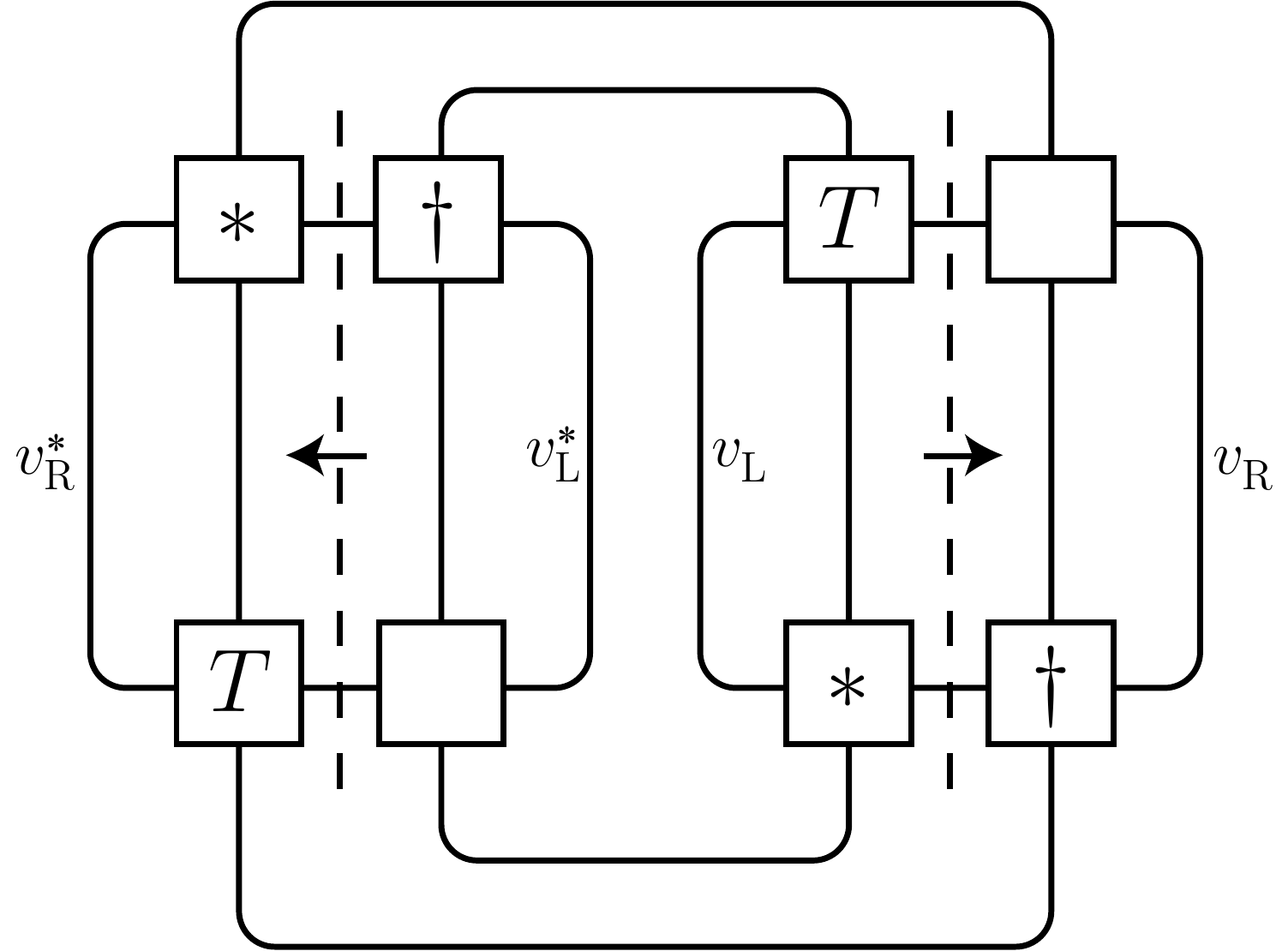}}\,\,,
\end{split}\end{equation}
where we used $T$ and $*$ to indicate that the tensors are those for $\hat Y^{T}$ and $\hat Y^*$. In addition, note that there is a further transposition in the bond space of the `block' of tensors on the left of the figure, as indicated by the flip of the arrow on the dashed line.

One can therefore evaluate $\nu (\hat Y)$ by contracting the tensor in Eq.~\eqref{eq:MPU_Contract}. 
However, as discussed in the main text one should choose the intervals $L$ and $R$ to be at least as large as the Lieb-Robinson length in order to obtain the correct index. This can be done by a direct generalization of the above discussion, but with the sites $x$ and $x+1$ replaced by  a collection of sites. As a technical remark, we also note that, for stability, it is important to restore the normalization factor in $\eta$ in the actual numerical evaluation, for if not the magnitude of the numerical values will diverge as the size of the quantum Hilbert space involved.

\subsection{Two-site MPUs}
To prepare for the construction of example MPUs used in the numerical computation of the chiral unitary index, we first discuss the construction of the building blocks used: random two-site MPUs.

Consider a general two-site unitary operator
\begin{equation}\begin{split}\label{eq:}
\hat U = \sum_{i_1,j_1=1}^{p_1} \sum_{i_2,j_2=1}^{p_2} 
U^{i_1,i_2}_{j_1,j_2} \, | i_1 i_2\rangle \langle j_1 j_2|,
\end{split}\end{equation}
where $U^{i_1,i_2}_{j_1,j_2}$ is a $p_1 p_2 $ dimensional unitary matrix. We use the notation that the upper indices are for the left-space, and the lower for the right. We seek to rewrite it in a MPU form, where we define tensors $M^{[x]}_{i_x j_x}$, $x=1,2$ and $i_x, j_x = 1,\dots, p_x$ such that 
\begin{equation}\begin{split}\label{eq:}
\hat U = \sum_{i_1,j_1=1}^{p_1} \sum_{i_2,j_2=1}^{p_2} 
M^{[1]}_{i_1,j_1} M^{[2]}_{i_2,j_2} \, | i_1 i_2\rangle \langle j_1 j_2|.
\end{split}\end{equation}
Since we have a two-site problem, $M^{[1]}_{i_1,j_1}$ is a $1\times \chi$ dimensional matrix for each pair of $i_1, j_1$, and $M^{[2]}_{i_2,j_2}$  is $\chi\times 1$ dimensional.  Similar to the canonicalization of MPS, we group the tensors into column and row (block-) vectors, and define a $p_1^2 \times p_2^2$ matrix $\tilde U$
\begin{equation}\begin{split}\label{eq:}
\tilde U \equiv& 
\left(
\begin{array}{c}
M^{[1]}_{1,1}\\  \vdots\\ M^{[1]}_{p_1,1}\\ M^{[1]}_{1,2}\\ \vdots\\ M^{[1]}_{p_1,p_1}
\end{array}
\right)
\left(
\begin{array}{cccccc}
M^{[2]}_{1,1} &  \dots & M^{[2]}_{p_2,1} & M^{[2]}_{1,2} &\dots & M^{[2]}_{p_2,p_2}
\end{array}
\right),
\end{split}\end{equation}
where by construction we have
\begin{equation}\begin{split}\label{eq:TFac}
(\tilde U)^{ab}_{cd} =& M^{[1]}_{a,b} M^{[2]}_{c,d} = U^{ac}_{bd}.
\end{split}\end{equation}
This shows that the tensors $M^{[x]}_{i,j}$ can be found by a matrix factorization of $\tilde U$, whose entries are fully determined by $U$. 
Note the crucial fact that $\tilde U$ is related to $U$ by a partial transpose, and generally cannot be transformed into one another via any row-column rearrangement (indeed their dimensions do not even match when $p_1\neq p_2$). This is important, for if the latter was true the MPU `Schmidt weights' would all be  $1/\sqrt{\chi}$, which is absurd if we are claiming full generality in our construction. 

To construct a generic two-site MPU, we (i) generate a random $p_1 p_2$-dimensional unitary matrix, (ii) rearrange the matrix elements as per Eq.~\eqref{eq:TFac}, and (iii) perform a matrix factorization (say SVD) to find the tensors  $M^{[x]}_{i,j}$. 
In practice, however, this procedure leads to an MPU representation of an FDLU with an undesirable large bond dimension. To see why, simply note that the number of non-zero singular values in step (iii) above will generically be $\min(p_1^2,p_2^2)$, and so the bond dimension of the resulting MPU will scale exponentially with the circuit depth and $p$.
To reduce the computation difficulty, we further engineer the MPU construction process to reduce the resultant bond dimension. Note that we lose the full generality of the constructed MPU once we bring the bond dimension down.

We will do this in two steps: First we will present a particular construction that, while controlling the bond dimension, does not represent a class of two-site MPUs containing the `generic' gates. Next we will generalize this simple construction to restore generality. Again, we emphasize that generality is only restored at the cost of allowing a bond dimension scaling as $p^2$.  

The first construction is loosely an analogue of designing a controlled-phase gate.
Let $\{\hat P_i ~|~ i = 1,\dots, \chi_1\}$ and $\{ \hat Q_j  ~|~ j = 1,\dots, \chi_2 \}$ be two sets of orthogonal projectors in $\mathcal H_1$ and $\mathcal H_2$ respectively, such that $\sum_{i=1}^{\chi_1} \hat P_i = \hat 1_1$ and $\sum_{j=1}^{\chi_2} \hat Q_j = \hat 1_2$. For (slightly) more generality we should choose the basis for the projectors (i.e. the vector subspaces) arbitrarily for the different gates. This can be achieved by multiplying the MPU below, defined with a simple projector basis, by a random on-site unitary $\hat U_1 \otimes \hat U_2$. 

We consider the matrices $\{ \hat L_{i,j},~\hat  R_{i,j}~|~ i=1,\dots, \chi_1;~ j=1,\dots,\chi_2\}$ satisfying
\begin{equation}\begin{split}\label{eq:}
\hat L_{i,j}^\dagger \hat L_{i',j'} = &
\left\{
\begin{array}{ccc}
\hat P_i & \text{for} & i=i', j=j'\\
0 & \text{for} & i\neq i' 
\end{array}
\right.;\\
\hat  R_{i,j}^\dagger \hat R_{i',j'} =&
\left\{
\begin{array}{ccc}
  \hat Q_j & \text{for} & i=i', j=j'\\
0 & \text{for}& j\neq j' 
\end{array}
\right.,
\end{split}\end{equation}
where the omitted cases (e.g.~$i=i'$, $j\neq j'$ for $\hat L_{i,j}$) are unconstrained. These requirements can be fulfilled by considering $\hat L_{i,j}$ that act only in the $\hat P_{i}$ subspace, and similarly for $\hat R_{i,j}$ in $\hat Q_j$.
Then we claim 
\begin{equation}\begin{split}\label{eq:}
\hat U = \sum_{i=1}^{\chi_1}\sum_{j=1}^{\chi_2} \hat L_{i,j} \otimes \hat R_{i,j}
\end{split}\end{equation}
is an MPU with bond-dimension $\chi_1 \chi_2$ (the actual bond dimension maybe even smaller after canonicalization). 
Indeed, check that
\begin{equation}\begin{split}\label{eq:}
\hat U^\dagger \hat U = & \sum_{i,i', j, j'} \left( \hat L_{i',j'}^\dagger \otimes \hat R_{i',j'}^\dagger\right)\left( \hat L_{i,j} \otimes \hat R_{i,j}\right)\\
=& \sum_{i, j }   \hat P_i   \otimes  \hat Q_j  = \hat 1.
\end{split}\end{equation}
Note that $\hat L_{i,j}$ is $j$-\emph{dependent} -- if not we will just end up with an MPU of bond dimension 1.

In the above construction, no matter how we partition our on-site Hilbert spaces we will never recover the fully generic two-site MPU. To improve this, we observe that the key points of the above construction are the following:
\begin{equation}\begin{split}\label{eq:}
\hat U = \sum_{i,j} \hat U_{i,j};~~~~~ 
\hat U_{i',j'}^\dagger  \hat U_{i,j} = \left( \hat P_i \otimes \hat Q_j \right) \delta_{i',i} \delta_{j',j}.
\end{split}\end{equation}
The second condition reduces to the usual unitary condition when $\chi_1 = \chi_2 = 1$, and the previous construction corresponds to the `trivial' solution to this condition, akin to restricting oneself to the on-site unitary subgroup of the two-site unitaries. 
It is now clear how to generate more general MPU while restricting their bond dimensions -- we simply use the more general solution, as discussed near Eq.~\eqref{eq:TFac}, for the second condition. Explicitly, we now consider
\begin{equation}\begin{split}\label{eq:MPU_chiRes}
\hat U = \sum_{i=1}^{\chi_1}\sum_{j=1}^{\chi_2} \sum_{\alpha=1}^{\min(r_i^2,\, r_j^2)}\hat L_{i,j}^{\alpha} \otimes \hat R_{i,j}^{\alpha},
\end{split}\end{equation}
where $r_i = {\rm rank} (\hat P_i)$ and $r_j = {\rm rank} (\hat Q_j)$, i.e.~the dimensions of the vector subspaces defined by the projectors, and we demand
\begin{equation}\begin{split}\label{eq:}
 \sum_{\alpha=1}^{\min(r_i^2,\,r_j^2)}
 (\hat L_{i,j}^{\alpha \dagger}  \hat L_{i,j}^{\alpha} )\otimes (\hat R_{i,j}^{\alpha \dagger} \hat R_{i,j}^{\alpha}) =& \hat P_i \otimes \hat Q_j.
\end{split}\end{equation}
This gives an MPU with bond dimension $\chi =\sum_{i,j} \min(r_i^2,r_j^2) $, which is restricted by how we partition the site Hilbert spaces. At the same time, we recover the fully general case when $\chi_1 = \chi_2=1$, at the cost of restoring the bond dimension $\min(p_1^2,p_2^2) $.

The discussion above can be readily generalized to MPUs that act on more than two sites, simply by a re-grouping of the DOF into two `super' sites. In particular, it will be interesting to explore if the above approach in reducing bond dimension, which manifestly preserves unitarity, offers any practical advantage in time evolution simulation of 1d quantum systems.

\subsection{Construction of MPUs used in Fig.~\ref{fig:GNVW}}
\begin{figure}[h]
\begin{center}
{\includegraphics[width=0.45 \textwidth]{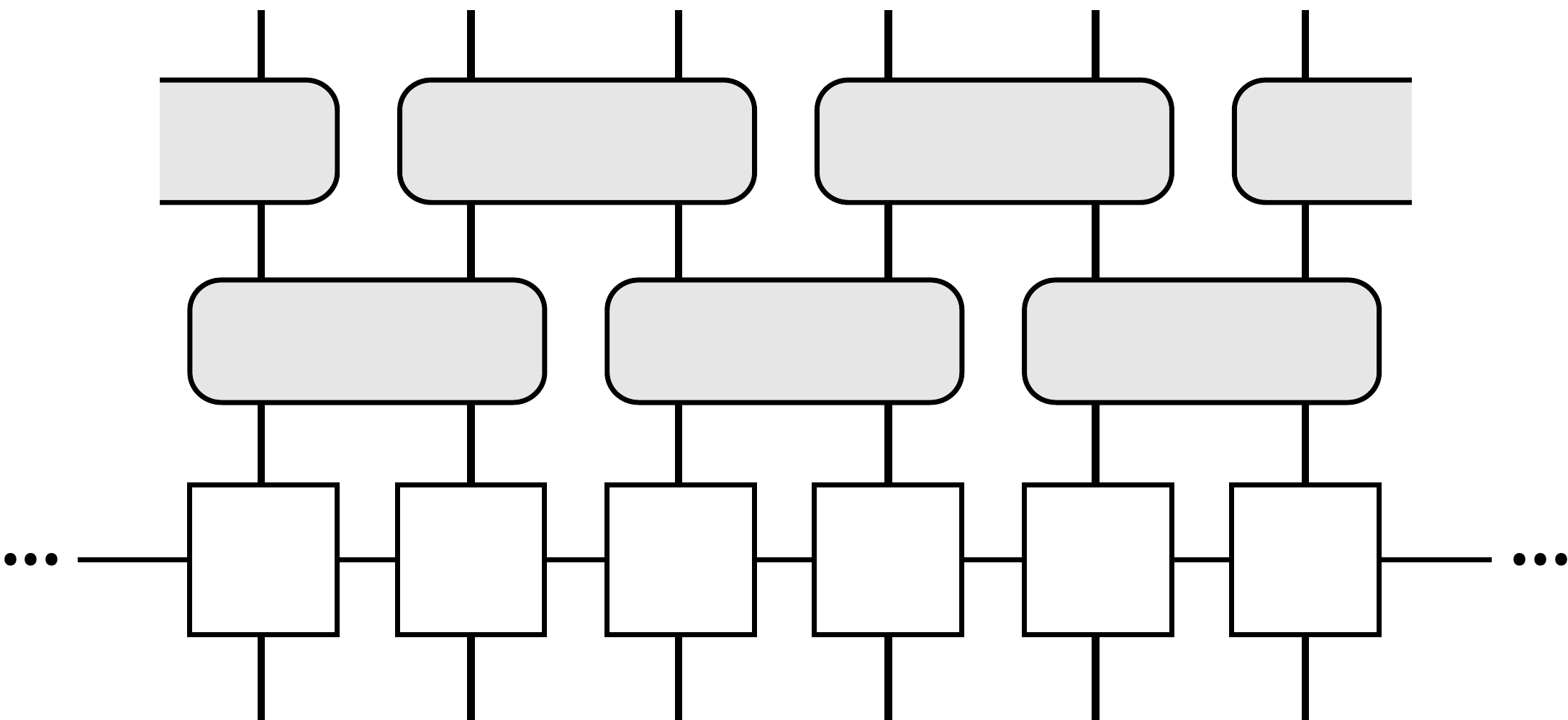}} 
\caption{{\bf `Dressing' of a `base' MPU with a known chiral unitary index.} The lowest layer (unshaded) represents a `base' MPU constructed by taking copies of the translation operators. The upper layers (shaded) represent a simple FDLU. Each shaded box corresponds to a \emph{different} random unitary operator, so the system is disordered.
\label{fig:MPU_Circuit}
 }
\end{center}
\end{figure}

To demonstrate the computability and quantization of the chiral unitary index $\nu$, we consider some example MPUs taking the form in Fig.~\ref{fig:MPU_Circuit}.

The lowest layer (unshaded boxes) represent a `base' MPU with a known index. Explicitly, we take it to be either the identity operator, or (copies of) the translation operators for different site Hilbert space dimensions $p$. As discussed in the main text, they can be represented as MPUs. 

The upper layers (shaded boxes) represent an FDLU multiplied to the `base' MPU. While we show an FDLU of depth two in Fig.~\ref{fig:MPU_Circuit}, in certain cases we considered FDLUs of depth four (see Table \ref{tab:MPU_Setting}).
Each shaded box is a random two-site unitary, for which we have detailed their MPU representation in the previous subsection.
Note also that, despite the appearance of the figure, the system is \emph{not} translationally invariant as each shaded box represents a different random gate. 

Due to the simple architecture of the circuit, one can also directly read off their Lieb-Robinson lengths: One simply follows the paths upward in Fig.~\ref{fig:MPU_Circuit} and try to go as far as possible in one direction. For instance, if the `base' is the identity operator, with our FDLU of depth two an operator local at site $0$ can only be nontrivial in the interval $[-2,2]$, so $\ell_\text{LR}=2$. In addition, the `Lieb-Robinson light-cone' is strict for such circuits, i.e.~the statement that $\hat Y \hat O_{x} \hat Y^\dagger$ is nontrivial only within a distance $\ell_\text{LR}$ from $x$ is exactly true.

The settings for the various example MPUs used in Fig.~\ref{fig:GNVW} are summarized in Table \ref{tab:MPU_Setting}. As can be seen in the last column of the table, generic two-site MPUs were used in the FDLU for all sub-figures other than (f). For (f), we use two-site MPUs of the form in Eq.~\eqref{eq:MPU_chiRes}, where for each gate (a shaded box in Fig.~\ref{fig:MPU_Circuit}) we randomly pick either the left or the right site as being labeled by `1', and the other by `2'. We consider $\chi_1 = 3$ and $\chi_2=1$, where the six-dimensional site Hilbert space for site $1$ is partitioned into $6=2+2+2$. 
Since the `base' MPU $\hat t^{(2)} \otimes \overline{ \hat  t^{(3)} }$ has bond dimension $6$, this brings the total MPU bond dimension down from $6^3 = 216$ to $6 \times (3\times 4) = 72$.
Finally, note that in Figs.~\ref{fig:GNVW} (b) and (c), the evaluated index for certain cuts converge earlier than the other. This is due to an even-odd effect, imprinted by the arbitrary choice of the `starting point' of the first layer of FDLU in Fig.~\ref{fig:MPU_Circuit}, and is merely an unimportant artifact of our simple circuit architecture. 

\begin{center}
\begin{table}
\caption{{\bf Settings for the MPU used in Fig.~\ref{fig:MPU_Circuit}.} 
$L$ denotes the total length of the ring used in the computation, and `\#. cut' denotes the number of cuts at which the chiral unitary index was computed.
$p$ denotes the dimension of the site Hilbert space, `base' and `depth' respectively denote the choice in the `base MPU' and the depth of the FDLU, as illustrated in Fig.~\ref{fig:MPU_Circuit}. $\ell_\text{LR}$ indicates the Lieb-Robinson length of the MPU. `Generic' indicates whether or not fully generic two-site MPUs are used in the FDLU. 
\label{tab:MPU_Setting}
}
\begin{tabular}{cccccccc}
~&$L$ & \#. cut & $p$ & Base & Depth & $\ell_\text{LR}$ & Generic  \\
\hline
(a) & $128$ & $10$ & $2$ & $\hat 1$  & $4$ & $4$ & $\checkmark$\\
(b) & $128$ & $10$ & $2$ & $\hat t^{(2)} $  & $4$ & $5$ & $\checkmark$\\
(c) & $128$ & $10$ & $2$ & $\overline{\hat t^{(2)} } $ & $4$ & $5$ & $\checkmark$\\
(d) & $64$ & $1$ & $2\times 2$ & ~$\hat t^{(2)} \otimes \overline{\hat  t^{(2)} }$~  & $2$ & $3$ & $\checkmark$\\
(e) & $64$ & $1$ & $2\times 2$ & ~$\hat t^{(2)} \otimes  \hat  t^{(2)} $~& $2$ & $3$ & $\checkmark$\\
(f) & $64$ & $1$ & $2\times 3$ & ~$\hat t^{(2)} \otimes\overline{  \hat  t^{(3)} }$~ & $2$& $3$  & $\times$
\end{tabular}
\end{table}
\end{center}

\section{Review of noninteracting fermionic AFAI}
\label{sec:AFAI}
In this appendix, we review the key properties of the free fermion AFAI proposed in Refs.~[\onlinecite{Demler,RudnerLevin,Lindner}]. 
In equilibrium systems, a chiral edge mode is anomalous, i.e.~it cannot exist as a purely 1d system. To see this, imagine a clean single-band 1d model, with the band structure $E_k$ satisfying $ d E_k/ d k >0$ for all values of $k$. This implies $E_{\pi} > E_{- \pi}$, violating the periodicity of the Brillouin zone. At the edge of a 2d bulk, however, the contradiction is resolved, as the chiral mode can terminate in the bulk bands, which will necessarily carry non-zero Chern numbers.

For a Floquet system with the time-periodic single-particle Hamiltonian $H(t) = H(t+T) $, however, the instantaneous Hamiltonian plays a secondary role, and the long-time behavior is governed by the single-particle Floquet operator $U_{\rm F} \equiv \mathcal T e^{- i \int_0^T \, dt H(t)  }$. The eigenvalues of $U_{\rm F}$ are phases, and can be expressed as $e^{- i \epsilon_n T}$. $\{ \epsilon_n \}$ are known as the quasi-energies, and are only defined modulo $2\pi/T$. Interestingly, since the energy direction is also periodic, the previous argument forbidding a stand alone chiral mode in 1d is now invalid. For instance, the straight chiral mode $\epsilon_k =  k /T$ gives $\epsilon_\pi = \pi/T = \epsilon_{-\pi} + 2\pi/T$, which seems to give a legitimate quasi-energy band.

Nonetheless, such a chiral mode is still anomalous.\cite{RudnerLevin} This can be seen by noting that the winding number $\mathcal W = \frac{i}{2\pi }  \int dk \,\text{Tr} \,\left( U^\dagger \partial_k U \right)$ is a topological invariant quantized to be an integer.\cite{RudnerLevin} While the single-band model with $\epsilon_k =  k /T$ gives $\mathcal W = 1$, the identity operator $U = 1$ corresponds to $\mathcal W = 0$. For a purely 1d system, there exists a smooth family of time evolution operators $U(t)$ interpolating between $U(0)=1$ and $U(T)  = U_{\rm F}$. Since the winding number is robust against smooth deformation, the winding number of any purely 1d Floquet system is necessarily zero.\cite{RudnerLevin} Hence, a chiral mode, signified by a non-zero $\mathcal W$, remains anomalous, and can only arise as the boundary of a 2d system.
In contrast to the equilibrium case, such chiral modes can terminate themselves, and so their presence no longer implies non-zero Chern numbers of the bulk bands. The bulk of the system is therefore Anderson localizable, giving rise to AFAI.

While the definition of the winding number $\mathcal W$ can be adapted for disordered systems,\cite{Lindner} its generalization to interacting systems is unclear. 
As a first step towards such generalization, it is instructive to look instead at the many-body Floquet operator corresponding to a chiral edge mode. For the straight chiral mode, one finds
\begin{equation}\begin{split}\label{eq:AFAIEdge}
\hat U_{\rm F} = \prod_{k} e^{- i \epsilon_k T \hat c_k^\dagger \hat c_k} = e^{- i \sum_{k} k \hat c_k^\dagger \hat c_k} = e^{- i \hat P},
\end{split}\end{equation}
which is nothing other than the unit translation operator. 
This simple observation bridges the fermionic AFAI model with the chiral bosonic model we constructed, which also features the (bosonic) translation operator at the edge. 
Although our current formalism does not immediately apply to fermionic systems, it is suggestive that the fermionic translation operator is as anomalous as its bosonic counterpart. 
Therefore, to rigorously establish the stability of the fermionic AFAI against introduction of interactions, it only remains to prove that the fermionic translation operator is anomalous, which we leave as an important open question.

\section{Chiral models of Majorana fermions}
\label{sec:Majorana}
We will again consider the four-step driving protocol defined on the checkerboard lattice, except that each site now hosts a complex fermion with creation operator $\hat c_{\vec x}^\dagger$, corresponding to Majorana fermions
\begin{equation}\begin{split}\label{eq:}
\hat \chi_{\vec x} \equiv \hat c_{\vec x} + \hat c_{\vec x}^\dagger ;~~~
\hat {\bar \chi}_{\vec x} \equiv \frac{1}{i} (\hat c_{\vec x} - \hat c_{\vec x}^\dagger ).
\end{split}\end{equation}
We will first consider a model that acts only on $\hat \chi_{\vec x}$ but not $\hat {\bar  \chi}_{\vec x}$. Similar to Eq.~\eqref{eq:Hs_Spin}, at each step $s$ of duration $T/4$ we `turn on' a collection of bonds defined by $\vec b_s$, and consider a Hamiltonian that exchanges $\hat \chi_{\vec r_{\rm A}} \leftrightarrow \hat \chi_{\vec r_{\rm B}+ \vec b_s}$. This is achieved by the noninteracting Hamiltonian
\begin{align}\label{eq:M_Hs}
\hat H&_{(s)} = \frac{ i  \pi  }{ T} \sum_{\vec r } \hat \chi_{\vec r_{\rm A}} \hat \chi_{\vec r_{\rm B} + \vec b_s},
\end{align}
for which the corresponding time-evolution operator $\hat U_{(s)}$ gives $\hat U_{(s)} \hat \chi_{\vec r_{\rm A}} \hat U^\dagger_{(s)} = -\hat \chi_{\vec r_{\rm B} + \vec b_s} $ and $\hat U_{(s)} \hat \chi_{\vec r_{\rm B} } \hat U^\dagger_{(s)} = \hat \chi_{\vec r_{\rm A}- \vec b_s} $. When the system is defined with PBC, a similar analysis gives $\hat U_{\rm F} \hat \chi_{\vec x} \hat U^\dagger_{\rm F} = \hat \chi_{\vec x}$ for all $\vec x$. As the system is noninteracting, the transformation of the complete set of Majorana operators fully specify the many-body Floquet operator, and  this implies $\hat U_{\rm F} = \hat 1$.

With OBC, a chiral edge is again exposed.  For instance, with the same boundary condition as defined near Eq.~\eqref{eq:Bose_Edge}, one finds
\begin{equation}\begin{split}\label{eq:M_trans_2}
\hat U_{\rm F}' \hat \chi_{(x,1)_{\rm A}} (\hat U'_{\rm F})^{ \dagger}
= - \hat \chi_{(x+1,1)_{\rm A}},
\end{split}\end{equation}
which acts as the Majorana translation operator along the edge up to sign change. Similar to the discussion on the experimental proposal in Sec.~\ref{sec:ExpP}, once disorder is introduced to render the bulk MBL, such sign is immaterial to the topological nature of the model. More concretely, we again consider appending to the driving protocol a disordering fifth step (and suitably rescaling the energy scale of the previous steps such that $\hat U_{(s)}$ for $s=1,\dots, 4$ are unchanged), with the Hamiltonian 
\begin{equation}\begin{split}\label{eq:}
\hat H_{(5)} = \sum_{\vec x \in \Lambda} i \frac{5 J_x}{2 T} \hat \chi_{\vec x}  \hat {\bar \chi}_{\vec x}.
\end{split}\end{equation}
This leads to a MBL Floquet Hamiltonian with $ \{ i \chi_{\vec x} \hat {\bar \chi}_{\vec x} \}$ being the l-bits.
In particular, for the special case of $J_x = \pi$, the corresponding time-evolution gives $\hat U_{(5)} \hat \chi_{\vec x} \hat U_{(5)}^\dagger = - \hat \chi_{\vec x} $, and therefore canceling the negative sign in Eq.~\eqref{eq:M_trans_2}. Again this gives a smooth path connecting the edge operator in Eq.~\eqref{eq:M_trans_2} to the Majorana translation operator while maintaining the MBL nature of the bulk, and so justifies the claim that the model above has its topological nature characterized by the anomaly of the Majorana translation operator.

Finally, we note that a pair of counter-propagating Majorana translation operators is trivial, i.e.~it can arise form the the finite-time evolution of a purely 1d local Hamiltonian. Such triviality follows directly from the corresponding argument for bosons in Fig.~\ref{fig:SWAP}c, where one simply replaces the (bosonic) SWAP gate by the Majorana counterpart discussed near Eq.~\eqref{eq:M_Hs}.

\section{Connections to cohomological classification}
\label{sec:Coho}
In this appendix, we review previous results in the cohomology classification of MBL Floquet SPT phases. We stress that the chiral bosonic phases cannot be captured by the devloped cohomology description. In addition, we will point out here that the proposed classification is plausibly invalid for a 1d system with ${\rm U}(1)$ symmetry. Interestingly, for this case the cohomology classification could be linked to the chiral bosonic phases in 2d.

For bosons, previous classification results argue that an MBL Floquet system in $d$ spaital dimensions with on-site unitary, abelian symmetry $G$ is classified by the group cohomology $H^{d+1}(G \times \mathbb Z, {\rm U}(1) )$, where the extra factor of $\mathbb Z$ corresponds to the discrete time-translation invariance of a Floquet system.\cite{ElseNayak,PotterMorimotoVishwanath} Applying the K{\"u}nneth formula one finds
\begin{equation}\begin{split}\label{eq:FSPT_X}
H^{d+1}(G \times \mathbb Z, {\rm U}(1)) = H^{d+1}(G, {\rm U}(1)) \times H^{d}(G, {\rm U}(1)) .
\end{split}\end{equation}
The first factor $H^{d+1}(G, {\rm U}(1)) $ is identical to the static SPT classification in the same dimension, and can be phrased as a `strong index'. Physically, one can interpret such Floquet phases as being directly connected to the static (MBL) SPT phases.
The second factor $H^{d}(G, {\rm U}(1))$ is present only because of the extra factor of $\mathbb Z$ on the left hand side, which suggests that it corresponds to some `intrinsically Floquet' physics. Indeed, each nontrivial entry of that factor corresponds to the pumping of one $(d-1)$-dimensional SPT phase to the boundary in a Floquet cycle, and can be viewed as a `pumped index' that is meaningful only when the time periodicity is maintained, akin to the corresponding discussion of `weak' SPTs protected by lattice translation invariance.

More concretely, we first restrict our attention to 1d problems, for which we expect a Floquet phase that `pumps' a 0d SPT (charges) to the edge in every cycle. As with the previously studied cases, one expects a physical picture in which, locally, symmetry charges are uniformly pushed to one direction in every cycle. While (symmetry) charge neutrality is nonetheless maintained in the bulk, charges accumulate at the edges. Another way to rephrase this is that, while the full system Floquet operator will necessarily realize the symmetries linearly, the two edge operators can each feature projective representations that `cancel' as a whole.
Note that the notion of an `edge operator' above invoked the MBL assumption.

To illustrate this explicitly, consider a simplified version of the models discussed in Refs.~[\onlinecite{Keyserlingk,PotterMorimotoVishwanath}].
Consider a driven 1d chain of spin-1/2's, $\Lambda$, and denote the site Pauli operators by $\hat X_x$, $\hat Y_x$ and $\hat Z_x$. Suppose the system has a $\mathbb Z_2$ symmetry generated by $\hat Z = \prod_{x \in \Lambda} \hat Z_x$, such that the Floquet operator $\hat U_{\rm F} = \prod_{x\in \Lambda}  (\hat X_{x} \hat X_{x+1})$ verifies $[\hat Z, \hat U_{\rm F}]=0$. With PBC, one simply finds $\hat U_{\rm F} = \hat 1$, so the Floquet Hamiltonian is trivial and the `strong index' is $0$. With OBC, however, one gets $\hat U_{\rm F}^{\rm OBC} = \hat X_{1} \hat X_{L}$, which features a projective representation on each edge ($[\hat Z, \hat X_{1} ] \neq 0$) and indicates the existence of a nontrivial MBL Floquet SPT phase. Breaking the discrete time-translation, say when one imagines doubling the period of the drive such that $\hat U_{\rm F}' = \hat U_{\rm F}^2$, one arrives back at a trivial edge. This implies a $\mathbb Z_2$ classification, as dictated by the `pumped' factor in Eq.~\eqref{eq:FSPT_X}, $H^{1}(\mathbb Z_2,{\rm U}(1)) = \mathbb Z_2$.

When $G$ is continuous, however, one has to be more careful in applying this equation, since one might need to specify further continuity conditions on top of the group multiplication structure.
For the equilibrium case, a proposed classification is via the notion of `Borel' cohomology,\cite{XieGuLiuWen} which has the property
\begin{equation}\begin{split}\label{eq:}
H^{d+1}_{\text{Borel}}(G,{\rm U}(1) ) \simeq
H^{d+2}(BG,\mathbb Z ),
\end{split}\end{equation}
where $BG$ is the classifying space. This is motivated by the corresponding relation for ordinary group cohomology of discrete groups: $H^{d+1}(G,{\rm U}(1)) \simeq H^{d+2}(G,\mathbb Z) \simeq H^{d+2}(BG,\mathbb Z)$.
However, this modified classification is known to be incomplete when $G={\rm U}(1)$, since it fails to capture the chiral phases in 2D (the so called `$E_8$' states).\cite{XieGuLiuWen} 
Nonetheless, the results are still meaningful as they provide a partial classification, in the sense that entries in $H^{d+1}_{\text{Borel}}(G,{\rm U}(1) )$ do correspond to distinct SPT phases.

From the equilibrium results, one expects that Floquet systems with continuous unitary symmetry $G$ are now (partially) classified by
\begin{equation}\begin{split}\label{eq:}
H^{d+1}_{\text{Borel}}(\mathbb Z \times G, {\rm U}(1)) 
= H^{d+1}_{\text{Borel}}(G, {\rm U}(1)) \times H^{d}_{\text{Borel}}(G, {\rm U}(1)),
\end{split}\end{equation}
with a similar interpretation as before. Of particular interest to us here is, as stated, when $G= {\rm U}(1)$, for which one finds\cite{XieGuLiuWen}
\begin{equation}\begin{split}\label{eq:}
H^{d-1}_{\text{Borel}}({\rm U}(1),{\rm U}(1))=
H^{d}(\mathbb C\rm{P}^{\infty}, \mathbb Z) = 
\left\{
\begin{array}{ccc}
\mathbb Z & \text{for} & d~\text{even}\\
0 & \text{for} & d~\text{odd}
\end{array}
\right.
.
\end{split}\end{equation}
As such, we have
\begin{equation}\begin{split}\label{eq:}
H^{0+1}_{\text{Borel}}(\mathbb Z \times {\rm U}(1), {\rm U}(1) ) =& \mathbb Z \times \{ 0 \}\, ;\\
H^{1+1}_{\text{Borel}}(\mathbb Z \times {\rm U}(1), {\rm U}(1) ) =& \{ 0 \}  \times \mathbb Z\,;\\
H^{2+1}_{\text{Borel}}(\mathbb Z \times {\rm U}(1), {\rm U}(1) ) =&  \mathbb Z \times \{ 0 \} \,;\\
\vdots~~~~~~~~~~~~&
\end{split}\end{equation}
i.e.~all of them are $\mathbb Z$, but the physical interpretation alternates between a `strong' and a `pumped' SPT.

Now we focus again on $d=1$, for which the $\mathbb Z$ classification above is expected to correspond to a `pumped' MBL Floquet SPT phase in 1d, in which particles (carrier of the ${\rm U}(1)$ charge) are unidirectionally pumped in one direction in every Floquet cycle. 
More explicitly, we relate the above $\mathbb Z$ classification to projective representations of ${\rm U}(1) \times \mathbb Z$. 
Let $(\phi, n)\in {\rm U}(1)\times \mathbb Z$, with group multiplication in additive notation, and let $[(\phi,n)]$ be the (projective) representation of $(\phi,n)$. (In the more common multiplicative notation, one writes $e^{i \phi} \in {\rm U}(1)$; in our notation here $\phi$ is defined $\mod 2 \pi$.)
Consider the factor system $\omega$ defined via:
\begin{equation}\begin{split}\label{eq:}
&[(\phi_1,n_1)] [(\phi_2,n_2)] \\
=& \omega((\phi_1,n_1), (\phi_2,n_2)) [(\phi_1+ \phi_2, n_1+n_2)],
\end{split}\end{equation}
with the following co-cycle:
\begin{equation}\begin{split}\label{eq:}
\omega_{\nu}((\phi_1,n_1), (\phi_2,n_2))  = \exp(i \nu \phi_1 n_2).
\end{split}\end{equation}
This corresponds to the following modification of the commutator (as ${\rm U}(1) \times \mathbb Z$ is abelian, projective representations are captured by nontrivial commutators):
\begin{equation}\begin{split}\label{eq:ProjRepMath}
&[(\phi,0)] [(0,1)] [(-\phi,0)] [(0,-1)] \\
=& \exp(i 2 \nu \phi)  [(\phi,1)] [(-\phi,-1)] \\
=& \exp(i  \nu \phi)  [(0,0)],
\end{split}\end{equation}
which is indeed projective iff $\nu\neq 0$.

Now let us rephrase the above using a more physical notation. Let $\hat Q$ be the Hermitian charge operator generating ${\rm U}(1)$, and using the bulk-boundary correspondence in Sec.~\ref{sec:Edge}, we write our Floquet operator, which generates the $\mathbb Z$ factor, for an open 1d spin chain as a product of three unitary operators 
\begin{equation}\begin{split}\label{eq:}
\hat U_{\rm F} = \hat y_{\rm L} \hat u_{\rm B} \hat y_{\rm R},
\end{split}\end{equation}
where we assume $\hat y_{\rm L}$ ($\hat y_{\rm R}$ ) has a finite support localized on the left (right) edge, and $\hat u_{\rm B}$ corresponds to the bulk operator. Following the above discussion, we consider a system for which the following commutation relations are realized:
\begin{equation}\begin{split}\label{eq:}
e^{i \phi \hat Q} \hat y_{\rm L} e^{- i \phi \hat Q} =& e^{i  \phi \nu} \hat y_{\rm L};\\
e^{i \phi \hat Q} \hat u_{\rm B} e^{- i \phi \hat Q} =&  \hat u_{\rm B};\\
e^{i \phi \hat Q} \hat y_{\rm R} e^{- i \phi \hat Q} =& e^{-i  \phi \nu} \hat y_{\rm R}.
\end{split}\end{equation}
Note that the first line above is nothing but a translation of Eq.~\eqref{eq:ProjRepMath} into the more physical notation. In addition, we assume $\hat u_{\rm B}$ is MBL with the local charge operators $\{ \hat Q_x\}$ being the l-bits, such that one can write $\hat u_{\rm B} = \exp(- i \sum_r F_r (\{ \hat Q_x\}))$ with $F_r$ being local real functions. As a consequence, there is a stronger version of the commutation relation: $e^{i \phi \hat Q_x} \hat u_{\rm B} e^{- i \phi \hat Q_x} = \hat u_{\rm B}$ for all sites $x$.

By the localization assumption, we can consider the local charge operator $\hat Q_{L}$, which  is the truncation of $\hat Q$ to the support of $\hat y_{\rm L}$. 
Now consider any state $\hat \rho$ of the system, with an average charge
$
q_{\rm L}\equiv \text{Tr}\left( \hat \rho \hat Q_{\rm L}  \right) 
$
on the left edge.
By assumption $[\hat Q_{\rm L}, \hat u_{\rm B}] =[\hat Q_{\rm L} , \hat y_{\rm R}] =  0$, and since $[ \hat Q - \hat Q_{L}, \hat y_{\rm L}] = 0$, the commutation relation with $\hat y_{\rm L}$ is refined to $ e^{i \phi \hat Q_{\rm L} } \hat y_{\rm L} e^{- i \phi \hat Q_{\rm L} } =  e^{i  \phi \nu}\hat y_{\rm L}$.
So under time evolution governed by $\hat U_{\rm F}$, one finds
\begin{equation}\begin{split}\label{eq:}
 \text{Tr}\left( \hat U_{\rm F} \hat \rho\hat U_{\rm F}^\dagger e^{ i \phi \hat Q_{L}} \right) 
 = & \text{Tr}\left( \hat \rho \,
\hat y_{\rm L}^\dagger  e^{i \phi \hat Q_{L}} \hat y_{\rm L}
 \right) \\
=&
\text{Tr}\left( \hat \rho \, e^{i \phi (\nu + \hat Q_{\rm L})}
 \right) =e^{ i \phi (\nu + q_{\rm L})},
\end{split}\end{equation}
implying an accumulation of charges on the left edge when $\nu\neq 0$, which is independent of the choice of $\hat \rho$.

As argued, such symmetry charge accumulation is common for all `pumped' bosonic MBL Floquet SPT phases in 1d, protected by a unitary abelian on-site symmetry. Unlike the previous case with $G=\mathbb Z_2$, however, the ${\rm U}(1)$ charges are unbounded, and so the nontrivial nature of the edge cannot be neutralized by adding subharmonic terms to the Floquet drive. In the long-time limit, therefore, infinite opposite charges will accumulate at the two ends of the chain, regardless of the initial state. Though we haven't proven that such systems are indeed anomalous, they are absurd on physical grounds -- while the site Hilbert spaces could be finite-dimensional in the bulk, the edge Hilbert spaces are necessarily infinite-dimensional in order for the Floquet operator to remain unitary.

It is therefore suggestive that the cohomology classification of MBL Flouqet phases might be invalid when $G = {\rm U}(1)$, or more generally when the group $G$ is continuous, despite we have not provided any concrete evidence leading to this claim. 
For instance, one might argue that we should start with a system having all site Hilbert spaces being infinite-dimensional, say in a quantum rotor model. In that setup, however, it is unclear why the bulk Floquet operator cannot be restricted into a finite dimensional subspace using ${\rm U}(1)$ conservation. 

Another possible resolution is that the discussion is not self-consistent. For instance, it could be that such a system can 
never be MBL, and hence does not exhibit a bulk-edge decoupling.
Given the symmetry group ${\rm U}(1)$ is abelian, it is rather surprising if there is such an obstruction to many-body localization. Yet, we have already seen a similar obstruction -- we argued that the chiral unitary operators $\hat Y$ ($\nu (\hat Y) \neq 0$) is anomalous, and hence cannot be localized.
In fact, the above discussion, concerning the unidirectional pumping of charges in a state-independent manner, can be viewed as a ${\rm U}(1)$-enriched version of our discussion of chiral unitary operators in 1d. 
This serves as a potential link between cohomological SPT classifications and chiral phases, and exploration of this link will be an interesting direction for future studies.

\end{appendix}

\bibliography{MBLCF}

\end{document}